\documentclass[aps,prd,twocolumn,nofootinbib,superscriptaddress,longbibliography]{revtex4-2}

\usepackage[T1]{fontenc}
\usepackage[utf8]{inputenc}
\usepackage{amsfonts}
\usepackage{amsmath}
\usepackage{amssymb}
\usepackage{amsthm}
\usepackage{booktabs}
\usepackage{bm}
\usepackage{graphicx}
\usepackage{mathrsfs}
\usepackage{mathtools}
\usepackage{slashed}
\usepackage{tikz}
\usepackage{xcolor}
\usetikzlibrary{decorations.markings}
\usepackage[
  hypertexnames=false,
  colorlinks=true,
  linkcolor=blue,
  citecolor=blue,
  urlcolor=blue
]{hyperref}

\newcommand{\sD}{\mathscr{D}}

\newcommand{\meff}{m_{\mathrm{eff}}}
\newcommand{\doilink}[1]{\href{https://doi.org/#1}{\nolinkurl{doi:#1}}}
\newcommand{\arxivlink}[1]{\href{https://arxiv.org/abs/#1}{\nolinkurl{arXiv:#1}}}

\numberwithin{equation}{section}
\renewcommand{\theequation}{\arabic{section}.\arabic{equation}}
\renewcommand{\theHequation}{\arabic{section}.\arabic{equation}}

\begin{document}

\title{Four-Fermion Condensates in Curved Spacetimes:\ A Functional Approach}

\author{Stephon Alexander}
\email{stephon\_alexander@brown.edu}
\affiliation{Department of Physics, Brown University, Providence, RI 02912, USA}

\author{Heliudson Bernardo}
\email{heliudson.bernardo@uleth.ca}
\affiliation{Department of Physics \& Astronomy and Quantum Horizons Alberta, University of Lethbridge, 4401 University Drive, Lethbridge AB, T1K 3M4, Canada}

\author{Jiatong Yan}
\email{jiatongyan1106@gmail.com}
\affiliation{Department of Physics, University of Wisconsin-Madison, 1150 University Avenue, Madison, WI 53706, USA}

\begin{abstract}
Four-fermion interactions appear in effective descriptions of particle
physics, many-body systems, and gravitational theories with fermions.
Although the same local operator may enter perturbative scattering, a
vacuum Nambu--Jona-Lasinio (NJL) instability, or finite-density
Bardeen--Cooper--Schrieffer (BCS) pairing, these regimes are distinguished
by their interaction channels, quadratic kernels, and quantum states.  We
give a pedagogical functional account of these distinctions and compute
the local one-loop contribution of a scalar-channel NJL mean field to the
energy-momentum tensor in curved spacetime.  We first display the
state data in the in-out functional and construct the closed-time-path
functional required for an in-in expectation value. For the NJL saddle, a Hubbard--Stratonovich field shifts the fermion mass,
and the parity-even Dirac determinant generates local volume,
curvature, and curvature-squared operators.  We find the covariant quantum effective action
before specializing to a spatially flat FLRW background and derive the
corresponding energy density and pressure. The constant-condensate limit agrees
with the direct flat-space mean-field calculation.  We also explain which
additional state and channel data are required for finite-density BCS
pairing and comment on renormalization conditions in curved spacetimes.
\end{abstract}

\maketitle

\section{Introduction}
\label{sec:introduction}

In semiclassical gravity, quantum matter affects the geometry through an
expectation value of its energy-momentum tensor.  This statement is simple
to write but contains two logically distinct inputs.  The local ultraviolet
part of the expectation value is fixed by the field content and the
covariant operator that governs its fluctuations, whereas its finite,
state-dependent part depends on the quantum state and on how that state is
evolved. The same separation appears in the curved-space effective
action, which in general depends both on the state and the bare action. This underlies the effective action approach, where quantum effects renormalize independent gravitational couplings, such as the Einstein--Hilbert and curvature-squared operators
\cite{BirrellDavies1982,ParkerToms2009,BuchbinderShapiro2021} and the cosmological constant \cite{Weinberg1989,Martin2012, Padilla:2015aaa, Bernardo:2022cck}.

Four-fermion interactions provide a useful setting in which to make these
distinctions explicit.  In four spacetime dimensions a local four-fermion
operator has mass dimension six and is naturally interpreted within an
effective field theory with a specified ultraviolet scale.  Its
nonperturbative scalar channel underlies the Nambu--Jona-Lasinio (NJL) mechanism
of dynamical mass generation
\cite{NambuJonaLasinio1961a,NambuJonaLasinio1961b,Klevansky1992}.  NJL-type
models have since been used as controlled low-energy models of chiral
symmetry breaking and dense fermionic matter, with their regulator and
matching prescription treated as part of the definition of the effective
theory \cite{InagakiMutaOdintsov1997,Buballa2005}.

The occurrence of the same local operator in distinct regimes can obscure
their physical differences.  At weak coupling and for scattering boundary
conditions, a four-fermion interaction produces a perturbative contact vertex.  With vacuum state
data and a sufficiently attractive scalar channel, a self-consistent
particle--antiparticle expectation value
\(\langle\bar\psi\psi\rangle\) can instead generate an NJL mass.  At finite
density, an attractive projected Cooper channel produces a
particle--particle expectation value \(\langle\psi\psi\rangle\) and a
gap near a Fermi surface, as explained by the Bardeen-Cooper-Schrieffer (BCS) theory 
\cite{BardeenCooperSchrieffer1957,Gorkov1958,Nambu1960}.  The latter case
does not follow from the sign of the coupling alone: the
Fermi surface, channel projection, and state-occupation data are essential
\cite{Polchinski1992,Shankar1994,RajagopalWilczek2000, Alford:2007xm}.

These mechanisms have several applications in cosmology.  Integrating out
nondynamical torsion produces local fermion contact interactions, commonly
in an axial-current channel
\cite{Hehl1976,FreidelMinicTakeuchi2005}.  Relating such an interaction to a
scalar NJL channel therefore requires a Fierz rearrangement and a stated
channel approximation.  Vacuum chiral condensates and torsion-induced
interactions have been investigated as ingredients of dark-energy,
inflationary, reheating, and nonsingular cosmologies
\cite{AlexanderVaid2006,Poplawski2011,Poplawski2012,Magueijo:2012ug, Weller2013,
QuintanarMacorra2015I,QuintanarMacorra2015,LucatProkopec2017,Tukhashvili2024, Alexander:2025whu, Alexander:2026}.  Finite-density or explicitly BCS-inspired
constructions form a related but distinct application
\cite{AlexanderCalcagni2009,AlexanderBiswas2009,Alexander2017,Alexander:2020wpm,Tong2024, Alexander:2024qml, Liang:2024xww}.
However, the current literature often does not emphasize the distinction between vacuum and finite-density condensation, and no unified approach to these two cases is provided.

Since both the NJL and BCS mechanisms are intrinsically quantum, in applications to cosmology one is unavoidably (although sometimes not explicitly) led to consider the quantum expectation value of the fermionic energy-momentum tensor. In this context, the difference between the in and out states becomes important. In asymptotically flat problems, the in-out effective action organizes scattering amplitudes
and local one-loop counterterms. A cosmological source, however, is an
expectation value in an initial state and must be computed with an in-in,
or closed-time-path, Schwinger-Keldysh functional. The latter doubles the fields and sources,
contains the initial density matrix, and glues the two histories at a final
time \cite{Schwinger1961,Keldysh1965,Jordan1986,CalzettaHu1987}. However, the local
short-distance coefficients are common to the two formalisms, provided the
same covariant regulator is used, whereas their finite state-dependent parts
need not agree. This point becomes especially important in a time-dependent
geometry or in the presence of a chemical potential. 

The main goal of this paper is to provide a self-contained  functional setup to compute the local expectation value of the energy-momentum tensor of a scalar-channel condensate in curved spacetimes. Within the path integral approach, we also aim to clarify the differences between the vacuum and finite-density condensation cases. We first review where scalar and fermionic states
enter the path integral, and then separate contact scattering, vacuum NJL
condensation, and finite-density BCS pairing.  We review the free scalar and
Dirac fermion calculations so that the normalization, signs, degeneracy factors,
and dimensional regularization are fixed before introducing the four-fermion
interaction.  We then derive the flat-space mean-field energy density and
pressure, construct the curved-space one-loop action, and specialize its
covariant metric variation to a spatially flat Friedmann-Lema\^itre-Robertson-Walker (FLRW) background.  A general
homogeneous derivative expansion is given alongside the constant-saddle
result computed explicitly from heat kernel methods. We hope that the level of scrutiny adopted here will help improve the contextualization of fermionic models in cosmology and will equip the community with theoretical tools for new applications.

This work is organized as follows. We explain the state-dependent functional setup underlying the calculations of this paper in Section~\ref{sec:functional-setup}. In Section~\ref{sec:free-fields}, we discuss the free-field
results. In Sec. \ref{sec:njl-one-loop}, we review the quantum effective action for the NJL model in flat spacetime and investigate its generalization to curved spaces. In
Section~\ref{sec:flrw}, we specialize the covariant answer to FLRW geometry,
and in Sec.~\ref{sec:discussion} we comment on our assumptions and future
directions. Many technical details are discussed in the appendices, including a first-principles field-theory derivation of the grand-canonical potential for the BCS pairing case. We use metric signature \((-+++)\) and units \(\hbar=c=1\).

\section{Functional Setup for Vacuum and Finite-Density Fermions}
\label{sec:functional-setup}

In the context of semiclassical gravity, the energy-momentum tensor that appears in the semiclassical Einstein
equations is an expectation value in a specified quantum state that typically satisfies some physically motivated requirements. The action for the theory alone does not fix the state in
which the expectation value is taken. The state enters through
endpoint wavefunctionals or, more generally, through an initial density
matrix.  We therefore begin by displaying these data explicitly.  This
will allow us to distinguish an in-out transition amplitude from the
in-in expectation value needed in cosmology before introducing any
four-fermion coupling.

For a Lorentzian generating functional coupled to a source \(J\), we
write
\begin{equation}
 Z[J;g]\equiv e^{iW[J;g]},
 \qquad
 W[J;g]\equiv-i\ln Z[J;g], 
 \label{eq:W-definition}
\end{equation}
where $J(x)$ is (schematically) the relevant source and $g$ is the spacetime metric. For fermions, the source represents
the pair \((\eta,\bar\eta)\), whereas on the closed-time-path contour it
denotes the doubled sources \((J_+,J_-)\) as we shall discuss. We define
the energy-momentum tensor by
\begin{equation}
 T_{\mu\nu}
 =-
 \frac{2}{\sqrt{-g}}
 \frac{\delta S_{\rm m}}{\delta g^{\mu\nu}},
 \label{eq:emt-definition-functional}
\end{equation}
where \(S_{\rm m}\) is the matter action.  Equation
\eqref{eq:emt-definition-functional} makes covariance manifest and is
the main reason for organizing the calculation in terms of generating
functionals.

\subsection{States and in-out transition amplitudes}
\label{subsec:states-in-out}

Consider first a real scalar field \(\Phi\) coupled to an external
source \(J\).  Let \(\varphi_i(\boldsymbol{x})\) and
\(\varphi_f(\boldsymbol{x})\) denote field configurations on some
initial and final Cauchy surfaces, respectively.  The transition
amplitude between an initial state \(|\Psi_i\rangle\) and a final state
\(|\Psi_f\rangle\) is
\begin{widetext}
\begin{align}
 Z_{\Psi_f,\Psi_i}[J;g]
 &=
 \int [d\varphi_fd\varphi_i]\,
 \Psi_f^*[\varphi_f]\Psi_i[\varphi_i]
 \int_{\Phi(t_i)=\varphi_i}^{\Phi(t_f)=\varphi_f}
 {\cal D}\Phi \exp\left\{iS_\Phi[g,\Phi]+i\int_{t_i}^{t_f}d^4x\sqrt{-g}\,J\Phi\right\}.
 \label{eq:lorentzian-state-functional}
\end{align}
\end{widetext}
Here \(S_\Phi\) is a generic local covariant scalar action; the
minimally coupled choice used for free fields is given in
Sec.~\ref{subsec:free-scalar-review}.
The two wavefunctionals $\Psi_{i,f}$ are the boundary data that specify which matrix
element the path integral computes.

For example, inserting a local functional \({\cal O}[\Phi](x)\) gives
the normalized in-out matrix element
\begin{align}
 \langle {\cal O}(x)\rangle_{\Psi_f,\Psi_i}^{\rm in\text{-}out}
 &\equiv
 \frac{\langle\Psi_f|\widehat{\cal O}(x)|\Psi_i\rangle}
 {\langle\Psi_f|\Psi_i\rangle}
 \nonumber\\
 &=
 \frac{1}{Z_{\Psi_f,\Psi_i}[0;g]}
 \int_{\Psi_i}^{\Psi_f}{\cal D}\Phi\,
 {\cal O}[\Phi](x)e^{iS_\Phi[g,\Phi]} .
 \label{eq:normalized-in-out-matrix-element}
\end{align}
The notation
\(\int_{\Psi_i}^{\Psi_f}{\cal D}\Phi\) in
Eq.~\eqref{eq:normalized-in-out-matrix-element} is shorthand for the
three integrations displayed explicitly in
Eq.~\eqref{eq:lorentzian-state-functional}: one integrates over the
initial configuration \(\varphi_i\), the final configuration
\(\varphi_f\), and all bulk histories satisfying
\(\Phi(t_i)=\varphi_i\) and \(\Phi(t_f)=\varphi_f\), with weights
\(\Psi_i[\varphi_i]\) and \(\Psi_f^*[\varphi_f]\).
For \({\cal O}=\Phi\), the same result follows from
\begin{equation}
 \langle\Phi(x)\rangle_{\Psi_f,\Psi_i}^{\rm in\text{-}out}
 =
 \left.
 \frac{1}{i}\frac{\delta_{\rm cov}\ln Z_{\Psi_f,\Psi_i}[J;g]}
 {\delta J(x)}
 \right|_{J=0},
\end{equation}
where $\delta_{\rm cov}/\delta J \equiv 1/\sqrt{-g} \; \delta/\delta J$. This illustrates that, as far as correlation functions are concerned, the source \(J\) is only a correlator-generating device, and should be
smooth and compactly supported or sufficiently rapidly decreasing. Note that if the source were to remain
nonzero at an asymptotic endpoint then it would change the Hamiltonian relative to
which the endpoint state is defined.

Choosing vacuum wavefunctionals and taking
\(t_i\rightarrow-\infty\), \(t_f\rightarrow+\infty\) gives the usual
vacuum-to-vacuum functional.  In this case, one can also adopt the Feynman
\(i\epsilon\) prescription as a compact way of encoding the vacuum boundary
conditions. For a quasifree scalar state on a Cauchy surface \(\Sigma\), the state
dependence can be displayed schematically as
\begin{align}
 \Psi_0[\varphi;\Sigma]
 &=
 {\cal N}_\Sigma
 \exp\Bigg[-\frac12
 \int_\Sigma d^3x\sqrt{h(x)}
 \int_\Sigma d^3y\sqrt{h(y)}
 \nonumber\\
 &\hspace{3.3cm}\times
 \varphi(x){\cal D}_\Sigma(x,y)\varphi(y)
 \Bigg],
 \label{eq:scalar-quasifree-wavefunctional}
\end{align}
where \(h_{ij}\) is the induced spatial metric and
\({\cal D}_\Sigma\) is fixed by a choice of positive-frequency
subspace \cite{Symanzik1981,LongShore1998}. A preferred ground state may only be identified for stationary spacetimes, whereas on a general curved spacetime
this is not generically possible.  Thus \({\cal D}_\Sigma\), or its density-matrix analogue,
is part of the physical specification of the problem.

Although the in-out matrix element is useful for scattering and for organizing
the local one-loop effective action, it cannot be used for the computation of expectation values on a given time slice in general. For instance, in a cosmological context, one is typically interested in the
expectation value of an observable in a state that evolves from the
initial Cauchy hypersurface. In particular, imposing an independently chosen
final state can make an in-out energy-momentum tensor complex and
acausal.  A semiclassical cosmological source instead requires the
same initial state on the bra and ket states.  This observation
leads naturally to the closed-time-path construction, discussed next.

\subsection{In-in expectation values and the closed-time-path contour}
\label{subsec:ctp-in-in}

Let \(\varrho_i\) be a normalized density operator on the initial
hypersurface.  The closed-time-path (CTP), or Schwinger--Keldysh,
functional can be written as
\begin{equation}
 Z_{\varrho_i}[J_+,J_-;g_+,g_-]= \operatorname{Tr}\!\left[\varrho_i\,U_{J_-,g_-}^{\dagger}(t_f,t_i)U_{J_+,g_+}(t_f,t_i)\right].
 \label{eq:ctp-functional}
\end{equation}
The \(+\) branch evolves the ket forward from \(t_i\) to \(t_f\), and
the \(-\) branch evolves the bra backward from \(t_f\) to \(t_i\).
The trace identifies the two field configurations at \(t_f\).  There
is therefore one initial density matrix and a final gluing condition
\cite{Schwinger1961,Keldysh1965,CalzettaHu1987}.

For the scalar field, this statement is explicit in
\begin{widetext}
\begin{align}
 Z_{\varrho_i}[J_+,J_-;g_+,g_-]
 &=
 \int [d\Phi_+ d\Phi_-]\,
 \varrho_i[\varphi_i^+,\varphi_i^-]\,
 \delta\!\left[\Phi_+(t_f)-\Phi_-(t_f)\right]
 \nonumber\\
 &\quad\times
 \exp\Bigg\{
 iS_\Phi[g_+,\Phi_+]-iS_\Phi[g_-,\Phi_-]
 +i\int d^4x\sqrt{-g_+}\,J_+\Phi_+
 -i\int d^4x\sqrt{-g_-}\,J_-\Phi_-
 \Bigg\}.
 \label{eq:scalar-ctp-density-functional}
\end{align}
\end{widetext}
For a pure initial state,
\(\varrho_i[\varphi_i^+,\varphi_i^-]
=\Psi_i[\varphi_i^+]\Psi_i^*[\varphi_i^-]\).  The relative signs in
Eq.~\eqref{eq:scalar-ctp-density-functional} follow from the opposite
orientation of the two branches.  At equal metrics and sources,
unitarity gives
\begin{equation}
 Z_{\varrho_i}[J,J;g,g]=\operatorname{Tr}\varrho_i=1.
 \label{eq:ctp-unitarity}
\end{equation}

Derivatives with respect to the two sources generate a matrix of connected real-time correlators.  If
\(s_+=+1\) and \(s_-=-1\), then
\begin{equation}
 G^{ab}(x,x')
 =
 \frac{1}{(is_a)(is_b)}
 \frac{\delta^2_{\rm cov}\ln Z_{\varrho_i}}
 {\delta J_a(x)\delta J_b(x')},
 \qquad a,b=\pm .
 \label{eq:ctp-propagator-matrix}
\end{equation}
The \(++\) entry reduces to the Feynman propagator in the vacuum when $\varrho_i$ is the vacuum density matrix, while
for an occupied state every component contains the occupation data
specified by \(\varrho_i\).  Sources and metrics on the two branches
are kept independent while differentiating and are set equal only in
the physical limit.

The connected functional on this contour is
\begin{equation}
 W_{\rm CTP}[J_+,J_-;g_+,g_-]
 \equiv
 -i\ln Z_{\varrho_i}[J_+,J_-;g_+,g_-],
 \label{eq:ctp-W-definition}
\end{equation}
where the subscript ``CTP'' labels the closed-time-path contour while the subscript \(\varrho_i\) on \(Z\) records
the initial state, which is held fixed when the bulk sources and
metrics are varied.

With the convention in Eq.~\eqref{eq:emt-definition-functional}, the
in-in energy-momentum tensor is
\begin{equation}
 \langle T_{\mu\nu}(x)\rangle_{\varrho_i}
 =-
 \frac{2}{\sqrt{-g(x)}}
 \left.
 \frac{\delta W_{\rm CTP}}
 {\delta g_+^{\mu\nu}(x)}
 \right|_{g_+=g_-=g\,,\,J_+=J_-=0} .
 \label{eq:ctp-emt-main}
\end{equation}
The initial density-matrix kernel and the geometric data on its initial
hypersurface are held fixed in this metric variation.  If the state
preparation itself is varied, its boundary contribution must be included
separately.
This is the causal expectation value that sources the semiclassical
Einstein equations \cite{Jordan1986,HuVerdaguer2008}.  For Hadamard
states, the local ultraviolet counterterms in the CTP and in-out
descriptions are determined by the same coincidence-limit differential
operator.  Their finite state-dependent and nonlocal parts need not
agree.  In a time-dependent background, the latter include, for
example, particle-production \cite{BirrellDavies1982,ParkerToms2009} and nonlocal effects \cite{CalzettaHu1987,HuVerdaguer2008} that cannot be inferred from a local in-out derivative expansion.

The Schwinger-Keldysh construction therefore emphasizes two questions that will
remain distinct below.  The short-distance operator determines which
local gravitational terms require renormalization, while the initial
density matrix determines the finite state-dependent source.  To apply
this separation to fermions, the endpoint states must be written in a
basis adapted to a first-order action.  Grassmann coherent states
provide precisely this basis.

\subsection{Fermionic functional integral and four-fermion regimes}
\label{subsec:fermionic-functional-regimes}

For each fermionic mode, let \(|\zeta\rangle\) and
\(\langle\bar\zeta|\) be Grassmann coherent states.  Their resolution
of the identity is \cite{Shankar1994}
\begin{equation}
 \mathbf 1
 =
 \int d\bar\zeta\,d\zeta\,
 e^{-\bar\zeta\zeta}
 |\zeta\rangle\langle\bar\zeta|,
 \label{eq:fermion-coherent-identity}
\end{equation}
where the exponential compensates the nonorthogonal overlap
\(\langle\bar\zeta|\zeta'\rangle=e^{\bar\zeta\zeta'}\)
\cite{Berezin1966}.  Time slicing with this identity gives, in the
presence of Grassmann sources \(\eta\) and \(\bar\eta\),
\begin{widetext}
\begin{align}
 Z_{\Psi_f,\Psi_i}[\eta,\bar\eta;g]
 &=
 \int [d\bar\zeta_fd\zeta_f
 d\bar\zeta_id\zeta_i]\,
 e^{-\bar\zeta_f\zeta_f-\bar\zeta_i\zeta_i}
 \langle\Psi_f|\zeta_f\rangle
 \langle\bar\zeta_i|\Psi_i\rangle
 \nonumber\\
 &\quad\times
 \int_{\psi(t_i)=\zeta_i}^{\bar\psi(t_f)=\bar\zeta_f}
 [d\psi d\bar\psi]\,
 \exp\left\{
 iS_\psi[g,\bar\psi,\psi]
 +i\int d^4x\sqrt{-g}\,
 (\bar\eta\psi+\bar\psi\eta)
 \right\}.
 \label{eq:fermion-coherent-functional}
\end{align}
\end{widetext}
Only \(\psi\) at the initial surface and \(\bar\psi\) at the final
surface are fixed because the Dirac action is first order.  The
coherent-state construction consequently contains the standard
endpoint contribution and the Gaussian factors displayed in
Eq.~\eqref{eq:fermion-coherent-functional}.  These ingredients are
fixed by the coherent-state decomposition and are the same for every
physical state.  The state dependence resides instead in the overlaps
with \(|\Psi_i\rangle\) and \(|\Psi_f\rangle\).

For a general mixed initial state, define the fermionic coherent-state
density-matrix kernel by
\begin{equation}
 \varrho_i^{(F)}[\bar\zeta_i^+,\zeta_i^-]
 \equiv
 \langle\bar\zeta_i^+|\widehat\varrho_i|\zeta_i^-\rangle .
 \label{eq:fermion-density-kernel}
\end{equation}
The corresponding CTP functional can be written compactly as
\begin{widetext}
\begin{align}
 Z_{\varrho_i}^{(F)}
 [\eta_\pm,\bar\eta_\pm;g_\pm]
 &=
 \int_{\rm CTP}[d\psi_\pm d\bar\psi_\pm]\,
 \varrho_i^{(F)}
 [\bar\psi_+(t_i),\psi_-(t_i)]
 \nonumber\\
 &\quad\times
 \exp\Bigg\{
 iS_\psi[g_+,\bar\psi_+,\psi_+]
 -iS_\psi[g_-,\bar\psi_-,\psi_-]
 +i\sum_{a=\pm}s_a
 \int d^4x\sqrt{-g_a}\,
 \big(\bar\eta_a\psi_a+\bar\psi_a\eta_a\big)
 \Bigg\},
 \quad s_\pm=\pm1 .
 \label{eq:fermion-ctp-density-functional}
\end{align}
\end{widetext}
Here \(\int_{\rm CTP}\) includes the coherent-state resolution factors
and the final-time trace gluing.  For
\(\widehat\varrho_i=|\Psi_i\rangle\langle\Psi_i|\), the kernel
factorizes into
\(\langle\bar\zeta_i^+|\Psi_i\rangle
 \langle\Psi_i|\zeta_i^-\rangle\).

It is important to keep this state prescription distinct from any eventual
quadratic approximation used in the action. A vacuum, a filled Fermi
sea, and a paired state correspond to different density-matrix
kernels. As we discuss later in this section, a shifted NJL mass or an off-diagonal BCS gap can appear in the
quadratic bulk operator only after the interacting theory is expanded
about the corresponding mean-field saddle. It is a dynamical
consequence of that saddle and completely compatible with the structure of the Schwinger-Keldysh functional integral.

We now introduce the scalar-channel four-fermion theory.  It is useful
to define the coupling
\begin{equation}
 \lambda\equiv\frac{1}{M_{\rm 4F}^2},
 \qquad [\lambda]=-2,
 \label{eq:lambda-fourfermion-definition}
\end{equation}
where \(M_{\rm 4F}\) is the scale suppressing the dimension-six
operator.  In the conventions used throughout this paper,
\begin{align}
 S_{\rm 4F}
 & =
 \int d^4x\sqrt{-g}\left[
 \bar\psi\sD_m\psi
 +\lambda(\bar\psi\psi)^2
 \right],
 \label{eq:sectionII-fourfermion}
 \\
 \sD_m
 &=-\gamma^\mu\nabla_\mu+m .
\end{align}
We use\footnote{For the \((+---)\) signature and an
explicit \(i\) factor in the kinetic operator, one may set
\(\gamma^\mu_{(-+++)}=i\gamma^\mu_{(+---)}\).  Then
\(-\gamma^\mu_{(-+++)}\nabla_\mu+m\) differs from
\(i\gamma^\mu_{(+---)}\nabla_\mu-m\) only by an overall minus sign. The spin sums must be consistently translated in this case. Note that the hat in the gamma matrix index serves as a reminder that it should be a flat index.} \(\bar\psi=i\psi^\dagger\gamma^{\hat 0}\) and
\(\{\gamma^\mu,\gamma^\nu\}=2g^{\mu\nu}\) with signature
\((-+++)\).

The local action in Eq.~\eqref{eq:sectionII-fourfermion} does not
select a unique physical problem. The four-fermion theory has three relevant regimes: the perturbative, non-perturbative, and finite-density cases. In the perturbative regime, the interaction gives a contact amplitude
of the schematic form
\begin{equation}
 i{\cal M}_{\rm tree}
 \sim
 i\lambda
 [\bar u(p_3)u(p_1)]
 [\bar u(p_4)u(p_2)],
 \label{eq:contact-amplitude}
\end{equation}
with exchange terms and channel factors fixed by the external species.
Here $u$ is the basis spinor for the mode expansion of $\psi$, as we discuss in Sec.~\ref{subsec:free-fermion-review} (see Eq. \eqref{eq:free-dirac-mode-expansion-main}).

In the vacuum NJL regime, an attractive scalar channel can instead
support a nonzero particle--antiparticle bilinear
\(\Pi=\langle\bar\psi\psi\rangle\).  The quadratic operator then has a
self-consistently shifted Dirac mass \cite{Klevansky1992,Buballa2005}.  At finite density, BCS pairing
involves a particle--particle expectation value and a Fermi surface.
It requires an attractive projected Cooper channel \cite{Polchinski1992,Shankar1994,RajagopalWilczek2000}.  A scalar
\((\bar\psi\psi)^2\) interaction may contribute to such a channel after
a Fierz projection, but the sign of the scalar term alone does not
determine whether that projected channel is attractive
\cite{Polchinski1992,Shankar1994}.  NJL and BCS therefore differ in
their bilinears, their state data, and their quadratic kernels. 

Before discussing each regime explicitly, we first review the functional, grand-canonical description of fermions necessary to describe the finite-density BCS case. 

\subsection{Thermal state preparation and the zero-temperature
grand-canonical partition function}
\label{subsec:zero-temperature-grand-canonical}

We now derive the grand-canonical functional from the density-matrix
construction of Secs.~\ref{subsec:ctp-in-in} and
\ref{subsec:fermionic-functional-regimes}.  Assume that the background
is stationary and that the number operator is conserved,
\([\widehat H,\widehat N]=0\). The importance of the latter condition is clarified below. Let us introduce the Gibbs operator 
\begin{equation}
 \widehat{\mathcal G}_{\beta,\mu}
 \equiv e^{-\beta\left(\widehat H-\mu\widehat N\right)},
 \label{eq:unnormalized-gibbs-operator}
\end{equation}
where $\mu$ is the chemical potential conjugated to $\widehat{N}$. The normalized thermal density operator and its normalization are
\begin{equation}
 \widehat\rho_{\beta,\mu}
 =
 \frac{\widehat{\mathcal G}_{\beta,\mu}}
 {Z(\beta,\mu)},
 \qquad
 Z(\beta,\mu)
 \equiv
 \operatorname{Tr}\widehat{\mathcal G}_{\beta,\mu}.
 \label{eq:grand-canonical-partition-function}
\end{equation}

In connection with Eq.~\eqref{eq:ctp-functional}, the normalized CTP functional is
\begin{equation}
 Z_{\varrho_{\beta,\mu}}[J_+,J_-]
 =
 \frac{\operatorname{Tr}\!\left[
 e^{-\beta\widehat{\mathcal K}_\mu}
 U_{J_-}^{\dagger}(t_f,t_i)
 U_{J_+}(t_f,t_i)
 \right]}
 {Z(\beta,\mu)} .
 \label{eq:normalized-thermal-ctp}
\end{equation}
When the two real-time sources and metrics are identified, unitarity
gives \(U_J^\dagger U_J=\mathbf1\), and hence
\begin{equation}
 Z_{\varrho_{\beta,\mu}}[J,J]=1.
 \label{eq:thermal-ctp-unitarity}
\end{equation}

The same formulas can be expressed directly in the field representation.
For the scalar functional in
Eq.~\eqref{eq:scalar-ctp-density-functional}, the unnormalized thermal
kernel is the Euclidean evolution kernel
\begin{align}
 {\cal R}_{\beta}^{(B)}
 [\varphi_i^+,\varphi_i^-]
 &=
 \left\langle\varphi_i^+\left|
 e^{-\beta\widehat H}
 \right|\varphi_i^-\right\rangle
 \nonumber\\
 &=
 \int_{\substack{\Phi_E(0)=\varphi_i^-\\
                  \Phi_E(\beta)=\varphi_i^+}}
 [d\Phi_E]\,e^{-S_E^{(B)}[\Phi_E]} .
 \label{eq:bosonic-thermal-kernel}
\end{align}
A real scalar carries no conserved particle number, so \(\mu=0\) in
this example.  Substitution into
Eq.~\eqref{eq:scalar-ctp-density-functional} appends an imaginary-time
segment of length \(\beta\) to the two real-time branches.  Closing the
trace identifies the endpoints of that segment and produces periodic
bosonic boundary conditions.

For fermions, the corresponding coherent-state kernel is
\begin{widetext}
\begin{align}
 {\cal R}_{\beta,\mu}^{(F)}
 [\bar\zeta_i^+,\zeta_i^-]
 &=
 \left\langle\bar\zeta_i^+\left|
 e^{-\beta\left(\widehat H-\mu\widehat N\right)}
 \right|\zeta_i^-\right\rangle
 \nonumber\\
 &=
 \int_{\psi_E(0)=\zeta_i^-}^{\bar\psi_E(\beta)=\bar\zeta_i^+}
 [d\psi_E d\bar\psi_E]\,
 \exp\Bigg[
 -S_E[\bar\psi_E,\psi_E]
 +\mu\int_0^\beta d\tau\int d^3x\,
 \bar\psi_E\gamma_E^0\psi_E
 \Bigg].
 \label{eq:fermionic-thermal-kernel}
\end{align}
\end{widetext}
The coherent-state trace identity
\begin{equation}
 \operatorname{Tr}\widehat A
 =
 \int d\bar\zeta\,d\zeta\,
 e^{-\bar\zeta\zeta}
 \langle-\bar\zeta|\widehat A|\zeta\rangle
 \label{eq:fermionic-coherent-trace}
\end{equation}
turns the endpoint identification into antiperiodic boundary
conditions,
\begin{equation}
 \psi_E(\beta)=-\psi_E(0),
 \qquad
 \bar\psi_E(\beta)=-\bar\psi_E(0).
 \label{eq:fermionic-antiperiodicity}
\end{equation}
Equivalently, the thermal contour consists of a forward real-time
branch, a backward real-time branch, and a vertical segment from
\(t_i\) to \(t_i-i\beta\).  At equal real-time sources, the forward and
backward evolutions cancel and only the vertical segment remains.
Consequently,
\begin{widetext}
\begin{align}
 Z(\beta,\mu)
 &=
 \operatorname{Tr}
 e^{-\beta(\widehat H-\mu\widehat N)}
 \nonumber\\
 &=
 \int_{\rm AP}[d\psi d\bar\psi]\,
 \exp\Bigg[
 -S_E[\bar\psi,\psi]
 +\mu\int_0^\beta d\tau\int d^3x\,
 \bar\psi\gamma_E^0\psi
 \Bigg],
 \label{eq:euclidean-mu-functional}
\end{align}
\end{widetext}
where AP denotes antiperiodic boundary conditions. We stress that, although these manipulations might be standard in flat spacetime, the functional representation of the grand-canonical partition function is obtained from the CTP density-matrix functional \cite{Matsubara1955,LandsmanVanWeert1987} also in curved spacetimes.

The equilibrium trace considered presupposes a stationary generator. In the more general case, e.g. a time-dependent FLRW background, one instead specifies the
initial density matrix on a Cauchy surface and evolves it along the CTP
contour. The thermal construction supplies one possible initial
kernel when a stationary preparation slice exists.
That is one of the reasons why functional methods are so useful for curved spacetime calculations.

In a static volume \({\cal V}\), the zero-temperature grand-potential
density is
\begin{equation}
 \Omega(\mu)
 =-
 \lim_{\beta\rightarrow\infty}
 \frac{1}{\beta{\cal V}}\ln Z(\beta,\mu) .
 \label{eq:zero-temp-grand-potential}
\end{equation}
If \(|n\rangle\) simultaneously diagonalizes \(\widehat H\) and
\(\widehat N\), which is the case when $[\hat{H}, \hat{N}] =0$, then
\begin{align}
 Z(\beta,\mu)
 =\sum_n e^{-\beta(E_n-\mu N_n)} \xrightarrow{\beta\rightarrow\infty}
 e^{-\beta(E_\star-\mu N_\star)},
\end{align}
where \(|\star\rangle\) minimizes \(E_n-\mu N_n\).  Hence
\begin{equation}
 \Omega=\frac{E_\star-\mu N_\star}{\cal V},
 \qquad
 n\equiv\frac{N_\star}{\cal V}
 =-\frac{\partial\Omega}{\partial\mu} .
 \label{eq:grand-potential-ground-state}
\end{equation}
For a homogeneous zero-temperature state, extensivity gives \cite{LandsmanVanWeert1987}
\begin{equation}
 p=-\Omega,
 \qquad
 n=-\frac{\partial\Omega}{\partial\mu},
 \qquad
 \rho=\Omega+\mu n .
 \label{eq:thermo-definitions}
\end{equation}
These relations already show that a vacuum term in \(\Omega\) contributes
with \(p_{\rm vac}=-\rho_{\rm vac}\), whereas occupied modes give a
different equation of state.
The limit \(\beta\rightarrow\infty\) prepares the ground state of
$\widehat H-\mu\widehat N$, which may then be used as the initial
density matrix in the Lorentzian CTP functional.

This construction also explains why a chemical-potential term may be
removed from a local bulk operator without removing finite-density
physics. So, including a chemical potential term in the action is not enough to study a finite-density state. In fact, starting with
\begin{equation}
 S
 =\int d^4x\,
 \bar\psi(-\gamma^\mu\partial_\mu+m+i\mu\gamma^0)\psi ,
\end{equation}
and using the field redefinition
\begin{equation}
 \psi=e^{i\mu t}\chi,
 \qquad
 \bar\psi=\bar\chi e^{-i\mu t}
 \label{eq:mu-field-redefinition}
\end{equation}
removes \(\mu\) from the Lorentzian bulk action. However, it simultaneously
rephases charged sources and, more importantly, the endpoint wavefunctional.  In Euclidean
time, the analogous transformation changes antiperiodicity into
\begin{equation}
 \chi(\beta)=-e^{-\beta\mu}\chi(0) .
 \label{eq:mu-euclidean-twist}
\end{equation}
Thus the chemical potential might survive as state or contour data even
when it is absent from a locally transformed bulk operator
\cite{Cohen2003}.

We now explain in which case the presence of $\mu$ in the fermionic action is relevant for $\beta \to \infty$. For a free Dirac field, the selected zero-temperature (but finite density) state at
\(\mu>m_{\rm eff}\) takes the form
\begin{equation}
 |\Omega_\mu\rangle
 =
 \prod_{\boldsymbol{k},s:\,E_k<\mu}
 c_{\boldsymbol{k}s}^{\dagger}|0\rangle,
 \qquad
 k_F=\sqrt{\mu^2-m_{\rm eff}^2} ,
 \label{eq:fermi-sea-state}
\end{equation}
where $c_{\mathbf{k}s}^\dagger$ is the creation operator for the fermion quanta and $k_F$ is the Fermi momentum. The exact onset condition follows directly from the spectral form of
the grand potential.  Let \(|0\rangle\) be the zero-density ground
state, with energy \(E_0\) and charge \(N_0=0\), and define
\begin{equation}
 \mu_c^+
 \equiv
 \inf_{N_n>0}\frac{E_n-E_0}{N_n},
 \qquad
 \mu_c^-
 \equiv
 \inf_{N_n<0}\frac{E_n-E_0}{|N_n|}.
 \label{eq:silver-blaze-thresholds}
\end{equation}
For
\begin{equation}
 -\mu_c^-<\mu<\mu_c^+,
 \label{eq:silver-blaze-interval}
\end{equation}
every charged state has \(E_n-\mu N_n>E_0\).  The state minimizing
\(\widehat H-\mu\widehat N\) is therefore still the vacuum.  At exactly
zero temperature, the grand potential and vacuum observables are
independent of \(\mu\) throughout this open interval, and
\(n=-\partial\Omega/\partial\mu=0\).

For the free charge-conjugation-symmetric Dirac theory, the least
positive- and negative-charge excitations each carry one unit of charge
and have minimum energy \(m_{\rm eff}\).  Hence
\(\mu_c^+=\mu_c^-=m_{\rm eff}\), and the vacuum equivalent interval is
\begin{equation}
 |\mu|<m_{\rm eff}.
 \label{eq:free-silver-blaze-interval}
\end{equation}
In this case, the field redefinition \eqref{eq:mu-field-redefinition} indeed removes $\mu$ from the physical problem considered. Here \(m_{\rm eff}\) is treated as a fixed parameter of the quadratic action.  In an interacting theory, the first state to appear may be
a bound state or collective excitation, so the onset is governed by the
minimum energy per unit conserved charge in
Eq.~\eqref{eq:silver-blaze-thresholds}. At nonzero temperature, the independence is no
longer exact because charged states acquire Boltzmann-suppressed
occupations. The absence of a physical response at zero temperature despite the explicit appearance
of \(\mu\) in the Dirac action was discussed in \cite{Cohen2003,Cohen2004}.

Let \(g_s\) count occupied spin and internal states; for one Dirac
species with only particle states filled, \(g_s=2\).  The normal-state
grand potential separates into vacuum and medium pieces, \cite{LandsmanVanWeert1987,Cohen2003,Cohen2004,Chernodub:2020yaf}
\begin{equation}
 \Omega(m_{\rm eff},\mu)
 =\Omega_{\rm vac}(m_{\rm eff})
 -g_s\int_{|\boldsymbol{k}|<k_F}
 \frac{d^3k}{(2\pi)^3}(\mu-E_k), 
 \label{eq:omega-normal-split}
\end{equation}
where $E_k$ denotes the dispersion relation of the quanta. It follows from the second equation in \eqref{eq:thermo-definitions} that
\begin{equation}
 n=\frac{g_s k_F^3}{6\pi^2},
 \label{eq:number-density-fermi}
\end{equation}
and, writing \(\rho_0\equiv\Omega_{\rm vac}\),
\begin{align}
 \rho&=\rho_0+\rho_{\rm med},
 &
 \rho_{\rm med}
 &=g_s\int_{|\boldsymbol{k}|<k_F}
 \frac{d^3k}{(2\pi)^3}E_k,
 \nonumber\\
 p&=-\rho_0+p_{\rm med},
 &
 p_{\rm med}
 &=g_s\int_{|\boldsymbol{k}|<k_F}
 \frac{d^3k}{(2\pi)^3}(\mu-E_k).
 \label{eq:rho-p-medium-integrals}
\end{align}
The medium pieces satisfy \(\rho_{\rm med}+p_{\rm med}=\mu n\).  For
\(\mu>m_{\rm eff}\) and $E_k = \sqrt{k^2 +m_{\rm eff}^2}$, the radial integrals give
\begin{align}
 \rho_{\rm med}
 &=\frac{g_s}{16\pi^2}
 \left[
 \mu k_F(2\mu^2-m_{\rm eff}^2)
 -m_{\rm eff}^4
 \ln\left(\frac{\mu+k_F}{m_{\rm eff}}\right)
 \right],
 \nonumber\\
 p_{\rm med}
 &=\frac{g_s}{48\pi^2}
 \left[
 \mu k_F(2\mu^2-5m_{\rm eff}^2)
 +3m_{\rm eff}^4
 \ln\left(\frac{\mu+k_F}{m_{\rm eff}}\right)
 \right].
 \label{eq:rho-p-medium-closed}
\end{align}
For \(0\leq\mu\leq m_{\rm eff}\), \(k_F=0\) and the particle-medium
contributions vanish.  A sufficiently negative chemical potential would
instead populate antiparticle states and must be treated with the
charge-conjugate branch \cite{LandsmanVanWeert1987,Cohen2003,Cohen2004}.  

Having explained how the free-theory, finite-temperature and finite-density results are derived in the functional approach, we will review the role of the four-fermion interacting term in the remainder of this section. We start with the non-perturbative Nambu-Jona-Lasinio case. This case is the main focus of this work, and more details about it and its relation to the perturbative limit are discussed in Sec. \ref{subsec:njl-review}.

\subsection{Vacuum NJL saddle}
\label{subsec:vacuum-njl-saddle}

The four-fermion interaction can be linearized by introducing a real
Hubbard--Stratonovich field \(\Theta\)
\cite{Stratonovich1958,Hubbard1959}:
\begin{align}
 &\exp\left[
 i\int d^4x\sqrt{-g}\,\lambda(\bar\psi\psi)^2
 \right]
 \nonumber\\
 &\qquad\propto
 \int[d\Theta]\,
 \exp\left[
 i\int d^4x\sqrt{-g}\left(
 -\frac{\Theta^2}{4\lambda}
 +\Theta\bar\psi\psi
 \right)
 \right].
 \label{eq:state-HS}
\end{align}
Varying the auxiliary action gives
\begin{equation}
 \Theta=2\lambda\bar\psi\psi .
\end{equation}
At the translationally invariant mean-field saddle we define
\begin{equation}
 \Pi\equiv\langle\bar\psi\psi\rangle,
 \qquad
 \Theta\equiv2\lambda\Pi,
 \qquad
 m_{\rm eff}(\Theta)\equiv m+\Theta .
 \label{eq:njl-notation-theta}
\end{equation}
In terms of \(\Theta\), the mean-field action is
\begin{equation}
 S_{\rm MF}
 =
 \int d^4x\sqrt{-g}\left[
 \bar\psi\sD_{m_{\rm eff}}\psi
 -\frac{\Theta^2}{4\lambda}
 \right].
 \label{eq:njl-mean-field-action-section2}
\end{equation}
The auxiliary contribution is therefore
\begin{equation}
 V_{\rm aux}(\Theta)
 =\frac{\Theta^2}{4\lambda}
 =\lambda\Pi^2 .
 \label{eq:njl-auxiliary-potential-theta}
\end{equation}

After integrating out the fermions, the vacuum in-out functional takes
the form
\begin{equation}
 W_{\rm NJL}[\Theta,g]
 =
 -i\operatorname{Tr}\ln\sD_{m_{\rm eff}}
 -\int d^4x\sqrt{-g}\,
 \frac{\Theta^2}{4\lambda} .
 \label{eq:njl-effective-action-section2}
\end{equation}
The mean-field equation is the stationary condition
\begin{equation}
 \frac{1}{\sqrt{-g}}
 \frac{\delta W_{\rm NJL}}{\delta\Theta(x)}=0.
 \label{eq:njl-saddle-functional}
\end{equation}
For a constant saddle in flat spacetime this reduces, in the present
one-Dirac-field normalization, to
\begin{equation}
 \frac{\Theta}{2\lambda}
 =
 2m_{\rm eff}
 \int\frac{d^3k}{(2\pi)^3}\frac{1}{E_k},
 \qquad
 E_k=\sqrt{\boldsymbol{k}^2+m_{\rm eff}^2} .
 \label{eq:njl-gap-section2}
\end{equation}
This is equivalent to \(\Theta=2\lambda\Pi\) because
\begin{equation}
 \Pi
 = \langle\bar\psi\psi\rangle = 2m_{\rm eff}
 \int\frac{d^3k}{(2\pi)^3}\frac{1}{E_k},
 \label{eq:njl-condensate-section2}
\end{equation}
as can be checked from a mode-expansion calculation.

Both of the previous integrals require a regulator and a matter-sector matching
condition. Self-consistency of the mean-field approach requires a critical value of the interaction coupling $\lambda$. To exhibit this critical behavior, consider a sharp three-momentum cutoff
\(\Lambda\) and the chiral limit \(m=0\). Eq. \eqref{eq:njl-condensate-section2} gives
\begin{equation}
 \frac{1}{2\lambda}
 =2\int^{\Lambda}\frac{d^3k}{(2\pi)^3}
 \frac{1}{\sqrt{\boldsymbol{k}^2+\Theta^2}} .
 \label{eq:njl-critical-cutoff-equation}
\end{equation}
At \(\Theta\rightarrow0\), the right-hand side approaches
\(\Lambda^2/(2\pi^2)\).  Thus, in this particular normalization and
regulator, a nonzero saddle first becomes possible at
\begin{equation}
 \lambda_c \sim \frac{\pi^2}{\Lambda^2} .
 \label{eq:njl-critical-coupling-example}
\end{equation}
The numerical coefficient is regulator and flavor-normalization
dependent. However, we conclude unambiguously that a sufficiently strong attractive
vacuum channel is needed for consistency.

The NJL saddle reorganizes the vacuum around massive
particle--antiparticle modes. As discussed next, the BCS saddle is different: the reference
state already has nonzero density, and the quadratic theory mixes
particles and holes.  We next formulate that construction explicitly
so that the distinction with the vacuum determinant in
Eq.~\eqref{eq:njl-effective-action-section2} is made manifest.

\subsection{Finite-density BCS saddle}
\label{subsec:finite-density-bcs-saddle}

Let \(\mu\) be the chemical potential conjugate to a conserved number
operator \(\widehat N\), and let 
\begin{equation}
 \widehat{\mathcal K}_\mu
 \equiv\widehat H-\mu\widehat N
 \label{eq:grand-canonical-hamiltonian-definition}
\end{equation}
be the grand-canonical Hamiltonian.  For a fermion of effective mass
\(m_{\rm eff}\), define
\begin{equation}
 E_k=\sqrt{k^2+m_{\rm eff}^2},
 \qquad
 \xi_{k,-}=E_k-\mu,
 \qquad
 \xi_{k,+}=E_k+\mu .
 \label{eq:bcs-xi-definition}
\end{equation}
The subscript \(-\) labels the particle branch near a Fermi surface,
and \(+\) labels the antiparticle branch.  When \(\mu>m_{\rm eff}\),
the normal-state Fermi momentum is
\begin{equation}
 k_F=\sqrt{\mu^2-m_{\rm eff}^2} .
\end{equation}

Introducing the charge-conjugate field
\begin{equation}
 \psi_C\equiv C\bar\psi^T,
 \qquad
 \Psi_{\rm NG}\equiv
 \begin{pmatrix}\psi\\ \psi_C\end{pmatrix},
 \label{eq:nambu-gorkov-spinor}
\end{equation}
where \(C\) is the charge-conjugation matrix, puts the quadratic action
in Nambu--Gorkov form
\cite{BardeenCooperSchrieffer1957,Gorkov1958,Nambu1960,PisarskiRischke1999},
\begin{equation}
  S = -\frac{1}{2}\int d^4 x \;\bar{\Psi}_{\rm NG} \mathcal{O}_{\rm BCS} \Psi_{\rm NG},
  \label{eq:action_NG_form}
\end{equation}
where $\mathcal{O}_{\rm BCS}$ has a block form with kinetic operators in the diagonal and pairing terms in the off-diagonal.

Using the Fourier convention
\(\psi(x)\propto e^{-ik\cdot x}\), the kinetic block in a local
orthonormal rest frame is
\begin{equation}
 D(k;\mu)
 \equiv
 i\gamma^\alpha k_\alpha+m_{\rm eff}
 +i\mu\gamma^{\hat 0}.
 \label{eq:bcs-D-definition}
\end{equation}
Indeed,
\(i\mu\bar\psi\gamma^{\hat0}\psi=\mu\psi^\dagger\psi\), so this term
implements the grand-canonical shift
\(\widehat H\rightarrow\widehat H-\mu\widehat N\).  The
charge-conjugate block is defined separately by
\begin{equation}
 \widetilde D(k;-\mu)
 \equiv
 C D^T(-k;\mu)C^{-1}
 =
 i\gamma^\alpha k_\alpha+m_{\rm eff}
 -i\mu\gamma^{\hat0},
 \label{eq:bcs-Dtilde-definition}
\end{equation}
where \(C\gamma^{\alpha T}C^{-1}=-\gamma^\alpha\).

Suppose that the microscopic interaction has an attractive projection
onto a Cooper channel specified by a matrix
\(\Gamma_{\rm pair}\) acting on the relevant spin, flavor, or internal
indices.  We denote the corresponding particle--particle auxiliary
field by \(\Delta_{\rm BCS}\). With \(G_C>0\) denoting the attractive coupling in that
channel, we define its normalization together with the gap by
\begin{equation}
 \Delta_{\rm BCS}
 \equiv
 -\frac{G_C}{2}
 \left\langle
 \psi^T C\Gamma_{\rm pair}\psi
 \right\rangle,
 \qquad
 (C\Gamma_{\rm pair})^T=-C\Gamma_{\rm pair}.
 \label{eq:bcs-pair-matrix-definition}
\end{equation}
The antisymmetry condition is required by Fermi statistics.  Spin,
flavor, color, or other projectors may be included in
\(\Gamma_{\rm pair}\); the overall sign of the condensate can be absorbed into the phase
convention for \(\Delta_{\rm BCS}\).
In a local
flat-space rest frame, the inverse propagator can be represented as
\begin{equation}
 {\cal O}_{\rm BCS}(k)
 =
 \begin{pmatrix}
 D(k;\mu) & \Delta_{\rm BCS}\Gamma_{\rm pair}\\
 \Delta_{\rm BCS}^{*}\Gamma_{\rm pair}^{\dagger}
 & \widetilde D(k;-\mu)
 \end{pmatrix} .
 \label{eq:bcs-ng-inverse-momentum}
\end{equation}
The off-diagonal entries are the essential difference from the NJL
mass shift.

After resolving the spinor and internal projectors, each independent
pairing sector reduces to a \(2\times2\) Bogoliubov block,
\begin{equation}
 \mathbb H_{k,s}
 =
 \begin{pmatrix}
 \xi_{k,s} & \Delta_{\rm BCS}\\
 \Delta_{\rm BCS}^{*} & -\xi_{k,s}
 \end{pmatrix},
 \qquad s=\pm .
 \label{eq:bcs-two-by-two-block-main}
\end{equation}
Its characteristic equation is
\begin{equation}
 \det(\mathbb H_{k,s}-\varepsilon\mathbf1_2)
 =
 \varepsilon^2-\xi_{k,s}^2
 -|\Delta_{\rm BCS}|^2 .
 \label{eq:bcs-characteristic-polynomial}
\end{equation}
The positive quasiparticle energies are therefore
\begin{align}
 {\cal E}_{-}(k)
 &=\sqrt{(E_k-\mu)^2+|\Delta_{\rm BCS}|^2},
 \nonumber\\
 {\cal E}_{+}(k)
 &=\sqrt{(E_k+\mu)^2+|\Delta_{\rm BCS}|^2} .
 \label{eq:bcs-branches}
\end{align}
In particular, \({\cal E}_{-}\) is gapped at the normal-state Fermi
surface.  For the positive eigenvalue, define the Bogoliubov amplitudes
by the explicit eigenvector equation
\begin{equation}
 \mathbb H_{k,s}
 \begin{pmatrix}
  u_{k,s}\\ v_{k,s}
 \end{pmatrix}
 =
 {\cal E}_{s}(k)
 \begin{pmatrix}
  u_{k,s}\\ v_{k,s}
 \end{pmatrix},
 \qquad s=\pm .
 \label{eq:bcs-eigenvector-equation}
\end{equation}
Equivalently,
\begin{align}
 \big[\xi_{k,s}-{\cal E}_{s}(k)\big]u_{k,s}
 +\Delta_{\rm BCS}v_{k,s}&=0,
 \nonumber\\
 \Delta_{\rm BCS}^*u_{k,s}
 -\big[\xi_{k,s}+{\cal E}_{s}(k)\big]v_{k,s}&=0.
 \label{eq:bcs-eigenvector-components}
\end{align}
Using Eq.~\eqref{eq:bcs-branches} and imposing canonical normalization
then gives, for either branch,
\begin{align}
 |u_{k,s}|^2
 &=\frac12\left(1+\frac{\xi_{k,s}}
 {{\cal E}_{s}(k)}\right),
 \nonumber\\
 |v_{k,s}|^2
 &=\frac12\left(1-\frac{\xi_{k,s}}
 {{\cal E}_{s}(k)}\right),
 \qquad
 |u_{k,s}|^2+|v_{k,s}|^2=1 .
 \label{eq:bcs-coherence-factors}
\end{align}

The BCS state itself can be defined in a finite box.  Let
\({\cal H}\) contain one representative of each unordered momentum
pair \(\{\boldsymbol{k},-\boldsymbol{k}\}\).  For a single particle
pairing block,
\begin{equation}
 |\Omega_{\rm BCS}\rangle
 =
 \prod_{\boldsymbol{k}\in{\cal H}}
 \left(
 u_{\boldsymbol{k}}
 +v_{\boldsymbol{k}}
 c_{\boldsymbol{k}\uparrow}^{\dagger}
 c_{-\boldsymbol{k}\downarrow}^{\dagger}
 \right)|0\rangle .
 \label{eq:bcs-finite-volume-state-main}
\end{equation}
Note that although Pauli exclusion truncates each factor, the product over
distinct momentum pairs is included.  

The state in Eq.~\eqref{eq:bcs-finite-volume-state-main} is selected by
the saddle of the grand-canonical functional introduced in
Sec.~\ref{subsec:zero-temperature-grand-canonical}. In Appendix
\ref{app:bcs-grand-potential}, we explain how to compute the mean-field grand-canonical generating functional. Here we state the
zero-temperature result and its consequences.

For brevity, write \(\Delta\equiv\Delta_{\rm BCS}\).  We first consider
the physical degrees of freedom participating in the specified Cooper
channel; unpaired sectors will be added separately below. Before
imposing the gap equation, their vacuum-referenced off-shell
mean-field grand-potential density is (see Eq. \eqref{eq:appF-minimal-zero-temperature-potential})
\begin{align}
 \overline\Omega_{\rm BCS}^{\rm MF}(\Delta;\mu)
 &=
 \Omega_{\rm vac}(m_{\rm eff})
 +\frac{|\Delta|^2}{G_C}
 +\delta\Omega_{\rm finite}(\Delta;\mu)
 \nonumber\\
 &\quad-
 \frac{g_{\rm pair}}{2}
 \int_{\mathbb R^3}\frac{d^3k}{(2\pi)^3}
 \nonumber\\
 &\qquad\times
 \left[
 {\cal E}_{-}(k;\Delta)
 +{\cal E}_{+}(k;\Delta)
 -2E_k
 \right].
 \label{eq:bcs-absolute-potential-main}
\end{align}
The physical potential is obtained by evaluating this
off-shell function at a stable homogeneous saddle. The quantity \(\Omega_{\rm vac}(m_{\rm eff})\) is the zero-density, \(\Delta=0\) grand potential of the same physical
degrees of freedom. The remaining finite counterterm contribution is denoted by $\delta \Omega_{\rm finite}$, and for a state-independent renormalization prescription it satisfies \(\delta\Omega_{\rm finite}(0;\mu)=0\).

The normal-state result in Eq.~\eqref{eq:omega-normal-split} follows
directly from Eq.~\eqref{eq:bcs-absolute-potential-main}.  For
\(\Delta=0\) and \(\mu>0\),
\({\cal E}_{-}=|E_k-\mu|\) and
\({\cal E}_{+}=E_k+\mu\).  Consequently,
\begin{equation}
 \frac12
 \left[
 |E_k-\mu|+E_k+\mu-2E_k
 \right]
 =
 (\mu-E_k)\Theta(\mu-E_k).
 \label{eq:bcs-normal-limit-step-identity-main}
\end{equation}
Since
\(\Theta(\mu-E_k)=\Theta(k_F-k)\), with
\(k_F=\sqrt{\mu^2-m_{\rm eff}^2}\) for
\(\mu>m_{\rm eff}\), Eq.~\eqref{eq:bcs-absolute-potential-main}
therefore becomes
\begin{equation}
 \overline\Omega_{\rm BCS}^{\rm MF}(0;\mu)
 =
 \Omega_{\rm vac}
 -g_{\rm pair}
 \int_{|\boldsymbol k|<k_F}
 \frac{d^3k}{(2\pi)^3}
 (\mu-E_k).
 \label{eq:bcs-normal-limit-potential-main}
\end{equation}
This is Eq.~\eqref{eq:omega-normal-split} when the compared normal
species are precisely the paired degrees of freedom, so that
\(g_{\rm pair}=g_s\).  The restriction to momenta within the Fermi surface emerges from the step function only in
the \(\Delta\to0\) limit.  For nonzero \(\Delta\), the integral extends
over the full regulated momentum domain ($\delta \Omega_{\rm finite}$ is such that the result converges, see Appendix \ref{app:bcs-grand-potential}).

Any charged degrees of freedom that do not participate in this Cooper
channel contribute a separate normal-state medium potential.  They are
not included implicitly in \(\Omega_{\rm vac}\), which contains only
the zero-density contribution of the degrees of freedom under
consideration.

The coupling \(G_C>0\) is the attractive coupling in the specified
Cooper channel, while \(g_{\rm pair}\) counts physical copies of one
resolved pairing block in the undoubled theory.  The factor \(1/2\)
in Eq.~\eqref{eq:bcs-absolute-potential-main} removes the artificial
Nambu--Gorkov duplication when the integral covers the full momentum
space.  For the scalar interaction in
Eq.~\eqref{eq:sectionII-fourfermion}, one may write
\(G_C=c_{\rm pair}\lambda\), but the Fierz coefficient
\(c_{\rm pair}\) is fixed only after the spin and internal
representation entering \(\Gamma_{\rm pair}\) has been specified.
The four-dimensional contact theory also requires a common physical
cutoff and matching prescription.

The mean-field gap $\Delta_\star$ is fixed by stationarity of the off-shell potential (see Appendix \ref{app:bcs-grand-potential} for details):
\begin{align}
 0&=
 \frac{\Delta_\star}{G_C}
 -
 \frac{g_{\rm pair}\Delta_\star}{4}
 \int_{\mathbb R^3}\frac{d^3k}{(2\pi)^3}
 \sum_{s=\pm}
 \frac{1}{{\cal E}_{s}(k;\Delta_\star)}
  \nonumber\\
 &\quad \left.
 +\frac{\partial\delta\Omega_{\rm finite}}
 {\partial\Delta^*}
 \right|_{\Delta_\star}.
 \label{eq:bcs-gap-equation-main}
\end{align}
A nonzero solution describes the paired phase only when it is a stable
minimum.  The point \(\Delta=0\) remains a well-defined normal-state
reference even when it is not the stable zero-temperature saddle.  If
\(m_{\rm eff}\) has a dynamical mean-field contribution, its stationarity
equation must be imposed together with
Eq.~\eqref{eq:bcs-gap-equation-main}.

The number density follows from
Eq.~\eqref{eq:grand-potential-ground-state}. At fixed
\(m_{\rm eff}\), \(G_C\), and microscopic matching data,
Eq.~\eqref{eq:bcs-absolute-potential-main} therefore gives
\begin{align}
 n_{\rm BCS}^{\rm paired}
 &=
 -\left.
 \frac{\partial\delta\Omega_{\rm finite}}{\partial\mu}
 \right|_{\substack{\Delta=\Delta_\star,\, G_C\, {\rm fixed}}}
 \nonumber\\
 &\quad+
 \frac{g_{\rm pair}}{2}
 \int\frac{d^3k}{(2\pi)^3}
 \left[
 -\frac{\xi_{k,-}}{{\cal E}_{-}(k;\Delta_\star)}
 +\frac{\xi_{k,+}}{{\cal E}_{+}(k;\Delta_\star)}
 \right].
 \label{eq:bcs-number-equation-main}
\end{align}
As a check, for \(\Delta_\star\to0\) and \(\mu>0\), the first vanishes while integrand of the second term becomes $-\operatorname{sgn}(E_k-\mu)+1$. Using
\begin{equation}
 \frac12
 \left[
 -\operatorname{sgn}(E_k-\mu)+1
 \right]
 =
 \Theta(\mu-E_k),
 \label{eq:bcs-number-step-identity-main}
\end{equation}
we have
\begin{equation}
 n_{\rm BCS}^{\rm paired}
 \longrightarrow
 \frac{g_{\rm pair}k_F^3}{6\pi^2}.
 \label{eq:bcs-number-normal-limit-main}
\end{equation}
This reproduces Eq.~\eqref{eq:number-density-fermi} when
\(g_{\rm pair}=g_s\).  Any unpaired degrees of freedom contribute
separately through
\(n_{\rm unpaired}
=-\partial\Omega_{\rm N}^{\rm unpaired}/\partial\mu\), so that the
total density is
\(n_{\rm BCS}=n_{\rm BCS}^{\rm paired}+n_{\rm unpaired}\).

\section{Free Scalar and Fermion Effective Actions}
\label{sec:free-fields}

The four-fermion mean-field calculation reduces the interacting theory
to a quadratic determinant in a background auxiliary field.  Before
generalizing to curved spacetimes, it is useful to review the corresponding free
calculations.  The scalar field fixes the relation between mode sums,
functional determinants, and covariant counterterms.  The Dirac field case introduces fermionic signs, spin degeneracy, and spin-connection
curvature that are needed for the NJL result on curved spacetimes.

\subsection{Scalar field}
\label{subsec:free-scalar-review}

For a minimally coupled real scalar field \(\Phi\) of mass \(m\),
\begin{equation}
 S_\Phi
 =-
 \int d^4x\sqrt{-g}\left[
 \frac12g^{\mu\nu}\partial_\mu\Phi\partial_\nu\Phi
 +\frac12m^2\Phi^2
 \right].
 \label{eq:free-scalar-action}
\end{equation}
Metric variation gives
\begin{equation}
 T_{\mu\nu}^{\Phi}
 =\partial_\mu\Phi\partial_\nu\Phi
 -g_{\mu\nu}\left[
 \frac12g^{\alpha\beta}\partial_\alpha\Phi\partial_\beta\Phi
 +\frac12m^2\Phi^2
 \right].
 \label{eq:free-scalar-emt}
\end{equation}
In flat spacetime we use the mode convention
\begin{align}
 \Phi(t,\boldsymbol{x})
 &=\int\frac{d^3k}{(2\pi)^{3/2}\sqrt{2\omega_k}}
 \left[
 a_{\boldsymbol{k}}
 e^{-i\omega_kt+i\boldsymbol{k}\cdot\boldsymbol{x}}
 \right.
 \nonumber\\
 &\hspace{3.0cm}\left.
 +a_{\boldsymbol{k}}^\dagger
 e^{i\omega_kt-i\boldsymbol{k}\cdot\boldsymbol{x}}
 \right],
 \label{eq:free-scalar-mode-expansion}
\end{align}
where $\omega_k=\sqrt{k^2+m^2}$ with
\([a_{\boldsymbol{k}},a_{\boldsymbol{p}}^\dagger]
=\delta^{(3)}(\boldsymbol{k}-\boldsymbol{p})\).  The vacuum mode sums
are \cite{BirrellDavies1982, ParkerToms2009,Martin2012}
\begin{align}
 \langle\rho\rangle_\Phi
 &=\frac12\int\frac{d^3k}{(2\pi)^3}\,\omega_k,
 \nonumber\\
 \langle p\rangle_\Phi
 &=\frac16\int\frac{d^3k}{(2\pi)^3}\,
 \frac{k^2}{\omega_k} .
 \label{eq:scalar-rho-p-bare}
\end{align}
These expressions are ultraviolet divergent.  A sharp spatial momentum cutoff
does not preserve the Lorentz-invariant vacuum equation of state, so we
use dimensional regularization for the covariant comparison. Before continuation to
\(n=d-1\) spatial dimensions, the pressure prefactor \(1/6\) is replaced
by \(1/[2(d-1)]\). Let
\(d=4-\epsilon\) be the continued spacetime dimension and
\(\mu_{\rm DR}\) the regularization scale
\cite{tHooftVeltman1972}.  We make use of the master integral
\begin{equation}
 \int\frac{d^n k}{(2\pi)^n}(k^2+m^2)^\alpha
 =\frac{(m^2)^{\alpha+n/2}}{(4\pi)^{n/2}}
 \frac{\Gamma(-\alpha-n/2)}{\Gamma(-\alpha)} .
 \label{eq:dimensional-master-integral-scalar}
\end{equation}
Setting \(n=d-1\) in Eq.~\eqref{eq:dimensional-master-integral-scalar} gives
\begin{align}
 \langle\rho\rangle_\Phi
 &=\frac{\mu_{\rm DR}^4}
 {2(4\pi)^{(d-1)/2}}
 \frac{\Gamma(-d/2)}{\Gamma(-1/2)}
 \left(\frac{m}{\mu_{\rm DR}}\right)^d,
 \nonumber\\
 \langle p\rangle_\Phi
 &=\frac{\mu_{\rm DR}^4}
 {4(4\pi)^{(d-1)/2}}
 \frac{\Gamma(-d/2)}{\Gamma(1/2)}
 \left(\frac{m}{\mu_{\rm DR}}\right)^d
 =-\langle\rho\rangle_\Phi .
 \label{eq:scalar-flat-dr-compact}
\end{align}
The last equality follows from
\(\Gamma(1/2)=-\Gamma(-1/2)/2\).  Dimensional regularization therefore
preserves the Lorentz-invariant form
\(\langle T_{\mu\nu}\rangle\propto g_{\mu\nu}\) at the regulated
level. 

The flat-space mode sums rely on translation invariance and a global
Fourier basis, neither of which is generally available in curved
spacetimes. Nonetheless, the short-distance part of the Green function is local and can be organized covariantly in powers of the
curvature and its derivatives.  We use the convention
\begin{align}
 R^\rho{}_{\sigma\mu\nu} = 
\partial_\mu\Gamma^\rho_{\nu\sigma}
-\partial_\nu\Gamma^\rho_{\mu\sigma}
+\Gamma^\rho_{\mu\lambda}\Gamma^\lambda_{\nu\sigma}
-\Gamma^\rho_{\nu\lambda}\Gamma^\lambda_{\mu\sigma},
 \label{eq:curvature-conventions}
\end{align}
where $\Gamma^{\rho}_{\nu\sigma}$ are the Christoffel symbols associated to the metric and \(\Box=g^{\mu\nu}\nabla_\mu\nabla_\nu\). The symbols \(R\) and
\(\nabla R\) in the schematic validity conditions below stand for
characteristic components of the full curvature tensor and its
derivatives in a local orthonormal frame.

As we saw in Sec. \ref{sec:functional-setup}, the endpoint states fix the inverse and its boundary conditions,
whereas the local ultraviolet expansion is determined by the
effective action which in turn generates a local expansion for the Green's function. Therefore, the free-theory results in Eq. \eqref{eq:scalar-rho-p-bare} can be generalized covariantly as follows. Define the operator
\begin{equation}
 K_\Phi(m^2)=-\Box+m^2 .
\end{equation}
The vacuum functional and its one-loop action are
\begin{equation}
 Z_\Phi[0;g]\propto[\det K_\Phi]^{-1/2},
 \qquad
 W_\Phi=\frac{i}{2}\operatorname{Tr}\ln K_\Phi .
 \label{eq:scalar-functional-determinant}
\end{equation}
If
\(G(m^2)=K_\Phi(m^2)^{-1}\), then
\begin{equation}
 \frac{\partial}{\partial m^2}
 \operatorname{Tr}\ln K_\Phi(m^2)
 =\operatorname{Tr}G(m^2) .
 \label{eq:scalar-mass-integration-identity}
\end{equation}

To find the vacuum functional, we make use of the heat-kernel methods \cite{Schwinger1951,DeWitt1964,Vassilevich2003,ParkerToms2009}. The local heat-kernel series is an asymptotic derivative expansion.  A
finite truncation is controlled schematically when
\begin{equation}
 \frac{|R|}{m^2}\ll1,
 \qquad
 \frac{|\nabla R|}{m^3}\ll1,
 \qquad\ldots,
 \label{eq:heat-kernel-validity}
\end{equation}
and should not be interpreted as a controlled large-mass expansion at
\(m=0\) \cite{Vassilevich2003,ParkerToms2009}.  In the conventions of
Appendix~\ref{app:scalar-green}, the first coefficients for a minimally
coupled scalar are
\begin{align}
 f_1^\Phi
 &=\frac16R,
 \nonumber\\
 f_2^\Phi
 &=\frac{1}{72}R^2
 -\frac{1}{180}R_{\mu\nu}R^{\mu\nu}
 +\frac{1}{180}R_{\mu\nu\rho\sigma}R^{\mu\nu\rho\sigma}
 +\frac{1}{30}\Box R .
 \label{eq:scalar-f1f2-compact}
\end{align}
The coincident Green-function recursion and the proper-time
reconstruction are given in that appendix.  Integrating
Eq.~\eqref{eq:scalar-mass-integration-identity} yields the local action
\begin{widetext}
\begin{align}
 W_\Phi^{\rm loc}
 &=\int d^dx\sqrt{-g}\,
 \frac{1}{2(4\pi)^{d/2}}
 \left(\frac{m}{\mu_{\rm DR}}\right)^{d-4}
 \left[
 m^4\Gamma\left(-\frac d2\right)
 +m^2f_1^\Phi\Gamma\left(1-\frac d2\right)
 +f_2^\Phi\Gamma\left(2-\frac d2\right)
 \right].
 \label{eq:scalar-W-compact}
\end{align}
\end{widetext}
For \(d=4-\epsilon\), this becomes
\begin{widetext}
\begin{align}
 W_\Phi^{\rm loc}
 &=\int d^dx\sqrt{-g}\Bigg\{
 \left[
 \frac{2}{\epsilon}-\gamma_{\rm EM}
 -\ln\left(\frac{m^2}{4\pi\mu_{\rm DR}^2}\right)
 \right]
 \left[
 \frac{m^4}{64\pi^2}
 -\frac{m^2f_1^\Phi}{32\pi^2}
 +\frac{f_2^\Phi}{32\pi^2}
 \right]
 +\frac{3m^4}{128\pi^2}
 -\frac{m^2f_1^\Phi}{32\pi^2}
 \Bigg\}.
 \label{eq:scalar-W-DR-compact}
\end{align}
\end{widetext}
Here \(\gamma_{\rm EM}\) is the Euler--Mascheroni constant.
The three terms in the first bracket of the integrand in Eq.~\eqref{eq:scalar-W-DR-compact} multiply the
spacetime volume, the Ricci scalar, and curvature-squared invariants.
They renormalize, respectively, the cosmological constant term, the
Einstein--Hilbert term, and the higher-curvature sector.

The scalar example discussed so far establishes how the functional methods for computing the local quantum effective action work.  The free Dirac
field follows the same steps, but the Grassmann determinant and
the curvature of the spinor connection modify the heat-kernel coefficients, as we discuss next.

\subsection{Fermion field}
\label{subsec:free-fermion-review}

For one four-component Dirac field of mass \(m\), the action in the
conventions introduced in Sec.~\ref{subsec:fermionic-functional-regimes}
is
\begin{equation}
 S_\psi
 =\int d^dx\sqrt{-g}\,
 \bar\psi(-\gamma^\mu\nabla_\mu+m)\psi .
 \label{eq:free-dirac-action}
\end{equation}
Metric variation gives the symmetric (Hilbert) energy-momentum tensor.
In flat spacetime it agrees, up to improvement terms and the field
equations, with the tensor obtained by the Belinfante--Rosenfeld
symmetrization of the canonical current
\cite{Belinfante1939,Belinfante1940,Rosenfeld1940,Hehl1976}:
\begin{align}
 T_{\mu\nu}^{\psi}
 &=\frac14\left[
 \bar\psi\gamma_\mu\nabla_\nu\psi
 -(\nabla_\nu\bar\psi)\gamma_\mu\psi
 \right]
 \nonumber\\
 &\quad+\frac14\left[
 \bar\psi\gamma_\nu\nabla_\mu\psi
 -(\nabla_\mu\bar\psi)\gamma_\nu\psi
 \right]
 \nonumber\\
 &\quad
 +g_{\mu\nu}\left[
 m\bar\psi\psi
 -\frac12\bar\psi\gamma^\alpha\nabla_\alpha\psi
 +\frac12(\nabla_\alpha\bar\psi)\gamma^\alpha\psi
 \right].
 \label{eq:free-dirac-emt}
\end{align}
The terms in the last line vanish on shell but are kept when defining the composite operator by
metric variation.

In flat spacetime, let \(E_k=\sqrt{k^2+m^2}\).  We use
\begin{align}
 \psi(x)
 &=\int\frac{d^3k}{(2\pi)^{3/2}\sqrt{2E_k}}
 \sum_{r=1}^{2}
 \left[
 c_{\boldsymbol{k}r}u(\boldsymbol{k},r)
 e^{-iE_kt+i\boldsymbol{k}\cdot\boldsymbol{x}}
 \right.
 \nonumber\\
 &\hspace{3.3cm}\left.
 +d_{\boldsymbol{k}r}^{\dagger}v(\boldsymbol{k},r)
 e^{iE_kt-i\boldsymbol{k}\cdot\boldsymbol{x}}
 \right],
 \label{eq:free-dirac-mode-expansion-main}
\end{align}
with
\begin{equation}
 \{c_{\boldsymbol{k}r},c_{\boldsymbol{p}s}^{\dagger}\}
 =\{d_{\boldsymbol{k}r},d_{\boldsymbol{p}s}^{\dagger}\}
 =\delta_{rs}\delta^{(3)}(\boldsymbol{k}-\boldsymbol{p}) .
\end{equation}
Define \(p^\mu=(E_k,\boldsymbol k)\), so that
\(p_\mu=(-E_k,\boldsymbol k)\) and \(p^2=-m^2\).  In our conventions, we have the following spin sums,
\begin{align}
 \sum_r u(\boldsymbol{k},r)\bar u(\boldsymbol{k},r)
 =-i\slashed p-m,
 \nonumber\\
 \sum_r v(\boldsymbol{k},r)\bar v(\boldsymbol{k},r)
 =-i\slashed p+m .
 \label{eq:free-dirac-spin-sums-main}
\end{align}
Appendix~\ref{app:fermion-modes} verifies that these
normalizations reproduce the equal-time anticommutator.

The two spin states and the fermionic zero-point sign give
\begin{align}
 \langle\rho\rangle_\psi^{\rm vac}
 &=-2\int\frac{d^3k}{(2\pi)^3}E_k,
 \nonumber\\
 \langle p\rangle_\psi^{\rm vac}
 &=-\frac23\int\frac{d^3k}{(2\pi)^3}\frac{k^2}{E_k} .
 \label{eq:free-dirac-bare-rho-p}
\end{align}
Consequently, dimensional regularization gives a relation between the scalar and fermionic cases in flat spacetime:
\begin{equation}
 \langle\rho\rangle_\psi^{\rm vac}
 =-4\langle\rho\rangle_\Phi^{\rm vac},
 \qquad
 \langle p\rangle_\psi^{\rm vac}
 =-4\langle p\rangle_\Phi^{\rm vac}
 =-\langle\rho\rangle_\psi^{\rm vac} .
 \label{eq:free-dirac-flat-emt}
\end{equation}
The proportionality factor \(-4\) is a flat-space artifact and stems from the Grassmann nature of the fermionic fields and the number of degrees of freedom encoded in the Dirac spinor. However, the relation between the scalar and Dirac field energy density and pressure is much more complicated in curved spacetimes because the spin connection has nonzero
bundle curvature.

The Gaussian integral over Grassmann gives
\begin{align}
 Z_\psi[0;g]\propto\det\sD_\psi,\quad 
 W_\psi=-i\operatorname{Tr}\ln\sD_\psi,
 \label{eq:free-dirac-determinant}
\end{align}
with $\sD_\psi=-\gamma^\mu\nabla_\mu+m$. To find the effective action, introduce
\begin{equation}
 \widetilde{\sD}_\psi
 =\gamma^5\sD_\psi\gamma^5
 =\gamma^\mu\nabla_\mu+m
\end{equation}
and define
\begin{equation}
 W_\psi^{(+)}
 =-\frac{i}{2}
 \operatorname{Tr}\ln(\sD_\psi\widetilde{\sD}_\psi) .
 \label{eq:parity-even-determinant}
\end{equation}
This symmetrized determinant determines the parity-even local
gravitational operators considered here.  The phase of the first-order
determinant and possible regularized multiplicative
anomalies are not relevant to our purposes
\cite{Hawking1977,ElizaldeVanzoZerbini1998}.

We now make use of the Schr\"odinger--Lichnerowicz identity, which relates the square of the covariant Dirac operator to the
connection Laplacian on the spin bundle and the scalar curvature \cite{Lichnerowicz1963,Vassilevich2003,ParkerToms2009},
\begin{equation}
 (\gamma^\mu\nabla_\mu)^2
 =\nabla_{\rm spin}^{2}-\frac14R,
 \label{eq:lichnerowicz-free}
\end{equation}
where we defined
\begin{equation}
 \nabla_{\rm spin}^{2}
 \equiv g^{\mu\nu}
 \left(\nabla_\mu\nabla_\nu
 -\Gamma^\rho_{\mu\nu}\nabla_\rho\right).
 \label{eq:spin-connection-laplacian}
\end{equation}

It follows that
\begin{equation}
 K_\psi
 \equiv\sD_\psi\widetilde{\sD}_\psi
 =-\nabla_{\rm spin}^2+\frac14R+m^2 .
 \label{eq:spinor-second-order-operator}
\end{equation}
See Appendix
\ref{app:dirac-heat-kernel} for more details about this step.  The
traced spinor heat-kernel coefficients through fourth adiabatic order
are \cite{Vassilevich2003,ParkerToms2009}
\begin{align}
 f_0^\psi&=4,
 \nonumber\\
 f_1^\psi&=-\frac13R,
 \nonumber\\
 f_2^\psi
 &=\frac{1}{360}\left(
 5R^2-8R_{\mu\nu}R^{\mu\nu}
 -7R_{\mu\nu\rho\sigma}R^{\mu\nu\rho\sigma}
 -12\Box R
 \right).
 \label{eq:dirac-f1f2-free}
\end{align}
We keep four-component Clifford traces while analytically continuing
the momentum and proper-time dimension to \(d=4-\epsilon\).  This
choice is part of the regularization scheme.

Using Eq.~\eqref{eq:spinor-second-order-operator}, the local
parity-even action is
\begin{widetext}
\begin{align}
 W_\psi^{\rm loc}
 =-\int d^dx\sqrt{-g}\,
 \frac{1}{2(4\pi)^{d/2}}
 \left(\frac{m}{\mu_{\rm DR}}\right)^{d-4}
 \left[
 m^4f_0^\psi\Gamma\left(-\frac d2\right)
 +m^2f_1^\psi\Gamma\left(1-\frac d2\right)
 +f_2^\psi\Gamma\left(2-\frac d2\right)
 \right].
 \label{eq:free-dirac-local-action}
\end{align}
\end{widetext}
Equivalently, its expansion around $d=4$ is
\begin{widetext}
\begin{align}
 W_\psi^{\rm loc}
 =-\int d^dx\sqrt{-g}\Bigg\{
 &\left[
 \frac{2}{\epsilon}-\gamma_{\rm EM}
 -\ln\left(\frac{m^2}{4\pi\mu_{\rm DR}^2}\right)
 \right]
 \left(
 \frac{f_0^\psi m^4}{64\pi^2}
 -\frac{m^2f_1^\psi}{32\pi^2}
 +\frac{f_2^\psi}{32\pi^2}
 \right)+\frac{3f_0^\psi m^4}{128\pi^2}
 -\frac{m^2f_1^\psi}{32\pi^2}
 \Bigg\}.
 \label{eq:free-dirac-local-action-laurent}
\end{align}
\end{widetext}
In flat spacetime only the volume term remains, and metric variation
again gives \(\langle p\rangle=-\langle\rho\rangle\), in agreement
with Eq.~\eqref{eq:free-dirac-flat-emt}.  

In the flat space limit, the NJL
saddle changes the determinant only through
\begin{equation}
 m\longrightarrow m_{\rm eff}=m+\Theta
 \label{eq:free-to-njl-mass-replacement}
\end{equation}
when \(\Theta\) is constant.  The free spinor determinant therefore
provides the one-loop kernel for the mean-field calculation. However, this changes dramatically in curved spacetimes because the
auxiliary potential and saddle equation that encode the four-fermion
dynamics have non-trivial metric dependence. In the next section we investigate the quantum effective action for the NJL model in curved spacetimes.

\section{One-Loop Action with Four-Fermion Terms}
\label{sec:njl-one-loop}

The discussion in Sec. \ref{subsec:fermionic-functional-regimes} separates three uses of a local
four-fermion operator: perturbative contact scattering, vacuum
particle--antiparticle condensation, and finite-density Cooper pairing.
We now focus on the second of these regimes.  Our purpose is first to
explain how the Nambu--Jona-Lasinio (NJL) saddle reorganizes perturbation
theory and then to compute the corresponding vacuum energy-momentum
tensor.  This order is useful because it makes clear which infinite
classes of diagrams are retained by the mean-field approximation before
the same approximation is implemented through a functional determinant.

\subsection{A short review of the NJL model}
\label{subsec:njl-review}

The NJL model was introduced as a field-theoretic realization of
dynamical symmetry breaking, with the fermion mass generated by a
nonzero scalar bilinear rather than inserted only as an explicit
parameter \cite{NambuJonaLasinio1961a,NambuJonaLasinio1961b}. In a
chirally symmetric realization, the scalar operator $(\bar{\psi}\psi)^2$ is accompanied by
the corresponding pseudoscalar channel $(\bar{\psi}\gamma_5 \psi)^2$.
In a multichannel model, the scalar-only saddle is a consistent
truncation when the state is parity even, has vanishing density and
currents, and the omitted channel equations admit the zero solution
without mixing with the scalar background. Torsion-induced or
Fierz-related interactions likewise require an explicit channel
projection before they can be identified with the scalar model used
here \cite{Hehl1976,PisarskiRischke1999,FreidelMinicTakeuchi2005} (see also the appendix in \cite{Alexander:2024qml}). Standard reviews of
the model, including its large-flavor and finite-density extensions,
are given in Refs.~\cite{Klevansky1992,Buballa2005}. 

Although the pseudoscalar coupling might be present in a
general chiral NJL model, a massive or dynamically massive
theory does not support an axial chemical potential as an ordinary equilibrium chemical
potential without additional ultraviolet subtractions because the chiral symmetry is broken
\cite{RuggieriChernodubLu2020}. So, the pseudoscalar channel is not used
here to define an axial-density state, and its homogeneous auxiliary
field is set to the parity-even saddle value. This should be distinguished from massless chiral-imbalanced models, where
the scalar and pseudoscalar auxiliary fields can be viewed as the radial
and angular components of a single chiral order parameter
\cite{Tong2024}.  In the present parity-even vacuum branch, however, the
explicit fermion mass and the absence of an axial-density source tilt this
chiral orientation toward the scalar direction, so the homogeneous
pseudoscalar auxiliary field is consistently set to zero. For discussions of pseudoscalar condensation in a cosmological context, see \cite{Tong2024}. 

In four spacetime dimensions a four-fermion coupling has mass dimension
minus two.  The model is therefore not perturbatively renormalizable in
the sense of requiring an infinite set of counterterms.  It is,
however, a consistent effective field theory once a physical cutoff,
the allowed operator basis, and matching conditions are specified.  In
this interpretation, the dimensionless expansion parameter is of the
form \(\lambda E^2\), where \(E\) is the characteristic energy and
\(\lambda\) is the scalar-channel, four-fermion coupling.  Results involving the
nontrivial saddle consequently depend on the regulator and matching
prescription, as expected for an effective interaction
\cite{Klevansky1992,Polchinski1992,Shankar1994,Buballa2005}.

In cosmological models \cite{AlexanderVaid2006,Poplawski2011,Poplawski2012,Magueijo:2012ug, Weller2013,
QuintanarMacorra2015I,QuintanarMacorra2015,LucatProkopec2017,Tukhashvili2024, Alexander:2025whu, Alexander:2026,AlexanderCalcagni2009,AlexanderBiswas2009,Alexander2017,Alexander:2020wpm,Tong2024, Alexander:2024qml, Liang:2024xww}, the mean-field approximation of the condensate is frequently used. In that context, ``mean-field approximation'' denotes a family of
self-consistent truncations akin to approximations historically used in condensed matter.
Weiss's molecular-field treatment of ferromagnetism is a foundational
statistical-mechanical example: the interactions of a magnetic moment
with the remaining system are represented by an internal field
proportional to the average magnetization \cite{Weiss1907}.  Related
factorizations were subsequently used by Bragg and Williams for order
in alloys \cite{BraggWilliams1934}, while Hartree developed a
self-consistent one-particle field for interacting atomic electrons
\cite{Hartree1928}.  For our purposes, what matters here is the precise NJL
truncation, which we now define.

Write the scalar bilinear as
\begin{equation}
 \bar\psi\psi=\Pi+\delta\mathcal O,
 \qquad
 \Pi\equiv\langle\bar\psi\psi\rangle .
 \label{eq:njl-mean-field-split}
\end{equation}
An exact useful identity is
\begin{equation}
 (\bar\psi\psi)^2
 =2\Pi\bar\psi\psi-\Pi^2+(\delta\mathcal O)^2.
 \label{eq:njl-mean-field-identity}
\end{equation}
This separates the background expectation value from its connected
fluctuations.  In this paper, the mean-field approximation is defined
by neglecting the last term in
Eq.~\eqref{eq:njl-mean-field-identity} and determining \(\Pi\)
self-consistently.  Equivalently, after a Hubbard--Stratonovich
transformation it is the saddle-point approximation for the auxiliary
scalar field.  For \(N\) identical fermion species with
\(\lambda\sim N^{-1}\), this saddle is the leading term of the
large-\(N\) expansion \cite{Klevansky1992,Buballa2005}.  For the one-field effective theory used below,
it is instead a truncation whose reliability depends on the size of the
omitted scalar and other collective
fluctuations \cite{Klevansky1992,Buballa2005}.

The same truncation has a diagrammatic interpretation that we explain next. Writing
\(\psi(x)=\int_p e^{-ip\cdot x}\psi(p)\), where
\(\int_p\equiv\int d^dp/(2\pi)^d\), we define the momentum-space kinetic operators and propagators by
\begin{align}
 \mathscr D_0(p)&\equiv i\slashed p+m, \quad S_0(p)\equiv\mathscr D_0^{-1}(p) =\frac{-i\slashed p+m}{p^2+m^2},
 \label{eq:njl-free-propagator}
 \\
 \mathscr D_\Theta(p)&\equiv \mathscr D_0(p)+\Theta\mathbf1_4, \quad S_\Theta(p)\equiv\mathscr D_\Theta^{-1}(p).
 \label{eq:njl-mean-field-propagator}
\end{align}
Expanding the inverse operator yields
\begin{align}
 S_\Theta
 &=S_0-S_0\Theta S_0
   +S_0\Theta S_0\Theta S_0-\cdots
 \nonumber\\
 &=\left(S_0^{-1}+\Theta\right)^{-1},
 \qquad
 \Theta=2\lambda\Pi[S_\Theta] .
 \label{eq:njl-dyson-series}
\end{align}
In standard Dyson notation \cite{Dyson1949},
\begin{equation}
 S_\Theta^{-1}=S_0^{-1}-\Sigma_{\rm dir},
 \qquad
 \Sigma_{\rm dir}\equiv-\Theta\mathbf1_4 .
 \label{eq:njl-dyson-self-energy}
\end{equation}
The local insertion is the self-consistent direct tadpole
contribution, commonly called the Hartree contribution in the NJL
literature \cite{Klevansky1992,Buballa2005}. Figure~\ref{fig:njl-dyson} displays the
geometric series in Eq.~\eqref{eq:njl-dyson-series}.

\begin{figure*}[t]
\begin{minipage}[t]{0.48\textwidth}
\centering
\begin{tikzpicture}[
 x=0.69cm,y=0.55cm,
 fermion/.style={draw=black,line width=0.75pt,
 postaction={decorate},decoration={markings,
 mark=at position 0.56 with {\arrow{>}}}}
]
 \draw[fermion,double,double distance=1.05pt] (0,0)--(1.5,0);
 \node[below] at (0.75,-0.25) {\scriptsize \(S_\Theta\)};
 \node at (1.9,0) {\(=\)};
 \draw[fermion] (2.25,0)--(3.75,0);
 \node[below] at (3.0,-0.25) {\scriptsize \(S_0\)};
 \node at (4.15,0) {\(-\)};
 \draw[fermion] (4.5,0)--(5.25,0);
 \node at (5.25,0) {\(\times\)};
 \draw[fermion] (5.25,0)--(6.0,0);
 \node[above] at (5.25,0.25) {\scriptsize \(\Theta\)};
 \node at (6.4,0) {\(+\)};
 \draw[fermion] (6.75,0)--(7.30,0);
 \node at (7.30,0) {\(\times\)};
 \draw[fermion] (7.30,0)--(7.95,0);
 \node at (7.85,0) {\(\times\)};
 \draw[fermion] (7.95,0)--(8.40,0);
 \node[above] at (7.30,0.25) {\scriptsize \(\Theta\)};
 \node[above] at (7.85,0.25) {\scriptsize \(\Theta\)};
 \node at (8.83,0) {\(\cdots\)};
\end{tikzpicture}
\caption{Direct mean-field resummation of the fermion propagator.  A
single line is \(S_0\), a double line is \(S_\Theta\), and each cross
is the scalar insertion \(\Theta\mathbf1_4\).  The value of the
insertion is fixed self-consistently by the last relation in
Eq.~\eqref{eq:njl-dyson-series}.}
\label{fig:njl-dyson}
\end{minipage}
\hfill
\begin{minipage}[t]{0.48\textwidth}
\centering
\begin{tikzpicture}[
 x=0.64cm,y=0.58cm,
 fermion/.style={draw=black,line width=0.75pt,
 postaction={decorate},decoration={markings,
 mark=at position 0.56 with {\arrow{>}}}},
 vertex/.style={circle,fill=black,inner sep=1.45pt}
]
 \node at (0,0) {\(\mathcal A_S\)};
 \node at (1.15,0) {\(=\)};
 \node[vertex] at (1.75,0) {};
 \node at (2.35,0) {\(+\)};
 \node[vertex] at (2.95,0) {};
 \node[vertex] at (4.05,0) {};
 \draw[fermion] (2.95,0) arc[start angle=180,end angle=0,
 x radius=0.55,y radius=0.42];
 \draw[fermion] (4.05,0) arc[start angle=0,end angle=-180,
 x radius=0.55,y radius=0.42];
 \node at (4.62,0) {\(+\)};
 \node[vertex] at (5.18,0) {};
 \node[vertex] at (6.28,0) {};
 \node[vertex] at (7.38,0) {};
 \draw[fermion] (5.18,0) arc[start angle=180,end angle=0,
 x radius=0.55,y radius=0.42];
 \draw[fermion] (6.28,0) arc[start angle=0,end angle=-180,
 x radius=0.55,y radius=0.42];
 \draw[fermion] (6.28,0) arc[start angle=180,end angle=0,
 x radius=0.55,y radius=0.42];
 \draw[fermion] (7.38,0) arc[start angle=0,end angle=-180,
 x radius=0.55,y radius=0.42];
 \node at (8.05,0) {\(+\cdots\)};
\end{tikzpicture}
\caption{Scalar RPA series.  Each black point is the reduced contact
kernel \(\mathcal V_S=2\lambda\), and each fermion loop is
\(\mathcal B_S(q)\).  The first three terms are those displayed in
Eq.~\eqref{eq:njl-bubble-chain}.}
\label{fig:njl-rpa-chain}
\end{minipage}
\end{figure*}

The fermion--antifermion four-point function contains a second
selected resummation.  With the normalization of
Eq.~\eqref{eq:njl-flat-action}, define the reduced scalar-composite
contact kernel and the dressed one-loop polarization by
\begin{align}
 \mathcal V_S&\equiv2\lambda,
 \label{eq:njl-scalar-vertex}
 \\
 \mathcal B_S(q)
 &\equiv-i\int\frac{d^dk}{(2\pi)^d}\,
 \operatorname{tr}_{D}\!\left[
 S_\Theta(k+q)S_\Theta(k)
 \right].
 \label{eq:njl-scalar-bubble}
\end{align}
Here \(\operatorname{tr}_{D}\) is the Dirac trace, and the definition
includes the closed-fermion-loop sign.  A full elementary
four-fermion amplitude additionally requires antisymmetrized external
legs and any flavor or color projectors.  Iterating the reduced kernel
in the scalar composite channel gives
\begin{align}
 \mathcal A_S(q)
 &=\mathcal V_S
 +\mathcal V_S\mathcal B_S(q)\mathcal V_S
 +\mathcal V_S\mathcal B_S(q)\mathcal V_S
     \mathcal B_S(q)\mathcal V_S+\cdots
 \nonumber\\
 &=\frac{\mathcal V_S}
 {1-\mathcal B_S(q)\mathcal V_S} .
 \label{eq:njl-bubble-chain}
\end{align}
This is the random-phase approximation (RPA), or scalar bubble-chain
resummation, built from mean-field fermion lines
\cite{Klevansky1992,Buballa2005}.  A zero of
\(1-\mathcal B_S(q)\mathcal V_S\) is a pole of the scalar
fermion--antifermion amplitude and signals a collective mode or an
instability of the saddle.  The series is displayed in
Fig.~\ref{fig:njl-rpa-chain}.

The two resummations organize different correlation functions.  The
self-consistent tadpole determines the fermion two-point function and
mass shift, whereas the RPA chain determines the scalar
fermion--antifermion four-point function.  With
\(\lambda\sim N^{-1}\), both are leading structures in the standard
large-\(N\) organization, but neither contains every Feynman diagram.
Crossed ladders, exchange terms outside the retained channel, vertex
corrections, and auxiliary-field loops occur beyond this truncation \cite{Klevansky1992}.

Having clarified the meaning of the mean-field approximation, we next apply it to compute the vacuum expectation value of the
flat-space energy-momentum tensor. This provides an operator-level
baseline for the subsequent curved-spacetime functional discussion.

\subsection{Flat-space NJL mean-field energy-momentum tensor}
\label{sec:njl-flat}

Cosmological applications often use a homogeneous scalar bilinear as a
source of dynamical mass or vacuum energy.
Before curvature is introduced, the flat-space calculation is the clean
place to fix the condensate normalization, the auxiliary contribution,
and the vacuum equation of state that such models inherit.

We use the conventions established in the free-field discussion of Sec. \ref{sec:free-fields}.  Up
to a boundary term, the scalar-channel action in \(d\) dimensions is
\begin{align}
 S_{\rm 4F}
 &=\int d^d x\left[
 -\frac12\bar\psi\gamma^\mu
       \overleftrightarrow{\partial_\mu}\psi
 +m\bar\psi\psi
 +\lambda(\bar\psi\psi)^2
 \right],
 \nonumber\\
 \lambda&\equiv\frac{1}{M_{\rm 4F}^2},
 \label{eq:njl-flat-action}
\end{align}
where
\(\bar\psi\gamma^\mu\overleftrightarrow{\partial_\mu}\psi
\equiv\bar\psi\gamma^\mu\partial_\mu\psi
-(\partial_\mu\bar\psi)\gamma^\mu\psi\), \(m\) is the bare fermion
mass, and \(M_{\rm 4F}\) is the effective-theory scale.  With the sign
in Eq.~\eqref{eq:njl-flat-action}, positive \(\lambda\) is the
attractive scalar convention used throughout this paper.  

Applying Eq.~\eqref{eq:njl-mean-field-identity}, we define
\begin{equation}
 \Pi\equiv\langle\bar\psi\psi\rangle,
 \qquad
 \Theta\equiv2\lambda\Pi,
 \qquad
 m_{\rm eff}\equiv m+\Theta .
 \label{eq:njl-Theta-definitions}
\end{equation}
The mean-field Lagrangian is then
\begin{align}
 \mathcal L_{\rm MF}
 &=-\frac12\bar\psi\gamma^\mu
       \overleftrightarrow{\partial_\mu}\psi
   +m_{\rm eff}\bar\psi\psi-V_{\rm aux},
 \nonumber\\
 V_{\rm aux}
 &\equiv\lambda\Pi^2
 =\frac{\Theta^2}{4\lambda},
 \label{eq:njl-flat-mean-field-lagrangian}
\end{align}
where \(\Theta\) denotes the particle--antiparticle NJL mass shift.

The quadratic theory in Eq.~\eqref{eq:njl-flat-mean-field-lagrangian}
has the one-particle energy
\begin{equation}
 E_k(\Theta)
 =\sqrt{\boldsymbol k^2+m_{\rm eff}^2(\Theta)}.
 \label{eq:njl-effective-energy}
\end{equation}
Using the mode normalization of Appendix~\ref{app:fermion-modes},
the empty massive vacuum gives
\begin{equation}
 \Pi(\Theta)
 =2m_{\rm eff}(\Theta)
 \int\frac{d^3k}{(2\pi)^3}
 \frac{1}{\sqrt{\boldsymbol k^2+m_{\rm eff}^2(\Theta)}}.
 \label{eq:njl-condensate-mode-integral}
\end{equation}
Together with Eq.~\eqref{eq:njl-Theta-definitions}, this becomes the
gap equation
\begin{equation}
 \Theta
 =4\lambda m_{\rm eff}(\Theta)
 \int\frac{d^3k}{(2\pi)^3}
 \frac{1}{\sqrt{\boldsymbol k^2+m_{\rm eff}^2(\Theta)}}.
 \label{eq:njl-gap-unregulated}
\end{equation}
This integral requires an effective-theory prescription.  For example,
with a sharp three-momentum cutoff \(\Lambda\), the chiral limit has
\(m_{\rm eff}(\Theta)=\Theta\).  On the nonzero branch, dividing the
gap equation by \(\Theta\) gives Eq. \eqref{eq:njl-critical-cutoff-equation} and a critical coupling of $\mathcal{O}(1/\Lambda^2)$. For comparison with the functional calculation, dimensional
regularization of the integral factor gives
\begin{align}
 I_d(m_{\rm eff})
 &\equiv
 \mu_{\rm DR}^{4-d}
 \int\frac{d^{d-1}k}{(2\pi)^{d-1}}
 \frac{1}{(\boldsymbol k^2+m_{\rm eff}^2)^{1/2}}
 \nonumber\\
 &=\frac{\mu_{\rm DR}^{4-d}}
 {(4\pi)^{(d-1)/2}}
 \frac{\Gamma(1-d/2)}{\Gamma(1/2)}
 \left(m_{\rm eff}^2\right)^{(d-2)/2},
 \label{eq:njl-gap-master-integral}
\end{align}
and hence
\begin{equation}
 \Theta=4\lambda m_{\rm eff}(\Theta)I_d(m_{\rm eff}(\Theta)).
 \label{eq:njl-gap-dimensional}
\end{equation}
Dimensional regularization is useful for matching the covariant
calculation, but by itself it does not supply the physical NJL matching
condition encoded by \(\Lambda\) in Eq.~\eqref{eq:njl-critical-coupling-example}.
In particular, analytic continuation discards the quadratic power term
that produced the numerical cutoff criterion.  A constraint on
\(\lambda\) in dimensional regularization actually requires a physical
matching condition.  At a reference scale \(\mu_\star\), one may, for
example, fix a chosen nonzero vacuum gap \(M_\star\) through
\begin{equation}
 \frac{1}{\lambda_R(\mu_\star)}
 =4I_R(M_\star;\mu_\star),
 \label{eq:njl-dr-gap-matching}
\end{equation}
where \(I_R\) is the subtracted version of
Eq.~\eqref{eq:njl-gap-master-integral}.  Equivalently, one may match the
renormalized scalar susceptibility,
\begin{equation}
 \chi_{S,R}^{-1}(0;\mu_\star)
 =\frac{1}{2\lambda_R(\mu_\star)}
 -\mathcal B_{S,R}(0;\mu_\star),
 \label{eq:njl-susceptibility-matching}
\end{equation}
and define criticality by \(\chi_{S,R}^{-1}=0\).  The resulting
dimensionless constraint on \(\lambda_R\mu_\star^2\) is scheme- and
matching-dependent. This is the expected EFT situation: the
subtraction scale is not a substitute for a physical ultraviolet
matching scale.

The gap equation equivalently follows by extremizing the flat-space mean-field
potential. Define (we assume \(m_{\rm eff}>0\) when writing fractional powers below)
\begin{equation}
 \mathcal P_d(m_{\rm eff})
 \equiv\frac{1}{(4\pi)^{d/2}}
 \left(\frac{m_{\rm eff}}{\mu_{\rm DR}}\right)^{d-4}.
 \label{eq:njl-Pd-definition}
\end{equation}
As can be seen from Eqs. \eqref{eq:free-dirac-local-action} and \eqref{eq:free-to-njl-mass-replacement}, the quantum effective potential is
\begin{equation}
 V_{\rm eff}^{\rm flat}(\Theta)
 =\frac{\Theta^2}{4\lambda}
 +2\mathcal P_d(m_{\rm eff})m_{\rm eff}^4
 \Gamma\left(-\frac d2\right).
 \label{eq:njl-flat-effective-potential}
\end{equation}
Since \(\partial m_{\rm eff}/\partial\Theta=1\), its stationary
condition is
\begin{equation}
 0=\frac{\Theta}{2\lambda}
 +2d\mathcal P_d(m_{\rm eff})m_{\rm eff}^3
 \Gamma\left(-\frac d2\right).
 \label{eq:njl-flat-potential-gap}
\end{equation}
Using
\(\Gamma(1-d/2)=-(d/2)\Gamma(-d/2)\),
Eq.~\eqref{eq:njl-flat-potential-gap} is precisely
Eq.~\eqref{eq:njl-gap-dimensional}.  This is the stationary form of the
self-consistency condition in Eq.~\eqref{eq:njl-dyson-series}.  With
the standard Dyson definition
\(S^{-1}=S_0^{-1}-\Sigma\), the present inverse-mass insertion is
\(\Sigma_{\rm dir}=-\Theta\mathbf1_4\), as stated in
Eq.~\eqref{eq:njl-dyson-self-energy}.  Thus the effective-potential
saddle and the direct Hartree equation determine the same quadratic
kernel within this truncation \cite{Klevansky1992,Buballa2005}.

We next compute the main observable needed in semiclassical gravity, $\langle \hat{T}_{\mu\nu}\rangle$. The mean-field energy-momentum tensor
operator is
\begin{align}
 T_{\mu\nu}^{\rm MF}
 &=\frac14\left(
 \bar\psi\gamma_\mu\overleftrightarrow{\partial_\nu}\psi
 +\bar\psi\gamma_\nu\overleftrightarrow{\partial_\mu}\psi
 \right)
 +\eta_{\mu\nu}\mathcal L_{\rm MF}.
 \label{eq:njl-flat-energy-momentum-operator}
\end{align}
The full functional computation of its expectation value for an arbitrary metric is revisited in the next section. Here we will collect the results from the operator approach in flat spacetime. The mode contractions are given in
Appendix~\ref{app:mean-field-emt-details}.  Before regularization,
they reduce to
\begin{align}
 \langle\rho\rangle_{\rm MF}
 &=-2\int\frac{d^{d-1}k}{(2\pi)^{d-1}}E_k
 +\frac{\Theta^2}{4\lambda},
 \nonumber\\
 \langle p\rangle_{\rm MF}
 &=-\frac{2}{d-1}
 \int\frac{d^{d-1}k}{(2\pi)^{d-1}}
 \frac{\boldsymbol k^2}{E_k}
 -\frac{\Theta^2}{4\lambda}.
 \label{eq:njl-flat-rho-p-unregulated}
\end{align}
These formulas display separately the fermion zero-point term and the
classical auxiliary contribution.  Applying the same dimensional
continuation to both components yields
\begin{align}
 \langle\rho\rangle_{\rm MF}
 &=-2\mu_{\rm DR}^{4-d}
 \frac{m_{\rm eff}^d}{(4\pi)^{(d-1)/2}}
 \frac{\Gamma(-d/2)}{\Gamma(-1/2)}
 +\frac{\Theta^2}{4\lambda},
 \nonumber\\
 \langle p\rangle_{\rm MF}
 &=+2\mu_{\rm DR}^{4-d}
 \frac{m_{\rm eff}^d}{(4\pi)^{(d-1)/2}}
 \frac{\Gamma(-d/2)}{\Gamma(-1/2)}
 -\frac{\Theta^2}{4\lambda}.
 \label{eq:njl-flat-rho-p-dimensional}
\end{align}
Thus a homogeneous vacuum saddle in flat spacetime has
\begin{equation}
 \langle p\rangle_{\rm MF}=-\langle\rho\rangle_{\rm MF}.
 \label{eq:njl-flat-vacuum-equation-of-state}
\end{equation}
This result is a regulated vacuum relation. We discuss the relation to the renormalized cosmological constant in Sec. \ref{subsec:njl-flrw-renormalization}.

The agreement between the mode sum and the stationary potential fixes
the flat-space normalization of the saddle.  We now retain the same
normalization while replacing the momentum modes approach by the covariant
spinor determinant (functional) one, as needed to generalize the results to curved spacetimes. 

\subsection{Vacuum NJL effective action in curved spacetime}
\label{sec:njl-curved}

We first construct the parity-even local vacuum action for a possibly
spacetime-dependent condensate. The assumptions at this stage are: one fermion loop,
the scalar-channel mean-field saddle, and no finite-density occupation data. State-dependent nonlocal terms and
particle production do not show up in this local action, so we can use the in-out effective action determinant to obtain the local counterterm coefficients \cite{Jordan1986,CalzettaHu1987,HuVerdaguer2008}.

The calculation is a generalization of the flat spacetime case discussed in Sec. \ref{subsec:vacuum-njl-saddle} and Sec. \ref{sec:njl-flat}. The Hubbard--Stratonovich identity can be written directly in terms of
the mass shift \(\Theta\):
\begin{align}
 &\exp\left[
 i\int d^d x\sqrt{-g}\,
 \lambda(\bar\psi\psi)^2\right]
 \nonumber\\
 &\qquad\propto
 \int\mathcal D\Theta\,
 \exp\left\{i\int d^d x\sqrt{-g}
 \left[-\frac{\Theta^2}{4\lambda}
 +\Theta\bar\psi\psi\right]\right\}.
 \label{eq:njl-HS-Theta}
\end{align}
Consequently,
\begin{align}
 S_{\rm aux}
 &=\int d^d x\sqrt{-g}
 \left[
 \bar\psi\mathscr D_M\psi
 -\frac{\Theta^2}{4\lambda}
 \right],
 \nonumber\\
 M(x)&\equiv m_{\rm eff}(\Theta(x))=m+\Theta(x),
 \nonumber\\
 \mathscr D_M
 &=-\gamma^\mu\nabla_\mu+M(x) .
 \label{eq:njl-curved-auxiliary-action}
\end{align}
At the saddle, variation with respect to \(\Theta\) reproduces
\(\Theta=2\lambda\langle\bar\psi\psi\rangle\).

For a general \(M(x)\), the parity-even determinant is reduced to
Laplace type using
\begin{align}
 \widetilde{\mathscr D}_M
 &=\gamma^5\mathscr D_M\gamma^5
 =\gamma^\mu\nabla_\mu+M,
 \nonumber\\
 K[M]
 &\equiv
 \mathscr D_M\widetilde{\mathscr D}_M
 =-\nabla_{\rm spin}^2+\frac14R+M^2
 -\gamma^\mu\nabla_\mu M.
 \label{eq:njl-curved-Laplace-operator}
\end{align}
The last term is present whenever the condensate varies and is the
source of derivative operators for \(\Theta\).  For the unfactored
operator \(K[M]\), denote the traced heat-kernel coefficients by
\(b_n^\psi[M]\).  Through the order needed to expose two-derivative
condensate terms,
\begin{align}
 b_0^\psi[M]&=4,
 \nonumber\\
 b_1^\psi[M]&=-4M^2-\frac13R,
 \nonumber\\
 b_2^\psi[M]
 &={}2M^4+\frac13RM^2+2(\nabla M)^2
 -\frac23\Box M^2
 \nonumber\\
 &\quad+\frac1{360}\left(
 5R^2-8R_{\mu\nu}R^{\mu\nu}
 \right.
 \nonumber\\
 &\hspace{1.8cm}\left.
 -7R_{\mu\nu\rho\sigma}R^{\mu\nu\rho\sigma}
 -12\Box R\right).
 \label{eq:njl-varying-M-heat-kernel}
\end{align}
Equivalently,
\begin{align}
 b_2^\psi[M]
 ={}&f_2^\psi-M^2f_1^\psi
 +\frac12M^4f_0^\psi
 \nonumber\\
 &+2(\nabla M)^2-\frac23\Box M^2.
 \label{eq:njl-varying-M-factorization}
\end{align}
Because \(\nabla_\mu M=\nabla_\mu\Theta\), the integrated
\(\Box M^2\) term is a boundary term, while
\(2(\nabla\Theta)^2\) produces the local kinetic counterterm.  In the
present one-Dirac-field convention its pole contribution is
\begin{equation}
 W_{\rm div}^{(+)}
 \supset-\frac{1}{8\pi^2\epsilon}
 \int d^4x\sqrt{-g}\,(\nabla\Theta)^2 .
 \label{eq:njl-Theta-kinetic-divergence}
\end{equation}
The pole and scale dependence of the kinetic coefficient are therefore
fixed locally; its complete finite renormalized value requires a
two-point matching condition. Eqs.~\eqref{eq:njl-curved-Laplace-operator}--\eqref{eq:njl-Theta-kinetic-divergence} are derived in Appendix~\ref{app:dirac-heat-kernel}.

For constant \(M=m_{\rm eff}\), the derivative terms vanish and the
mass dependence can be factored from the proper-time series.  The
remaining traced coefficients are
\begin{align}
 f_0^\psi&=4,
 \qquad
 f_1^\psi=-\frac13R,
 \nonumber\\
 f_2^\psi
 &=\frac1{360}\left(
 5R^2-8R_{\mu\nu}R^{\mu\nu}
 -7R_{\mu\nu\rho\sigma}R^{\mu\nu\rho\sigma}
 -12\Box R\right).
 \label{eq:njl-spinor-heat-kernel-coefficients}
\end{align}
Defining \(\mathcal P_d(m_{\rm eff})\) as in
Eq.~\eqref{eq:njl-Pd-definition}, the regulated local action is
\begin{align}
 W_{\rm MF}^{\rm reg}
 &=-\int d^d x\sqrt{-g}\,
 \frac{\Theta^2}{4\lambda}
 \nonumber\\
 &\quad
 -\frac12\int d^d x\sqrt{-g}\,
 \mathcal P_d(m_{\rm eff})
 \Bigg[
 m_{\rm eff}^4f_0^\psi\Gamma\left(-\frac d2\right)
 \nonumber\\
 &\hspace{1.2cm}
 +m_{\rm eff}^2f_1^\psi
     \Gamma\left(1-\frac d2\right)
 +f_2^\psi\Gamma\left(2-\frac d2\right)
 \Bigg].
 \label{eq:njl-curved-regulated-action}
\end{align}
This expression contains the volume, Einstein--Hilbert, and
quadratic-curvature sectors generated by the fermion loop.

It is useful to make this separation explicit.  Define
\begin{align}
 U_{\rm reg}(\Theta)
 &\equiv\frac{\Theta^2}{4\lambda}
 +2\mathcal P_d(m_{\rm eff})m_{\rm eff}^4
     \Gamma\left(-\frac d2\right),
 \label{eq:njl-U-regulated}
 \\
 F_{\rm reg}(\Theta)
 &\equiv\frac{\mathcal P_d(m_{\rm eff})m_{\rm eff}^2}{6}
     \Gamma\left(1-\frac d2\right),
 \label{eq:njl-F-regulated}
 \\
 \mathcal S_\psi
 &\equiv5R^2-8R_{\mu\nu}R^{\mu\nu}
 -7R_{\mu\nu\rho\sigma}R^{\mu\nu\rho\sigma}
 -12\Box R.
 \label{eq:njl-Spsi-definition}
\end{align}
Equation~\eqref{eq:njl-curved-regulated-action} then becomes
\begin{align}
 W_{\rm MF}^{\rm reg}
 &=\int d^d x\sqrt{-g}\Bigg[
 -U_{\rm reg}(\Theta)
 +F_{\rm reg}(\Theta)R
 \nonumber\\
 &\hspace{1.2cm}
 -\frac{\mathcal P_d(m_{\rm eff})}{720}
 \Gamma\left(2-\frac d2\right)\mathcal S_\psi
 \Bigg].
 \label{eq:njl-curved-separated-action}
\end{align}
We can also write $\mathcal{S}_\psi$ in terms of the Euler density and Weyl tensor. In four dimensions the Weyl tensor is the trace-free part of the
Riemann tensor,
\begin{align}
 C_{\mu\nu\rho\sigma}
 ={}&R_{\mu\nu\rho\sigma}
 -\frac12\Big(
 g_{\mu\rho}R_{\nu\sigma}
 -g_{\mu\sigma}R_{\nu\rho}
 \nonumber\\
 &\hspace{1.45cm}
 -g_{\nu\rho}R_{\mu\sigma}
 +g_{\nu\sigma}R_{\mu\rho}\Big)
 \nonumber\\
 &+\frac{R}{6}\Big(
 g_{\mu\rho}g_{\nu\sigma}
 -g_{\mu\sigma}g_{\nu\rho}\Big),
 \label{eq:weyl-tensor-definition}
\end{align}
and the Euler density is
\begin{equation}
 E_4\equiv
 R_{\mu\nu\rho\sigma}R^{\mu\nu\rho\sigma}
 -4R_{\mu\nu}R^{\mu\nu}+R^2 .
 \label{eq:euler-density-definition}
\end{equation}
Their contractions obey
\begin{equation}
 C_{\mu\nu\rho\sigma}C^{\mu\nu\rho\sigma}
 =R_{\mu\nu\rho\sigma}R^{\mu\nu\rho\sigma}
 -2R_{\mu\nu}R^{\mu\nu}+\frac13R^2 .
 \label{eq:weyl-square-identity}
\end{equation}
Hence, it follows algebraically that
\begin{equation}
 \mathcal S_\psi=-18C_{\mu\nu\rho\sigma}C^{\mu\nu\rho\sigma}
 +11E_4-12\Box R.
 \label{eq:njl-Spsi-Weyl-Euler}
\end{equation}
As we shall see, this identity is useful for specializing the results to a spatially flat FLRW
spacetime. The simplification below is made after subtracting the
poles and taking the exactly four-dimensional local action.  Varying a
dimensionally regulated pole before the \(d\to4\) limit can retain
pole-times-evanescent contributions, and anomaly-induced terms may also
reside in the nonlocal action; these are distinct from the constant local
coefficient varied here \cite{Duff1994}.

For a constant local saddle, differentiating
Eq.~\eqref{eq:njl-curved-regulated-action} gives
\begin{align}
 0
 &=\frac{\Theta}{2\lambda}
 +\frac{\mathcal P_d(m_{\rm eff})}{2}
 \Bigg[
 df_0^\psi m_{\rm eff}^3
       \Gamma\left(-\frac d2\right)
 \nonumber\\
 &\quad +(d-2)f_1^\psi m_{\rm eff}
       \Gamma\left(1-\frac d2\right)
 \nonumber\\
 &\quad
 +(d-4)\frac{f_2^\psi}{m_{\rm eff}}
       \Gamma\left(2-\frac d2\right)
 \Bigg].
 \label{eq:njl-curved-local-gap}
\end{align}
The coefficients \(f_n^\psi\) are held fixed under this derivative.
The denominator in the last term follows from the
mass dependence of the common prefactor:
\begin{equation}
 \frac{\partial\mathcal P_d(m_{\rm eff})}{\partial\Theta}
 =\frac{d-4}{m_{\rm eff}}\,
 \mathcal P_d(m_{\rm eff}),
 \qquad
 \frac{\partial m_{\rm eff}}{\partial\Theta}=1 .
 \label{eq:njl-Pd-Theta-derivative}
\end{equation}
Moreover,
\begin{equation}
 \lim_{d\to4}(d-4)
 \Gamma\left(2-\frac d2\right)=-2,
 \label{eq:njl-evanescent-gamma-limit}
\end{equation}
so the $(d-4)$ factor and the pole leave a finite contribution.
The singular behavior at \(m_{\rm eff}=0\) indicates that the local
massive heat-kernel expansion is no longer controlled in that case \cite{Vassilevich2003,ParkerToms2009}.
Equation~\eqref{eq:njl-curved-local-gap} is the local background-field
equation that is obtained by varying an unconstrained \(\Theta(x)\) and only
then evaluating it at a constant value.  If one restricts the ansatz to
a single spacetime-constant variational parameter before varying, the
condition is instead the spacetime integral of the displayed bracket;
these two procedures to obtain the gap equation are not equivalent on a general time-dependent
geometry.
When the curvature vanishes,
Eq.~\eqref{eq:njl-curved-local-gap} reduces to
Eq.~\eqref{eq:njl-flat-potential-gap}.  On a general time-dependent
geometry, however, a constant \(\Theta\) need not solve this algebraic
condition at every time.  A dynamical saddle requires the derivative
operators discussed next.

A minimal parity-even local truncation that keeps the potential, the
leading two-derivative condensate dynamics, and independent
pure-gravity terms through four derivatives of the metric can be
organized as
\begin{align}
 W[g,\Theta]
 &=W_{\rm loc}[g,\Theta]+W_{\rm bdy}+W_{\rm nonlocal},
 \nonumber\\
 W_{\rm loc}[g,\Theta]
 &=\int d^4x\sqrt{-g}\Bigg[
 -U(\Theta)
 -\frac12Z(\Theta)(\nabla\Theta)^2
 \nonumber\\
 &\hspace{1.0cm}
 +F(\Theta)R+\alpha_C C^2
 +\beta_R R^2
 +c_EE_4
 \Bigg].
 \label{eq:njl-general-local-action}
\end{align}
Here \(C^2\) denotes the Weyl tensor squared defined in
Eq.~\eqref{eq:weyl-square-identity}.  The functions \(U\),
\(F\), and \(Z\) are renormalized coefficient functions, whereas
\(\alpha_C\), \(\beta_R\), and \(c_E\) are the displayed
constant four-derivative gravitational couplings.  The
constant-background specialization fixes the regulated functions in
Eqs.~\eqref{eq:njl-U-regulated} and \eqref{eq:njl-F-regulated}.
The varying-background coefficient in Eq.~\eqref{eq:njl-varying-M-heat-kernel} fixes the pole and logarithmic
scale dependence of \(Z(\Theta)\); a finite renormalized value still
requires a specified two-point subtraction and matching condition.
Fermion loops and matter counterterms both contribute to that value.
The terms \(W_{\rm bdy}\) and \(W_{\rm nonlocal}\) encode boundary
and state-dependent nonlocal contributions, respectively. For a varying \(\Theta\), the complete
four-derivative matter expansion also permits condensate-dependent
coefficients multiplying the curvature-squared invariants and mixed
operators such as \(R(\nabla\Theta)^2\) and \((\Box\Theta)^2\).  Those
operators require additional derivative matching and are not retained
in Eq.~\eqref{eq:njl-general-local-action}; the curvature-squared term
computed in Eq.~\eqref{eq:njl-curved-separated-action} is used below
only at the constant saddle where its coefficient is constant.

Equation~\eqref{eq:njl-general-local-action} therefore defines the
two-derivative dynamical condensate truncation, supplemented by constant
pure-gravity four-derivative couplings. We next vary this covariant action
before imposing the FLRW metric; this order is essential for obtaining
both the energy density and the pressure.

\section{Energy-Momentum Tensor in a Spatially Flat FLRW Background}
\label{sec:flrw}

We now specialize the covariant local action to a spatially flat
Friedmann--Lema\^itre--Robertson--Walker (FLRW) spacetime.  We first
derive the result for general coefficient functions \(U\), \(F\), and
\(Z\).  The explicitly calculated one-loop constant-saddle result is
then recovered as a particular case. 

\subsection{Covariant local source}
\label{subsec:njl-covariant-local-source}

The local expectation value is defined by metric variation,
\begin{equation}
 \langle T_{\mu\nu}\rangle_{\rm loc}
 =-\frac{2}{\sqrt{-g}}
 \frac{\delta W_{\rm loc}}{\delta g^{\mu\nu}},
 \label{eq:njl-energy-momentum-definition}
\end{equation}
with \(\Theta\) held fixed during the variation.  For the terms with
two derivatives in Eq.~\eqref{eq:njl-general-local-action}, this gives
\begin{align}
 \langle T_{\mu\nu}\rangle_{(2)}
 &=Z\nabla_\mu\Theta\nabla_\nu\Theta
 -g_{\mu\nu}\left[
 \frac12Z(\nabla\Theta)^2+U\right]
 \nonumber\\
 &\quad -2F G_{\mu\nu}
 +2\left(\nabla_\mu\nabla_\nu-g_{\mu\nu}\Box\right)F.
 \label{eq:njl-general-covariant-energy-momentum}
\end{align}
All functions in this equation are evaluated at \(\Theta\).  The
derivative term
\(2(\nabla_\mu\nabla_\nu-g_{\mu\nu}\Box)F\) is required whenever the
induced Einstein--Hilbert coefficient depends on the condensate and
vanishes for constant \(F\).
The four-derivative terms define an additional separately conserved
contribution \(\langle T_{\mu\nu}\rangle_{(4)}\) when their
coefficients are constant.

The condensate equation obtained from the same local action is
\begin{equation}
 Z\Box\Theta
 +\frac12Z'(\nabla\Theta)^2
 -U'+F'R=0,
 \label{eq:njl-general-Theta-equation}
\end{equation}
where a prime denotes differentiation with respect to \(\Theta\). Note that equations~\eqref{eq:njl-general-covariant-energy-momentum} and
\eqref{eq:njl-general-Theta-equation} are not independent:
diffeomorphism invariance relates the divergence of the former to the
latter.  We now make this relation explicit in the FLRW case.

\subsection{Energy density, pressure, and local energy-momentum conservation}
\label{subsec:njl-flrw-rho-p}

In this section, we calculate the energy density and pressure of the fermionic field in the mean-field approximation. We start with some comments about the renormalized physical couplings. Because \(\lambda\) and \(\Theta\) are dimensionful, their limiting
values must be stated relative to a fixed matching scale
\(\mu_\star\).  Define
\begin{equation}
 g_{\rm 4F}(\mu_\star)
 \equiv\lambda(\mu_\star)\mu_\star^2,
 \qquad
 \widehat\Theta\equiv\frac{\Theta}{\mu_\star},
 \qquad
 \widehat m\equiv\frac{m}{\mu_\star}.
 \label{eq:njl-dimensionless-limits}
\end{equation}
For a cutoff EFT one may choose \(\mu_\star=\Lambda_{\rm UV}\), so
\(g_{\rm 4F}=\lambda\Lambda_{\rm UV}^2\).  The weak-coupling
free-field branch is
\(g_{\rm 4F}\to0\) at fixed \(\widehat m\), with the self-consistent
solution \(\widehat\Theta\to0\) and
\(m_{\rm eff}/\mu_\star\to\widehat m\).  By contrast,
\(\widehat\Theta\to0\) at fixed \(g_{\rm 4F}\) is the kinematic
zero-condensate limit; it need not solve the gap equation in the broken
phase.  The distinction is visible in the dimensionless auxiliary combination
\begin{equation}
 \frac{1}{\mu_\star^4}\frac{\Theta^2}{4\lambda}
 =\frac{\widehat\Theta^2}{4g_{\rm 4F}}.
 \label{eq:njl-dimensionless-auxiliary-term}
\end{equation}
It is natural that geometric limits should likewise be taken at fixed
\(H/\mu_\star\) and \(\dot H/\mu_\star^2\). 

Consider
\begin{equation}
 ds^2=-dt^2+a(t)^2\delta_{ij}dx^idx^j,
 \qquad
 H\equiv\frac{\dot a}{a},
 \qquad
 \Theta=\Theta(t),
 \label{eq:njl-flat-flrw-metric}
\end{equation}
where \(H\) is the Hubble parameter.  The curvature scalar and Einstein
tensor are
\begin{align}
 R&=6(\dot H+2H^2),
 \nonumber\\
 G_{00}&=3H^2,
 \qquad
 G_{ij}=-(2\dot H+3H^2)g_{ij}.
 \label{eq:njl-flrw-curvature}
\end{align}
Other relevant tensor components are collected in
Appendix~\ref{app:flrw-geometry}.

Writing
\(\langle T_{00}\rangle=\rho\) and
\(\langle T_{ij}\rangle=p\,g_{ij}\),
Eq.~\eqref{eq:njl-general-covariant-energy-momentum} gives
\begin{align}
 \rho_{(2)}
 &=\frac12Z\dot\Theta^2+U
 -6FH^2-6H\dot F,
 \label{eq:njl-general-flrw-rho}
 \\
 p_{(2)}
 &=\frac12Z\dot\Theta^2-U
 +2F(2\dot H+3H^2)
 +2\ddot F+4H\dot F.
 \label{eq:njl-general-flrw-p}
\end{align}
Here
\(\dot F=F'\dot\Theta\) and
\(\ddot F=F''\dot\Theta^2+F'\ddot\Theta\).  The terms proportional to
\(\dot F\) and \(\ddot F\) are the metric-variation terms that would
be missed if \(F\) were to be treated as a constant.

The homogeneous form of Eq.~\eqref{eq:njl-general-Theta-equation} is
\begin{equation}
 Z(\ddot\Theta+3H\dot\Theta)
 +\frac12Z'\dot\Theta^2
 +U'-F'R=0.
 \label{eq:njl-flrw-Theta-equation}
\end{equation}
Direct differentiation of Eqs.~\eqref{eq:njl-general-flrw-rho} and
\eqref{eq:njl-general-flrw-p} yields the off-shell identity
\begin{align}
 \dot\rho_{(2)}+3H(\rho_{(2)}+p_{(2)})
 &=\dot\Theta\Big[
 Z(\ddot\Theta+3H\dot\Theta)
 \nonumber\\
 &\hspace{1.2cm}
 +\frac12Z'\dot\Theta^2
 +U'-F'R\Big].
 \label{eq:njl-flrw-off-shell-conservation}
\end{align}
The right-hand side vanishes on
Eq.~\eqref{eq:njl-flrw-Theta-equation}.  Thus the local condensate and
its energy-momentum tensor must be evolved consistently; imposing an
arbitrary \(\Theta(t)\) while omitting its equation would generally
violate conservation.

For completeness, spatially flat FLRW is conformally flat, so the
constant-coefficient \(C^2\) variation vanishes.  A
constant-coefficient Euler term is topological, and \(\Box R\) is a
boundary term. More importantly, an
independent finite \(R^2\) coupling does contribute to the conservation equation.  If
\(W_{\rm loc}\supset\beta_R\int\sqrt{-g}\,R^2\), then
\begin{align}
 \rho_{R^2}
 &=-36\beta_R\left(
 2H\ddot H+6H^2\dot H-\dot H^2\right),
 \label{eq:njl-flrw-rho-R2}
 \\
 p_{R^2}
 &=12\beta_R\left(
 2\dddot H+12H\ddot H+9\dot H^2+18H^2\dot H
 \right).
 \label{eq:njl-flrw-p-R2}
\end{align}
These two terms obey
\(\dot\rho_{R^2}+3H(\rho_{R^2}+p_{R^2})=0\) identically. However, they should
be added to Eqs.~\eqref{eq:njl-general-flrw-rho} and
\eqref{eq:njl-general-flrw-p} when \(\beta_R\) is retained.

The general formulas now make it possible to identify exactly which
terms are fixed by the constant-background fermion determinant and
which require additional derivative matching.

\subsection{Explicit constant-saddle one-loop result}
\label{subsec:njl-flrw-constant-saddle}

Within the local-potential truncation actually evaluated in
Sec.~\ref{sec:njl-curved}, set
\(\dot\Theta=0\) and use
\(U=U_{\rm reg}\), \(F=F_{\rm reg}\).  The quantum-induced combination
in Eq.~\eqref{eq:njl-Spsi-Weyl-Euler} has no four-dimensional spatially
flat FLRW bulk variation when its coefficient is constant: the Weyl
tensor vanishes, the Euler term is topological, and the remaining term
is a boundary term\footnote{This statement applies to the subtracted,
exactly four-dimensional local action. It does not remove anomaly-induced
or state-dependent nonlocal contributions.}. The independent renormalized higher-curvature
couplings, such as \(\beta_R\) above, remain possible and should not be
set to zero. Such terms would contribute to modifications of gravity.

Retaining the
dimensionally regulated \(f_0^\psi\) and \(f_1^\psi\) sectors, the local action gives rise to the following fermionic energy density and pressure:
\begin{align}
 \rho_{\rm MF}^{\rm reg}
 &=2\mathcal P_d(m_{\rm eff})m_{\rm eff}^4
 \Gamma\left(-\frac d2\right)
 +\frac{\Theta^2}{4\lambda}
 \nonumber\\
 &\quad
 -\mathcal P_d(m_{\rm eff})m_{\rm eff}^2
 \Gamma\left(1-\frac d2\right)H^2,
 \label{eq:njl-flrw-rho-constant}
 \\
 p_{\rm MF}^{\rm reg}
 &=-2\mathcal P_d(m_{\rm eff})m_{\rm eff}^4
 \Gamma\left(-\frac d2\right)
 -\frac{\Theta^2}{4\lambda}
 \nonumber\\
 &\quad
 +\frac{\mathcal P_d(m_{\rm eff})m_{\rm eff}^2}{3}
 \Gamma\left(1-\frac d2\right)
 (2\dot H+3H^2).
 \label{eq:njl-flrw-p-constant}
\end{align}
The coefficients are dimensionally regulated and are to be understood
as analytically continued local coefficients before the
four-dimensional subtraction conditions are imposed.

Several limits provide immediate checks.  For \(H=\dot H=0\),
Eqs.~\eqref{eq:njl-flrw-rho-constant} and
\eqref{eq:njl-flrw-p-constant} reduce to the flat-space result, 
Eq.~\eqref{eq:njl-flat-rho-p-dimensional}.  Along the perturbative
branch \(g_{\rm 4F}(\mu_\star)\to0\) at fixed dimensionless masses and
geometric ratios, the saddle has \(\widehat\Theta\to0\) and
\(m_{\rm eff}/\mu_\star\to\widehat m\), recovering the free-fermion
case.  Finally,
for constant \(U\) and \(F\),
\begin{equation}
 \dot\rho_{\rm MF}^{\rm reg}
 +3H\left(\rho_{\rm MF}^{\rm reg}
 +p_{\rm MF}^{\rm reg}\right)=0,
 \label{eq:njl-flrw-constant-conservation}
\end{equation}
as which should be due to the Bianchi identity.  A globally constant
\(\Theta\) is nevertheless a solution of the displayed two-derivative
condensate equation only when \(U'-F'R=0\) is compatible with the
background.  If the determinant-generated four-derivative term is
retained, the local saddle condition also contains the \(f_2^\psi\)
contribution displayed in Eq.~\eqref{eq:njl-curved-local-gap}; otherwise
Eqs.~\eqref{eq:njl-general-flrw-rho}--\eqref{eq:njl-flrw-Theta-equation}
must be used.

\subsection{Renormalization and physical interpretation}
\label{subsec:njl-flrw-renormalization}

Renormalization in curved spacetimes is well established, and we shall not delve into its nuances here (for details, see e.g. \cite{Wald1977,Stelle1977,BirrellDavies1982,BuchbinderShapiro2021}). Our goal in this section is to highlight some of the structure needed to renormalize the curved-space NJL theory, which comes from the effective action in Eq. \eqref{eq:njl-curved-separated-action}. 

First, the gravitational action must contain the local operators needed to
absorb the fermion-loop poles.  A convenient renormalized basis is
\begin{align}
 S_{\rm grav}^{\rm ren}
 &=\int d^4x\sqrt{-g}\Bigg[
 \frac{R-2\Lambda_{\rm ren}}
 {16\pi G_{\rm ren}} +\alpha_{C,{\rm ren}}C^2
 \nonumber\\
 &\hspace{2.5cm}
  +\beta_{R,{\rm ren}}R^2
 +c_{E,{\rm ren}}E_4
 \Bigg],
 \label{eq:njl-renormalized-gravity-action}
\end{align}
Here \(\Lambda_{\rm ren}\) is the geometric cosmological constant and
has mass dimension two.  Its associated vacuum-energy density is
\begin{equation}
 \rho_{\Lambda,{\rm ren}}
 \equiv\frac{\Lambda_{\rm ren}}{8\pi G_{\rm ren}},
 \qquad [\rho_{\Lambda,{\rm ren}}]=4 .
 \label{eq:njl-vacuum-energy-cosmological-constant}
\end{equation}
Equivalently, one may write the volume term as
\(-\rho_{\Lambda,{\rm ren}}\).
The volume divergence in the effective action induces the renormalization of
\(\rho_{\Lambda,{\rm ren}}\), or equivalently the ratio
\(\Lambda_{\rm ren}/G_{\rm ren}\). The volume, \(R\), and \(f_2^\psi\) poles shift the corresponding
couplings.  Once the term proportional to \(F R\) has been absorbed
into the renormalized gravitational sector, it may be kept on the
geometric side of the semiclassical equation or displayed as part of
the matter source. 

Second, there is also a matter-sector renormalization problem because
\(m_{\rm eff}=m+\Theta\) is varied.  The divergent powers
\begin{align}
 m_{\rm eff}^4
 &=m^4+4m^3\Theta+6m^2\Theta^2
 +4m\Theta^3+\Theta^4,
 \nonumber\\
 m_{\rm eff}^2R
 &=\left(m^2+2m\Theta+\Theta^2\right)R
 \label{eq:njl-Theta-divergent-powers}
\end{align}
require the auxiliary-field operator basis allowed by the symmetries.
Schematically,
\begin{align}
 S_{\Theta,{\rm ct}} &=-\int d^4x\sqrt{-g}\Bigg[\sum_{n=0}^{4}\delta c_n\Theta^n + \delta\xi_1R\Theta+\delta\xi_2R\Theta^2
 \nonumber\\
 &\hspace{2.7cm}
 +\delta Z_\Theta(\nabla\Theta)^2+\cdots
 \Bigg].
 \label{eq:njl-Theta-counterterms}
\end{align}
The derivative counterterm is the reason that the function
\(Z(\Theta)\) in Eq.~\eqref{eq:njl-general-local-action} requires an
independent renormalization condition.  Since the four-dimensional NJL
interaction is an effective theory, a finite gap equation additionally
requires a cutoff and matching prescription, or an explicitly stated renormalization scheme (such as the one discussed in Appendix \ref{app:bcs-grand-potential}).

After these subtractions, finite local terms remain scheme dependent up
to the measured or matched values of the renormalized couplings.  The
flat-space equality between the mode and determinant calculations, the
covariant operator basis, and the conservation identities are robust
checks of our curved-space results.

\section{Discussion and Conclusion}
\label{sec:discussion}

In this work, we revisited the functional setup for applications of fermionic condensation in cosmological contexts, 
with emphasis on the nonperturbative, vacuum NJL model in curved spacetimes. We used the functional construction to explain the difference between the in-out and in-in effective actions. We also explained how the same four-fermion operator gives rise to distinct physical regimes:
perturbative scattering, vacuum NJL, and finite-density BCS. The free scalar and
Dirac examples then provided the mode-sum and heat-kernel benchmarks
needed to interpret the local ultraviolet terms calculated in curved spacetimes. 

For the vacuum scalar channel, the mean-field replacement generates the
mass shift
\(\Theta=2\lambda\langle\bar\psi\psi\rangle\) and the effective mass
\(m_{\rm eff}=m+\Theta\).  We showed that the self-consistent direct
tadpole series, the effective potential for the gap field, and the mode
calculation give the same regulated gap equation. The flat-space
energy density and pressure satisfy
\(\langle p\rangle=-\langle\rho\rangle\).  In curved spacetime, the
parity-even determinant produces the expected local volume,
Einstein--Hilbert, and quadratic-curvature operators.  Varying this
action before imposing the FLRW ansatz gives the energy density and
pressure in Eqs.~\eqref{eq:njl-general-flrw-rho} and
\eqref{eq:njl-general-flrw-p}; their conservation follows from the
condensate equation, as expected from the absence of gravitational anomalies.

The scope of these results is fixed by several assumptions.  The
explicit loop coefficients were calculated at one-loop order, in the
scalar mean-field channel, using a local derivative expansion.  The
potential and curvature coefficients were then specialized to a
constant vacuum saddle.  At leading two-derivative order, a
time-dependent condensate can be described by the coefficient functions
\(U(\Theta)\), \(F(\Theta)\), and \(Z(\Theta)\).  The
varying-background heat kernel fixes the divergent and logarithmic parts
of \(Z\), while its finite renormalized value requires a two-point
matching condition.  Nonlocal vacuum
polarization, particle production, collective fluctuations, and finite
state-dependent terms are outside this truncation and beyond the scope of this work.

Within these assumptions, the formalism is useful for cosmological
models whose low-energy fermion dynamics contains an attractive scalar
particle--antiparticle channel.  This includes NJL-like chiral
condensates in curved spacetime
\cite{GeyerGrandaOdintsov1996,InagakiMutaOdintsov1997},
fermionic dark energy and dark matter models, nonsingular-cosmology effective theories, and
other local four-fermion models for which a homogeneous scalar saddle
is justified. In the first-order formulation of gravity, torsion provides another important source of
four-fermion interactions, but integrating out torsion commonly
produces an axial-current channel.  Applying the scalar formulas then
requires an explicit Fierz transformation and a justified channel
projection; the axial interaction should not be identified with
\((\bar\psi\psi)^2\) without this additional step
\cite{Hehl1976,FreidelMinicTakeuchi2005,Magueijo:2012ug}.

We also clarified that a finite-density BCS state is a different regime choice, resulting in a distinct functional determinant compared to the NJL case.  It requires a
physical chemical potential, occupied modes, an attractive projected
Cooper channel, and a Nambu--Gorkov kernel.  Its
curved-spacetime energy-momentum tensor would also depend on the
initial density matrix and the real-time evolution of the pairing
field. These data are absent from the vacuum NJL saddle considered in
Secs.~\ref{sec:njl-one-loop} and
\ref{sec:flrw}.

Several extensions follow naturally.  A first step is to match the
finite derivative coefficient \(Z(\Theta)\) and compute the nonlocal
form factors in a specified renormalization scheme.  It would then be possible to evolve
the condensate and the semiclassical Einstein equation
self-consistently on the closed time path, which is relevant for nonsingular models. Further directions include
curved-spacetime finite-density BCS pairing, collective scalar and
pseudoscalar fluctuations beyond mean field, and multichannel
four-fermion interactions. We leave these developments for future work.

\section{Acknowledgements}

The research at the University of Lethbridge is supported by Quantum Horizons Alberta and NSERC through Discovery Grant RGPIN-2026-05926.

\appendix
\renewcommand{\thesection}{\Alph{section}}
\renewcommand{\theequation}{\Alph{section}.\arabic{equation}}
\renewcommand{\theHequation}{\Alph{section}.\arabic{equation}}
\onecolumngrid
\section{Scalar Green Function, Local Momentum Expansion, and Heat Kernel}
\label{app:scalar-green}

In this appendix we review the derivation of the coefficients \eqref{eq:scalar-f1f2-compact} used in Sec.~\ref{sec:free-fields} from the scalar Green's function equation. All curvatures in the Riemann-normal-coordinate expansion are evaluated at the origin \(x'\), and
\(y^\mu=x^\mu-x'^\mu\).

\paragraph{Determinant and mass-integration identity:} For the scalar Laplace-type operator
\begin{equation}
K(m^2)=-\nabla^2+m^2,
\qquad
K_x(m^2)G(x,x';m^2)
=
\frac{\delta^{(d)}(x-x')}{\sqrt{-g(x)}},
\label{eq:appA-green-definition}
\end{equation}
we have \(G=K^{-1}\) and therefore
\begin{equation}
\frac{\partial}{\partial m^2}\operatorname{Tr}\ln K(m^2)
=
\operatorname{Tr}G(m^2).
\label{eq:appA-mass-derivative}
\end{equation}
Equivalently, after absorbing the \(m^2\)-independent integration
functional into the renormalized local gravitational couplings,
\begin{equation}
\operatorname{Tr}\ln K(m^2)
=
-\int_{m^2}^{\infty}d\mathcal M^2\,
\operatorname{Tr}G(\mathcal M^2).
\label{eq:appA-mass-integration}
\end{equation}
Thus the scalar effective action can be reconstructed from the
coincidence limit of the Green function,
\begin{equation}
W_\Phi
=
-\frac{i}{2}\int d^d x\sqrt{-g}
\int_{m^2}^{\infty}d\mathcal M^2\,
G(x,x;\mathcal M^2),
\label{eq:appA-W-from-G}
\end{equation}
with the Lorentzian boundary condition understood as in
Sec.~\ref{sec:functional-setup}.

\paragraph{Riemann normal coordinates and rescaling:} Let $y^\mu$ denote Riemann normal coordinates centered at $x'$, so that $y^\mu(x')=0$, $g_{\mu\nu}(x')=\eta_{\mu\nu}$, and $\Gamma^\rho_{\mu\nu}(x')=\partial_\alpha g_{\mu\nu}(x')=0$. The metric and inverse metric through the order needed for the local four-derivative coefficients are
\begin{align}
g_{\mu\nu}
&=
\eta_{\mu\nu}
-\frac13R_{\mu\alpha\nu\beta}y^\alpha y^\beta
-\frac16R_{\mu\alpha\nu\beta;\gamma}
y^\alpha y^\beta y^\gamma
\nonumber\\
&\quad
+\left(
-\frac1{20}R_{\mu\alpha\nu\beta;\gamma\delta}
+\frac2{45}R_{\alpha\mu\beta\lambda}
R^\lambda{}_{\gamma\nu\delta}
\right)y^\alpha y^\beta y^\gamma y^\delta
+O(y^5),
\label{eq:appA-rnc-metric}
\\
g^{\mu\nu}
&=
\eta^{\mu\nu}
+\frac13R^\mu{}_{\alpha}{}^\nu{}_{\beta}
y^\alpha y^\beta
+\frac16R^\mu{}_{\alpha}{}^\nu{}_{\beta;\gamma}
y^\alpha y^\beta y^\gamma
\nonumber\\
&\quad
+\left(
\frac1{20}R^\mu{}_{\alpha}{}^\nu{}_{\beta;\gamma\delta}
+\frac1{15}R^\mu{}_{\alpha\lambda\beta}
R^\lambda{}_{\gamma}{}^\nu{}_{\delta}
\right)y^\alpha y^\beta y^\gamma y^\delta
+O(y^5).
\label{eq:appA-rnc-inverse}
\end{align}
For example, multiplying the quadratic terms in
Eqs.~\eqref{eq:appA-rnc-metric} and
\eqref{eq:appA-rnc-inverse} gives
\begin{equation}
\left(g_{\mu\nu}g^{\nu\rho}\right)_{y^2}
=
\frac13R_{\mu\alpha}{}^\rho{}_{\beta}y^\alpha y^\beta
-\frac13R_{\mu\alpha}{}^\rho{}_{\beta}y^\alpha y^\beta
=0,
\end{equation}
and the quartic inverse coefficient follows by including the product of
the two quadratic pieces.  This supplies a direct check of the
\(1/15\) curvature-squared term in Eq.~\eqref{eq:appA-rnc-inverse}.

Let \(g_{\rm P}=|\det g_{\mu\nu}|\) and introduce the Bunch--Parker
rescaling
\begin{equation}
G(x,x')=g_{\rm P}(x)^{-1/4}\overline G(x,x').
\label{eq:appA-bp-rescaling}
\end{equation}
Using
\begin{equation}
\nabla^2F
=
\frac1{\sqrt{g_{\rm P}}}
\partial_\mu\!\left(
\sqrt{g_{\rm P}}g^{\mu\nu}\partial_\nu F
\right),
\end{equation}
the rescaled equation is expanded as \cite{Martin2012}
\begin{align}
\Bigg[
-\eta^{\mu\nu}\partial_\mu\partial_\nu
+m^2-\frac16R
+\frac13R_\alpha{}^\nu y^\alpha\partial_\nu
&-\frac13R^\mu{}_{\alpha}{}^\nu{}_{\beta}
y^\alpha y^\beta\partial_\mu\partial_\nu
+\mathcal V_3+\mathcal V_4+O(\nabla^5g)
\Bigg]\overline G
=\delta^{(d)}(y).
\label{eq:appA-rescaled-operator}
\end{align}
Here \(\mathcal V_n\) contains \(n\) derivatives of the metric.  The
full normal-coordinate operators are those of the local momentum-space
expansion in \cite{DeWitt1964,Christensen1978,BunchParker1979}.  The terms
displayed explicitly are sufficient to see the cancellation that fixes
\(\overline G_2\) below.

\paragraph{Local momentum-space iteration:} Fourier transforming in the local tangent space,
\begin{equation}
\overline G(y)
=
\int\frac{d^d k}{(2\pi)^d}e^{ik\cdot y}
\sum_{n=0}^{\infty}\overline G_n(k),
\qquad
P(k)\equiv\frac1{k^2+m^2},
\end{equation}
orders the solution by derivatives of the metric.  The first two terms
are
\begin{equation}
\overline G_0=P,
\qquad
\overline G_1=0.
\label{eq:appA-G0-G1}
\end{equation}
At second order, Eq.~\eqref{eq:appA-rescaled-operator} gives
\begin{align}
(k^2+m^2)\overline G_2
&=
\frac16R\,P
+\frac13R_\alpha{}^\nu
\partial^\alpha(k_\nu P)+\frac13R^\mu{}_{\alpha}{}^\nu{}_{\beta}
\partial^\alpha\partial^\beta(k_\mu k_\nu P).
\end{align}
The last two terms cancel by the Riemann symmetries and the fact that
\(P=P(k^2)\).  Hence
\begin{equation}
{
\overline G_2=\frac16R\,P^2.
}
\label{eq:appA-G2}
\end{equation}
After the Riemann symmetries and contracted Bianchi identity are used,
the complete third- and fourth-order recursion on
\(P=P(k^2)\) reduces to
\begin{align}
 P^{-1}\overline G_3
 &=\frac{i}{6}R_{;\alpha}\partial^\alpha P,
 \nonumber\\
 P^{-1}\overline G_4
 &=\frac{R}{6}\overline G_2
 +a_{\alpha\beta}\partial^\alpha\partial^\beta P,
 \label{eq:appA-G3-G4-recursion}
\end{align}
where \(a_{\alpha\beta}\) is displayed explicitly in
Eq.~\eqref{eq:appA-aab}.  As a representative cancellation, the
remaining third-order derivative terms obey
\begin{align}
 &\left(-\frac13R^\nu{}_{\alpha;\beta}
 +\frac16R_{\alpha\beta}{}^{;\nu}\right)
 y^\alpha y^\beta\partial_\nu\overline G_0
+\frac16R^\mu{}_{\alpha}{}^\nu{}_{\beta;\gamma}
 y^\alpha y^\beta y^\gamma
 \partial_\mu\partial_\nu\overline G_0=0,
 \label{eq:appA-third-order-cancellation}
\end{align}
because \(\overline G_0\) depends only on
\(\eta_{\alpha\beta}y^\alpha y^\beta\).  At fourth order the same
reduction collects every surviving contraction into
\(a_{\alpha\beta}\); no term in \(\mathcal V_3\) or \(\mathcal V_4\)
is being discarded.
The odd term is a total momentum derivative,
\begin{equation}
\overline G_3
=
\frac{i}{6}R_{;\alpha}
P\,\partial^\alpha P.
\label{eq:appA-G3}
\end{equation}
It vanishes at coincidence after a translation-invariant momentum
integration, but it is needed for the off-diagonal Green function.

At fourth order the curvature contractions can be collected into
\begin{align}
a_{\alpha\beta}
&=
\frac2{15}R_{\lambda\beta}R^\lambda{}_{\alpha}
-\frac7{60}R^\lambda{}_{\alpha\nu\beta}R^\nu{}_{\lambda}
-\frac1{60}R_{\lambda\mu\rho\alpha}
R^{\lambda\mu\rho}{}_{\beta}-\frac3{40}R_{;\alpha\beta}
-\frac1{40}\Box R_{\alpha\beta}.
\label{eq:appA-aab}
\end{align}
The iterative equation then yields
\begin{equation}
{
\overline G_4
=
\frac{R^2}{36}P^3
+a_{\alpha\beta}P\,\partial^\alpha\partial^\beta P.
}
\label{eq:appA-G4}
\end{equation}
The cubic power \(P^3\) in the first term follows both from the
iteration and from dimensional analysis: \(R^2P^3\) has the same mass
dimension as \(\overline G\), whereas an \(R^2P^2\) term would not.

The identities
\begin{align}
P\,\partial^\alpha P
&=\frac12\partial^\alpha(P^2),
\nonumber\\
P\,\partial^\alpha\partial^\beta P
&=
\frac13\partial^\alpha\partial^\beta(P^2)
-\frac23\eta^{\alpha\beta}P^3,
\label{eq:appA-P-identities}
\end{align}
and the contracted coefficient
\begin{equation}
a^\alpha{}_{\alpha}
=
\frac1{60}R_{\mu\nu}R^{\mu\nu}
-\frac1{60}R_{\mu\nu\rho\sigma}R^{\mu\nu\rho\sigma}
-\frac1{10}\Box R
\label{eq:appA-atrace}
\end{equation}
give the off-diagonal form
\begin{align}
f_1(x,x')
&=
\frac16R
+\frac1{12}R_{;\alpha}y^\alpha
-\frac13a_{\alpha\beta}y^\alpha y^\beta,
\nonumber\\
f_2(x,x')
&=
\frac1{72}R^2-\frac13a^\alpha{}_{\alpha}+O(y).
\end{align}
Taking \(x'\to x\) therefore gives \eqref{eq:scalar-f1f2-compact}:
\begin{equation}
{
f_1=\frac16R,
\qquad
f_2=
\frac1{72}R^2
-\frac1{180}R_{\mu\nu}R^{\mu\nu}
+\frac1{180}R_{\mu\nu\rho\sigma}R^{\mu\nu\rho\sigma}
+\frac1{30}\Box R.
}
\label{eq:appA-f1-f2}
\end{equation}

\paragraph{Proper-time reconstruction:} Let \(\sigma(x,x')\) be Synge's world function, equal to one half of
the squared geodesic distance in the Euclidean continuation, and let
\begin{equation}
 \Delta_{\rm VM}(x,x')
 =\frac{\det[-\nabla_\mu\nabla_{\nu'}\sigma(x,x')]}
 {\sqrt{g_E(x)g_E(x')}}
 \label{eq:appA-van-vleck-definition}
\end{equation}
be the Van Vleck--Morette determinant, where
\(g_E=\det g^E_{\mu\nu}\).  Both quantities approach their flat-space
values, \(\sigma\to\tfrac12\delta_{\mu\nu}y^\mu y^\nu\) and
\(\Delta_{\rm VM}\to1\), at coincidence.  The proper-time
representation then follows from the Schwinger--DeWitt construction
\cite{Schwinger1951,DeWitt1964,BunchParker1979}.

After Wick rotation, the local Green function is equivalently
\begin{align}
G_E(x,x')
&\sim
\frac{\Delta_{\rm VM}^{1/2}(x,x')}{(4\pi)^{d/2}}
\int_0^\infty ds\,s^{-d/2}
e^{-m^2s-\sigma/(2s)}
\left[a_0(x,x')+s a_1(x,x')+s^2a_2(x,x')+\cdots\right].
\label{eq:appA-proper-time-G}
\end{align}
At coincidence \(a_0=1\), \(a_1=f_1\), and \(a_2=f_2\).  More
explicitly, the term labelled by \(n\) in the Green function is
proportional to
\begin{equation}
G_{E,n}(x,x;\mathcal M^2)
\propto
f_n\,\Gamma\!\left(n+1-\frac d2\right)
(\mathcal M^2)^{d/2-n-1}.
\label{eq:appA-Gn-after-s}
\end{equation}
Substitution in Eq.~\eqref{eq:appA-W-from-G}, followed by analytic
continuation in \(d\), then uses
\begin{align}
-\int_{m^2}^{\infty}d\mathcal M^2\,
(\mathcal M^2)^{d/2-n-1}
&=
\frac{(m^2)^{d/2-n}}{d/2-n}, \quad
-\frac{\Gamma(n+1-d/2)}{d/2-n}
&=
\Gamma(n-d/2).
\label{eq:appA-mass-gamma-recurrence}
\end{align}
Equivalently, the determinant proper-time integral is
\begin{equation}
\int_0^\infty ds\,s^{n-d/2-1}e^{-m^2s}
=
\Gamma\!\left(n-\frac d2\right)(m^2)^{d/2-n},
\end{equation}
and produces
\begin{align}
W_\Phi
&=
\int d^d x\sqrt{-g}\,
\frac1{2(4\pi)^{d/2}}
\left(\frac{m}{\mu_{\rm DR}}\right)^{d-4}
\Bigg[
m^4\Gamma\!\left(-\frac d2\right)
+m^2f_1\Gamma\!\left(1-\frac d2\right)
+f_2\Gamma\!\left(2-\frac d2\right)
\Bigg],
\label{eq:appA-scalar-action}
\end{align}
which is the result used for the scalar benchmark in
Sec.~\ref{sec:free-fields}.

\section{Grassmann Gaussian, Dirac Square, and Spinor Heat-Kernel Trace}
\label{app:dirac-heat-kernel}

This appendix supplies the intermediate functional and Clifford-algebra
steps behind the free and interacting Dirac calculations in
Secs.~\ref{sec:free-fields} and \ref{sec:njl-curved}, including the calculation of the coefficients in Eqs. \eqref{eq:dirac-f1f2-free} and \eqref{eq:njl-varying-M-heat-kernel}.

\paragraph{Completion of the Grassmann square:} With endpoint data understood as in Sec.~\ref{sec:functional-setup},
introduce Grassmann
sources in
\begin{equation}
Z[\eta,\bar\eta]
=
\int [d\psi d\bar\psi]\,
\exp\left\{
i\int d^d x\sqrt{-g}\,
\left(\bar\psi\sD\psi+\bar\eta\psi+\bar\psi\eta\right)
\right\}.
\end{equation}
The shifts
\begin{equation}
\psi=\chi-\sD^{-1}\eta,
\qquad
\bar\psi=\bar\chi-\bar\eta\sD^{-1}
\end{equation}
give
\begin{equation}
\bar\psi\sD\psi+\bar\eta\psi+\bar\psi\eta
=
\bar\chi\sD\chi-\bar\eta\sD^{-1}\eta.
\label{eq:appB-grassmann-square}
\end{equation}
Translation invariance of the Berezin measure then yields
\begin{align}
Z[\eta,\bar\eta]
&=
Z[0,0]\exp\Bigg[
-i\int d^d x\sqrt{-g_x}\int d^d y\sqrt{-g_y}\;\bar\eta(x)G_F(x,y)\eta(y)
\Bigg],
\label{eq:appB-source-functional}
\end{align}
where $G_F=\sD^{-1}$ and $Z[0,0]\propto\det\sD.$

\paragraph{Parity-even determinant and Lichnerowicz reduction:} For
\begin{equation}
\sD_m=-\gamma^\mu\nabla_\mu+m,
\qquad
\widetilde{\sD}_m=\gamma^5\sD_m\gamma^5
=\gamma^\mu\nabla_\mu+m,
\end{equation}
the parity-even part of the determinant is defined by
\begin{equation}
W_\psi^{(+)}
=
-\frac{i}{2}\operatorname{Tr}\ln
\left(\sD_m\widetilde{\sD}_m\right).
\label{eq:appB-parity-even}
\end{equation}
Since \(\nabla_\mu\gamma^\nu=0\), the first-order product is
\begin{align}
\sD_m\widetilde{\sD}_m
&=
\left(-\gamma^\mu\nabla_\mu+m\right)
\left(\gamma^\nu\nabla_\nu+m\right)
\nonumber\\
&=
-\gamma^\mu\gamma^\nu\nabla_\mu\nabla_\nu+m^2.
\label{eq:appB-dirac-product}
\end{align}
Splitting the two gamma matrices into symmetric and antisymmetric
parts gives
\begin{align}
\gamma^\mu\gamma^\nu\nabla_\mu\nabla_\nu
&=
g^{\mu\nu}\nabla_\mu\nabla_\nu
+\frac14[\gamma^\mu,\gamma^\nu]
[\nabla_\mu,\nabla_\nu],
\nonumber\\
[\nabla_\mu,\nabla_\nu]
&=
\frac14R_{\mu\nu\rho\sigma}\gamma^\rho\gamma^\sigma.
\label{eq:appB-spin-commutator}
\end{align}
The Clifford contraction
\begin{equation}
R_{\mu\nu\rho\sigma}
\gamma^\mu\gamma^\nu\gamma^\rho\gamma^\sigma
=-2R\,\mathbf1_4
\end{equation}
in the curvature convention of the manuscript gives
\begin{equation}
\left(\gamma^\mu\nabla_\mu\right)^2
=
\nabla_{\rm spin}^2-\frac14R.
\label{eq:appB-lichnerowicz}
\end{equation}
Consequently,
\begin{equation}
{
K_\psi
\equiv
\sD_m\widetilde{\sD}_m
=
-\nabla_{\rm spin}^2+\frac14R+m^2.
}
\label{eq:appB-spinor-laplace}
\end{equation}

\paragraph{Traced heat-kernel coefficients:} Write the mass-independent part of Eq.~\eqref{eq:appB-spinor-laplace}
as
\begin{equation}
P=-\left(\nabla_{\rm spin}^2+E\right),
\qquad
E=-\frac14R\,\mathbf1_4.
\end{equation}
For a Laplace-type operator the first coefficients are
\begin{align}
a_0&=\mathbf1,
\nonumber\\
a_1&=E+\frac16R\,\mathbf1,
\nonumber\\
a_2&=\frac1{360}\Big[
60\Box E+60RE+180E^2
+12\Box R+5R^2
\nonumber\\
&\hspace{2.2cm}
-2R_{\mu\nu}R^{\mu\nu}
+2R_{\mu\nu\rho\sigma}R^{\mu\nu\rho\sigma}
+30\Omega_{\mu\nu}\Omega^{\mu\nu}
\Big].
\label{eq:appB-universal-a2}
\end{align}
The spin-bundle curvature is
\cite{FreedmanVanProeyen2012}
\begin{equation}
\Omega_{\mu\nu}
\equiv[\nabla_\mu,\nabla_\nu]
=
\frac14R_{\mu\nu ab}\gamma^{ab},
\qquad
\gamma^{ab}=\frac12[\gamma^a,\gamma^b].
\end{equation}
Using
\begin{equation}
\operatorname{tr}_{s}
\left(\gamma^{ab}\gamma^{cd}\right)
=
4\left(\eta^{ad}\eta^{bc}-\eta^{ac}\eta^{bd}\right),
\end{equation}
one finds
\begin{equation}
\operatorname{tr}_{s}
\left(\Omega_{\mu\nu}\Omega^{\mu\nu}\right)
=
-\frac12R_{\mu\nu\rho\sigma}R^{\mu\nu\rho\sigma}.
\label{eq:appB-Omega-trace}
\end{equation}
Taking the spin trace in Eq.~\eqref{eq:appB-universal-a2} therefore
gives
\begin{align}
f_0^\psi&=4,
\nonumber\\
f_1^\psi&=-\frac13R,
\nonumber\\
f_2^\psi
&=
\frac1{360}\left(
5R^2
-8R_{\mu\nu}R^{\mu\nu}
-7R_{\mu\nu\rho\sigma}R^{\mu\nu\rho\sigma}
-12\Box R
\right).
\label{eq:appB-spinor-coefficients}
\end{align}
The \(f_0^\psi\) factor reproduces the four flat-space spinor
components.  The last term of Eq.~\eqref{eq:appB-universal-a2} is the
reason that the curvature-squared coefficient is not four copies of a
scalar coefficient.

\paragraph{Spacetime-dependent scalar mass:} The dynamical condensate used in Sec.~\ref{sec:njl-curved} requires the
same calculation without assuming a constant mass.  Let
\begin{equation}
 M(x)=m+\Theta(x),
 \qquad
 \sD_M=-\gamma^\mu\nabla_\mu+M,
 \qquad
 \widetilde{\sD}_M=\gamma^\mu\nabla_\mu+M .
\end{equation}
Acting on a test spinor and using the product rule gives
\begin{align}
 \sD_M\widetilde{\sD}_M
 ={}&-\gamma^\mu\gamma^\nu\nabla_\mu\nabla_\nu
 -\gamma^\mu\nabla_\mu M
 \nonumber\\
 &-M\gamma^\mu\nabla_\mu+M\gamma^\mu\nabla_\mu+M^2
 \nonumber\\
 ={}&-\nabla_{\rm spin}^2+\frac14R+M^2
-\gamma^\mu\nabla_\mu M .
 \label{eq:appB-varying-M-square}
\end{align}
The two terms proportional to \(M\gamma^\mu\nabla_\mu\) cancel, which
implies that no first-order derivative acts on the test spinor in the final
Laplace-type operator.

In the convention \(P=-(\nabla_{\rm spin}^2+E)\), we have now
\begin{equation}
 E[M]=-\frac14R\,\mathbf1_4-M^2\mathbf1_4
 +\gamma^\mu\nabla_\mu M .
 \label{eq:appB-varying-M-endomorphism}
\end{equation}
The spin traces required by
Eq.~\eqref{eq:appB-universal-a2} are
\begin{align}
 \operatorname{tr}_s E
 &=-R-4M^2,
 \nonumber\\
 \operatorname{tr}_s E^2
 &=4\left(\frac14R+M^2\right)^2
 +4(\nabla M)^2,
 \nonumber\\
 \operatorname{tr}_s\Box E
 &=-\Box R-4\Box M^2 .
 \label{eq:appB-varying-M-traces}
\end{align}
Terms linear in a single gamma matrix vanish under the spin trace.
Combining these expressions with
Eq.~\eqref{eq:appB-Omega-trace} gives the unfactored coefficients
\begin{align}
 b_0^\psi[M]&=4,
 \nonumber\\
 b_1^\psi[M]&=-4M^2-\frac13R,
 \nonumber\\
 b_2^\psi[M]
 &=2M^4+\frac13RM^2+2(\nabla M)^2
 -\frac23\Box M^2+\frac1{360}\left(
 5R^2-8R_{\mu\nu}R^{\mu\nu}
 -7R_{\mu\nu\rho\sigma}R^{\mu\nu\rho\sigma}
 -12\Box R\right).
 \label{eq:appB-varying-M-coefficients}
\end{align}
Equivalently,
\begin{equation}
 b_2^\psi[M]
 =f_2^\psi-M^2f_1^\psi+\frac12M^4f_0^\psi
 +2(\nabla M)^2-\frac23\Box M^2 .
 \label{eq:appB-varying-M-factorized}
\end{equation}
The integrated \(\Box M^2\) term is a boundary contribution under the
conditions used in the main text.  Since the parity-even fermion
determinant carries the prefactor \(-\tfrac12\), the pole of
\(\Gamma(2-d/2)\) gives
\begin{equation}
 W_{\rm div}^{(+)}
 \supset-\frac{1}{8\pi^2\epsilon}
 \int d^4x\sqrt{-g}\,(\nabla\Theta)^2 .
 \label{eq:appB-varying-M-kinetic-pole}
\end{equation}
This fixes the local divergent and logarithmic parts of the condensate
wave-function coefficient. Note that a finite renormalized coefficient still
requires a two-point renormalization prescription.

\section{Fermion Modes, Canonical Normalization, and NJL Diagonalization}
\label{app:fermion-modes}

For completeness, here we discuss the flat-space mode calculation using our conventions
introduced above. The vacuum NJL calculation below is evaluated at zero chemical potential, which is consistent because of the discussion in Sec. \ref{subsec:zero-temperature-grand-canonical}.

\paragraph{Mode equations and spin sums:} The mean-field equation is
\begin{equation}
\left(-\gamma^\mu\partial_\mu+\meff\right)\psi=0,
\qquad
E_k=\sqrt{\boldsymbol k^2+\meff^2}.
\label{eq:appD-dirac-equation}
\end{equation}
For the positive- and negative-frequency waves
\begin{align}
\psi_+(x)&=u(\boldsymbol k,r)e^{-iE_kt+i\boldsymbol k\cdot\boldsymbol x},
\nonumber\\
\psi_-(x)&=v(\boldsymbol k,r)e^{iE_kt-i\boldsymbol k\cdot\boldsymbol x},
\end{align}
Eq.~\eqref{eq:appD-dirac-equation} gives
\begin{align}
\left(iE_k\gamma^0-i k_i\gamma^i+\meff\right)u(\boldsymbol k,r)&=0,
\nonumber\\
\left(-iE_k\gamma^0+i k_i\gamma^i+\meff\right)v(\boldsymbol k,r)&=0.
\label{eq:appD-mode-equations}
\end{align}
Multiplying either operator by the same expression with
\(\meff\to-\meff\) gives
\begin{equation}
E_k^2-\boldsymbol k^2-\meff^2=0.
\end{equation}

Define \(p^\mu=(E_k,\boldsymbol k)\), so that
\(p_\mu=(-E_k,\boldsymbol k)\) and \(p^2=-\meff^2\).  A compatible
normalization is encoded in the projectors
\begin{align}
\sum_{r=1}^2u(\boldsymbol k,r)\bar u(\boldsymbol k,r)
&=-i\slashed p-\meff,
\nonumber\\
\sum_{r=1}^2v(\boldsymbol k,r)\bar v(\boldsymbol k,r)
&=-i\slashed p+\meff,
\qquad p^2=-\meff^2.
\label{eq:appendix-spinor-projectors}
\end{align}
Since \(u^\dagger=i\bar u\gamma^{\hat0}\), and similarly for \(v\),
the spatial momentum and mass terms cancel between the two projectors:
\begin{equation}
\sum_r\left[
u(\boldsymbol k,r)u^\dagger(\boldsymbol k,r)
+v(-\boldsymbol k,r)v^\dagger(-\boldsymbol k,r)
\right]
=2E_k\mathbf1_4.
\label{eq:appD-projector-sum}
\end{equation}

\paragraph{Equal-time anticommutator:} Before fixing the normalization, write
\begin{align}
\psi(x)
&=
\int\frac{d^3k}{(2\pi)^{3/2}N_k}\sum_r
\Big[
c_{\boldsymbol k r}u(\boldsymbol k,r)
e^{-iE_kt+i\boldsymbol k\cdot\boldsymbol x}
+d_{\boldsymbol k r}^\dagger v(\boldsymbol k,r)
e^{iE_kt-i\boldsymbol k\cdot\boldsymbol x}
\Big].
\label{eq:appD-mode-general-N}
\end{align}
The operators obey
\begin{equation}
\{c_{\boldsymbol k r},c_{\boldsymbol p s}^\dagger\}
=
\{d_{\boldsymbol k r},d_{\boldsymbol p s}^\dagger\}
=
\delta_{rs}\delta^{(3)}(\boldsymbol k-\boldsymbol p),
\end{equation}
with all other anticommutators zero.  At equal time,
\begin{align}
\{\psi(\boldsymbol x),\psi^\dagger(\boldsymbol y)\}
&=
\int\frac{d^3k}{(2\pi)^3N_k^2}
\sum_r\Big[
u(\boldsymbol k,r)u^\dagger(\boldsymbol k,r)+v(-\boldsymbol k,r)v^\dagger(-\boldsymbol k,r)
\Big]
e^{i\boldsymbol k\cdot(\boldsymbol x-\boldsymbol y)}
\nonumber\\
&=
\int\frac{d^3k}{(2\pi)^3}
\frac{2E_k}{N_k^2}\mathbf1_4
e^{i\boldsymbol k\cdot(\boldsymbol x-\boldsymbol y)}.
\end{align}
Comparison with
\(\delta^{(3)}(\boldsymbol x-\boldsymbol y)\mathbf1_4\)
fixes
\begin{equation}
{N_k^2=2E_k.}
\label{eq:appD-Nk}
\end{equation}
Thus the normalized expansion is
\begin{align}
\psi(x)
&=
\int\frac{d^3k}{(2\pi)^{3/2}\sqrt{2E_k}}\sum_r
\Big[
c_{\boldsymbol k r}u(\boldsymbol k,r)e^{-iE_kt+i\boldsymbol k\cdot\boldsymbol x}+d_{\boldsymbol k r}^\dagger v(\boldsymbol k,r)e^{iE_kt-i\boldsymbol k\cdot\boldsymbol x}
\Big].
\label{eq:appD-normalized-mode}
\end{align}

\paragraph{Scalar mean field and diagonal quadratic action:} The NJL mean-field Lagrangian is
\begin{equation}
\mathcal L_{\rm MF}
=
-\bar\psi\gamma^\mu\partial_\mu\psi
+\meff\bar\psi\psi
-\frac{\Theta^2}{4\lambda},
\qquad
\lambda\equiv M_{\rm 4F}^{-2},
\qquad
\Theta\equiv2\lambda\Pi,
\qquad
\meff=m+\Theta,
\qquad
\Pi\equiv\langle\bar\psi\psi\rangle.
\label{eq:appD-LMF}
\end{equation}
Thus \(\Theta\) is the scalar NJL mass shift. Using Eq.~\eqref{eq:appD-normalized-mode}, the number-diagonal part of
the scalar bilinear is
\begin{equation}
[\bar\psi\psi]_{\rm diag}
=
\int\frac{d^3k}{(2\pi)^3}\frac{\meff}{E_k}
\sum_r\left[
-c_{\boldsymbol k r}^\dagger c_{\boldsymbol k r}
+d_{\boldsymbol k r}d_{\boldsymbol k r}^\dagger
\right].
\label{eq:appD-bilinear-diag}
\end{equation}
The temporal and spatial kinetic pieces reduce to
\begin{align}
[\bar\psi\gamma^0\partial_0\psi]_{\rm diag}
&=
\int\frac{d^3k}{(2\pi)^3}E_k
\sum_r\left[
-c_{\boldsymbol k r}^\dagger c_{\boldsymbol k r}
+d_{\boldsymbol k r}d_{\boldsymbol k r}^\dagger
\right],
\nonumber\\
[\bar\psi\gamma^i\partial_i\psi]_{\rm diag}
&=
-\int\frac{d^3k}{(2\pi)^3}\frac{\boldsymbol k^2}{E_k}
\sum_r\left[
-c_{\boldsymbol k r}^\dagger c_{\boldsymbol k r}
+d_{\boldsymbol k r}d_{\boldsymbol k r}^\dagger
\right].
\label{eq:appD-kinetic-diag}
\end{align}
The coefficient of the number-diagonal quadratic part is therefore
proportional to
\begin{equation}
-E_k+\frac{\boldsymbol k^2+\meff^2}{E_k}=0.
\end{equation}
The anomalous \(c^\dagger d^\dagger\) coefficient contains
\begin{equation}
\bar u(\boldsymbol k,r)
\left(-iE_p\gamma^0+i p_i\gamma^i+\meff\right)
v(\boldsymbol p,s)=0,
\end{equation}
and the \(dc\) coefficient vanishes by the positive-frequency equation
in Eq.~\eqref{eq:appD-mode-equations}.  Hence the massive scalar-channel
NJL basis already diagonalizes the quadratic action.  This is distinct
from a particle-particle BCS gap (introduced in
Sec.~\ref{sec:functional-setup} and analyzed further in
Appendix~\ref{app:bcs-grand-potential}), for which a Nambu--Gorkov/Bogoliubov
rotation is required.

Finally, in the empty massive vacuum only
\(\langle0|d_{\boldsymbol k r}d_{\boldsymbol p s}^\dagger|0\rangle\)
survives.  Equation~\eqref{eq:appD-bilinear-diag} gives
\begin{equation}
{
\Pi\equiv\langle\bar\psi\psi\rangle
=
2\meff\int\frac{d^3k}{(2\pi)^3}\frac1{E_k}.
}
\label{eq:appD-condensate}
\end{equation}
Together with the auxiliary-field saddle
\(\Pi=\Theta/(2\lambda)=M_{\rm 4F}^2\Theta/2\), this is the
vacuum NJL gap equation Eq. \eqref{eq:njl-condensate-mode-integral} used in Sec.~\ref{sec:njl-flat}.

\section{Flat-Space Mean-Field Energy-Momentum Tensor from Modes}
\label{app:mean-field-emt-details}

In this appendix, we evaluate the mean-field energy-momentum tensor of
Sec.~\ref{sec:njl-flat} explicitly.  

\paragraph{A representative contraction:} Consider the first spatial derivative term in the isotropic pressure.
Using Eq.~\eqref{eq:appD-normalized-mode}, only the antiparticle
contraction survives in the vacuum:
\begin{align}
\frac1{2(d-1)}\sum_{i=1}^{d-1}
\langle0|\bar\psi\gamma_i\partial_i\psi|0\rangle
&=
\frac1{2(d-1)}\sum_{i,r}
\int\frac{d^{d-1}k}{(2\pi)^{d-1}2E_k}
(-ik_i)\bar v(\boldsymbol k,r)\gamma_i v(\boldsymbol k,r).
\end{align}
The bilinear identity
\begin{equation}
\bar v(\boldsymbol k,r)\gamma_i v(\boldsymbol k,r)
=-2ik_i
\end{equation}
and the two spin states give
\begin{equation}
\frac1{2(d-1)}\sum_i
\langle0|\bar\psi\gamma_i\partial_i\psi|0\rangle
=
-\frac1{d-1}
\int\frac{d^{d-1}k}{(2\pi)^{d-1}}
\frac{\boldsymbol k^2}{E_k}.
\label{eq:appE-spatial-one-side}
\end{equation}
The term with \(\partial_i\bar\psi\) gives the same contribution.  Thus
the symmetrized derivative part of the pressure is twice
Eq.~\eqref{eq:appE-spatial-one-side}.

The corresponding temporal contraction is
\begin{align}
\frac12\langle0|
\bar\psi\gamma_0\partial_0\psi
-(\partial_0\bar\psi)\gamma_0\psi
|0\rangle
&=
-2\int\frac{d^{d-1}k}{(2\pi)^{d-1}}E_k.
\label{eq:appE-temporal-contraction}
\end{align}

\paragraph{Trace, mass, and auxiliary potential:} The two trace-kinetic bilinears reduce on the vacuum to
\begin{align}
\frac12\langle\bar\psi\gamma^\alpha\partial_\alpha\psi\rangle
&=
\int\frac{d^{d-1}k}{(2\pi)^{d-1}}
\frac{E_k^2-\boldsymbol k^2}{E_k}
=
\meff^2\int\frac{d^{d-1}k}{(2\pi)^{d-1}}\frac1{E_k},
\nonumber\\
\frac12\langle(\partial_\alpha\bar\psi)\gamma^\alpha\psi\rangle
&=
-\meff^2\int\frac{d^{d-1}k}{(2\pi)^{d-1}}\frac1{E_k}.
\label{eq:appE-trace-kinetic}
\end{align}
The mass term is
\begin{equation}
\meff\langle\bar\psi\psi\rangle
=
2\meff^2\int\frac{d^{d-1}k}{(2\pi)^{d-1}}\frac1{E_k}.
\end{equation}
In \(T_{00}\) and \(T_{ii}\), this contribution cancels the
corresponding trace-kinetic combination, as follows directly from the
on-shell Dirac equation.  The surviving unregulated expressions are
therefore
\begin{align}
\rho_{\rm MF}
=
-2\int\frac{d^{d-1}k}{(2\pi)^{d-1}}E_k
+V_{\rm aux},
\quad
p_{\rm MF}
=
-\frac2{d-1}
\int\frac{d^{d-1}k}{(2\pi)^{d-1}}
\frac{\boldsymbol k^2}{E_k}
-V_{\rm aux},
\quad
V_{\rm aux}
=
\lambda\Pi^2
=\frac{\Theta^2}{4\lambda}
=\frac{M_{\rm 4F}^2\Theta^2}{4}.
\label{eq:appendix-flat-emt-integrals}
\end{align}

\paragraph{Dimensional regularization:} To analytically continue to arbitrary dimensions, one makes use of the master integral
\begin{equation}
\int\frac{d^n k}{(2\pi)^n}(\boldsymbol k^2+\meff^2)^\alpha
=
\frac{(\meff^2)^{\alpha+n/2}}{(4\pi)^{n/2}}
\frac{\Gamma(-\alpha-n/2)}{\Gamma(-\alpha)}
\label{eq:appE-master-integral}
\end{equation}
which is first defined in its convergence domain and then analytically
continued.  With \(n=d-1\) and \(\alpha=1/2\),
\begin{equation}
I_E
\equiv
\mu_{\rm DR}^{4-d}
\int\frac{d^{d-1}k}{(2\pi)^{d-1}}E_k
=
\mu_{\rm DR}^{4-d}
\frac{\meff^d}{(4\pi)^{(d-1)/2}}
\frac{\Gamma(-d/2)}{\Gamma(-1/2)}.
\label{eq:appE-energy-integral}
\end{equation}
Moreover,
\begin{align}
I_p
&\equiv
\mu_{\rm DR}^{4-d}
\int\frac{d^{d-1}k}{(2\pi)^{d-1}}
\frac{\boldsymbol k^2}{E_k}
\nonumber\\
&=
\mu_{\rm DR}^{4-d}
\int\frac{d^{d-1}k}{(2\pi)^{d-1}}
\left(E_k-\frac{\meff^2}{E_k}\right)
\nonumber\\
&=-(d-1)I_E,
\label{eq:appE-pressure-identity}
\end{align}
where the last line uses
\(\Gamma(1/2)=-\Gamma(-1/2)/2\).  Substitution into
Eq.~\eqref{eq:appendix-flat-emt-integrals} gives
\begin{align}
\rho_{\rm MF}
&=
-2\mu_{\rm DR}^{4-d}
\frac{\meff^d}{(4\pi)^{(d-1)/2}}
\frac{\Gamma(-d/2)}{\Gamma(-1/2)}
+\frac{M_{\rm 4F}^2\Theta^2}{4},
\nonumber\\
p_{\rm MF}
&=
+2\mu_{\rm DR}^{4-d}
\frac{\meff^d}{(4\pi)^{(d-1)/2}}
\frac{\Gamma(-d/2)}{\Gamma(-1/2)}
-\frac{M_{\rm 4F}^2\Theta^2}{4}.
\label{eq:appE-final-emt}
\end{align}
Thus \(p_{\rm MF}=-\rho_{\rm MF}\), and the operator calculation
reproduces the functional determinant result without using a
finite-density occupation prescription.

\section{BCS grand potential}
\label{app:bcs-grand-potential}

In this appendix, we compute the finite-temperature BCS grand potential starting from the grand-canonical partition function. The
result is Eq.~\eqref{eq:bcs-absolute-potential-main}, whose gap
equation, number density, and normal-state limit were discussed in Sec. \ref{subsec:finite-density-bcs-saddle}. Our goal is to show how these results are obtained from first principles, as presented in Sec. \ref{subsec:zero-temperature-grand-canonical}.

\paragraph{Projected functional and mean-field reduction:} We begin with the physical degrees of freedom that participate in the
Cooper channel specified in
Sec.~\ref{subsec:finite-density-bcs-saddle}. Their Euclidean kinetic operator is
\begin{equation}
 D_E(\mu)
 =\gamma_E^\alpha\nabla_\alpha+m_{\rm eff}-\mu\gamma_E^0,
 \label{eq:appF-minimal-euclidean-dirac-operator}
\end{equation}
where the chemical-potential term follows from the functional in
Eq.~\eqref{eq:euclidean-mu-functional}. We have
\begin{align}
 Z(\beta,\mu)
 &=\int_{\rm AP}[d\psi d\bar\psi]\,
 \exp\Bigg\{
 -\int_0^\beta d\tau\int_{\cal V}d^3x\,
 \bar\psi D_E(\mu)\psi
 +\frac{G_{C, 0}}{4}
 \int_0^\beta d\tau\int_{\cal V}d^3x\,
 \bigl(\bar\psi\Gamma_{\rm pair}C\bar\psi^T\bigr)
 \bigl(\psi^TC\Gamma_{\rm pair}\psi\bigr)
 \Bigg\}.
 \label{eq:appF-minimal-projected-partition-function}
\end{align}
The label AP on the measure refers to the antiperiodic boundary conditions, Eq.~\eqref{eq:fermionic-antiperiodicity}. Note that in
the Euclidean integral, \(\psi\) and \(\bar\psi\) are independent
Grassmann variables and
\(\bar\psi\Gamma_{\rm pair}C\bar\psi^T\) is therefore the Euclidean
continuation of the conjugate pair operator. The factor \(1/4\) in
Eq.~\eqref{eq:appF-minimal-projected-partition-function} assigns one
factor \(1/2\) to each unordered pair bilinear.  Here \(G_{C,0}\) is
the bare coupling after projection onto this channel.  Its relation to
the scalar coupling in Eq.~\eqref{eq:sectionII-fourfermion} depends on
the Fierz coefficient and the internal projectors.  The renormalized
coupling \(G_C\) used in the main text will be defined below, after the
dimensionally regulated determinant has been separated into its pole
and finite parts.

The normalization of the auxiliary field is fixed by the exact
complex Hubbard--Stratonovich identity
\cite{Stratonovich1958,Hubbard1959},
\begin{align}
 &\exp\Bigg\{
 \frac{G_{C,0}}{4}
 \int_0^\beta d\tau\int_{\cal V}d^3x\,
 \bigl(\bar\psi\Gamma_{\rm pair}C\bar\psi^T\bigr)
 \bigl(\psi^TC\Gamma_{\rm pair}\psi\bigr)
 \Bigg\}
 \nonumber\\
 &\quad\propto
 \int [d\Delta d\Delta^*]\,
 \exp\Bigg\{-\int_0^\beta d\tau\int_{\cal V}d^3x
 \left[
 \frac{|\Delta|^2}{G_{C,0}}
 +\frac{\Delta^*}{2}\psi^TC\Gamma_{\rm pair}\psi
 +\frac{\Delta}{2}\bar\psi\Gamma_{\rm pair}C\bar\psi^T
 \right]\Bigg\}.
 \label{eq:appF-minimal-hubbard-stratonovich-identity}
\end{align}
The proportionality factor is the complex Gaussian normalization.  It
is independent of \(\Delta\), \(\mu\), and the fermion fields, and we
absorb it into the functional measure.  It cancels from the
partition-function ratio below; the remaining additive normalization
is fixed by the vacuum matching condition. Varying with respect to \(\Delta^*\) and taking the mean-field expectation
value gives
\begin{equation}
 \Delta_{\rm BCS}
 =-\frac{G_{C,0}}{2}
 \left\langle
 \psi^TC\Gamma_{\rm pair}\psi
 \right\rangle.
 \label{eq:appF-minimal-gap-normalization}
\end{equation}

Substituting Eq.~\eqref{eq:appF-minimal-hubbard-stratonovich-identity} into Eq. \eqref{eq:appF-minimal-projected-partition-function} leaves an
exact functional integral over \(\Delta\). At fixed \(\Delta\), the resulting background functional is
\begin{align}
 Z_{\rm MF}(\Delta;\beta,\mu)
 &\equiv
 \exp\left[-\beta{\cal V}
 \frac{|\Delta|^2}{G_{C,0}}\right]
 \int_{\rm AP}[d\psi d\bar\psi]
 \exp\Bigg\{-\int_0^\beta d\tau\int_{\cal V}d^3x
 \left[
 \bar\psi D_E(\mu)\psi
 +\frac{\Delta^*}{2}\psi^TC\Gamma_{\rm pair}\psi
 +\frac{\Delta}{2}\bar\psi\Gamma_{\rm pair}C\bar\psi^T
 \right]\Bigg\}.
 \label{eq:appF-minimal-fixed-background-partition-function}
\end{align}
The exact paired-sector partition function still includes the
functional integral over the auxiliary field, while
\(Z_{\rm MF}(\Delta;\beta,\mu)\) keeps its homogeneous value fixed.
Approximating the remaining integral by a stable saddle gives the mean-field result
\begin{equation}
 Z(\beta,\mu)
 \simeq Z_{\rm MF}(\Delta_\star;\beta,\mu).
 \label{eq:appF-minimal-mean-field-approximation}
\end{equation}
The Hubbard--Stratonovich transformation and the fermion integral at
fixed \(\Delta\) are exact.  Neglecting fluctuations about
\(\Delta_\star\) is the mean-field approximation.

To connect the fixed-background functional with Sec.~\ref{subsec:finite-density-bcs-saddle},
we now rewrite its fermionic exponent in the Nambu--Gorkov variables
defined there. To do so, let us write Eq. \eqref{eq:action_NG_form} in terms of $\psi$. The adjoint of \(\psi_C=C\bar\psi^T\) is
\begin{equation}
 \bar\psi_C=-\psi^TC^{-1},
 \qquad
 \bar\Psi_{\rm NG}
 =\begin{pmatrix}\bar\psi&\bar\psi_C\end{pmatrix}.
 \label{eq:appF-minimal-charge-conjugate-adjoint}
\end{equation}
By the Euclidean continuation of
Eq.~\eqref{eq:bcs-Dtilde-definition}, the charge-conjugate block is
\(CD_E^T(\mu)C^{-1}\).  Consequently, with the spacetime integrations
understood,
\begin{align}
 \int d^4x \;\bar\psi_C\left[CD_E^T(\mu)C^{-1}\right]\psi_C
 =-\int d^4x\; \psi^TD_E^T(\mu)\bar\psi^T=\int d^4 x\;\bar\psi D_E(\mu)\psi.
 \label{eq:appF-minimal-diagonal-identity}
\end{align}
The last equality follows by interchanging the Grassmann fields and integration by parts.  Hence the two diagonal entries in Eq. \eqref{eq:action_NG_form} are two
representations of the same quadratic form.

For the off-diagonal terms we choose the phase of \(C\) so that
\(C^T=-C\) and \(C^{-1}=-C\), and use the Hermitian representative
\(\Gamma_{\rm pair}^\dagger=\Gamma_{\rm pair}\) of the pairing
channel.  The two terms are then
\begin{align}
 \bar\psi\,\Delta\Gamma_{\rm pair}\psi_C
 =\Delta\bar\psi\Gamma_{\rm pair}C\bar\psi^T, \quad \bar\psi_C\,\Delta^*\Gamma_{\rm pair}^\dagger\psi
 =\Delta^*\psi^TC\Gamma_{\rm pair}\psi.
 \label{eq:appF-minimal-off-diagonal-identities}
\end{align}
Combining Eqs.~\eqref{eq:appF-minimal-diagonal-identity} and
\eqref{eq:appF-minimal-off-diagonal-identities} yields
\begin{align}
 \frac12\bar\Psi_{\rm NG}{\cal O}_{\rm BCS}^{E}\Psi_{\rm NG}
 =\bar\psi D_E(\mu)\psi
 +\frac12\left[
 \Delta^*\psi^TC\Gamma_{\rm pair}\psi
 +\Delta\bar\psi\Gamma_{\rm pair}C\bar\psi^T
 \right].
 \label{eq:appF-minimal-explicit-link-to-main-nambu-action}
\end{align}
Here \({\cal O}_{\rm BCS}^{E}\) is the Euclidean continuation of the
operator \({\cal O}_{\rm BCS}\) defined in the main text.  Equation
\eqref{eq:appF-minimal-explicit-link-to-main-nambu-action} is therefore
the continuation of Eq.~\eqref{eq:action_NG_form}. The two
diagonal terms add to the original Dirac action, which explains the
overall factor \(1/2\) in the doubled representation.

The Nambu components are not independent integration variables.  To
evaluate the Gaussian using the independent fields \(\psi\) and
\(\bar\psi\), we use
\begin{align}
 \Psi_{\rm NG}
 =
 \begin{pmatrix}{\bf1}&0\\0&C\end{pmatrix}
 \begin{pmatrix}\psi\\\bar\psi^T\end{pmatrix}, \quad
 \bar\Psi_{\rm NG}
 =
 \begin{pmatrix}\psi^T&\bar\psi\end{pmatrix}
 \begin{pmatrix}0&-C^{-1}\\{\bf1}&0\end{pmatrix}.
 \label{eq:appF-minimal-independent-variable-map}
\end{align}
Direct multiplication gives
\begin{align}
 \begin{pmatrix}0&-C^{-1}\\{\bf1}&0\end{pmatrix}
 {\cal O}_{\rm BCS}^{E}
 \begin{pmatrix}{\bf1}&0\\0&C\end{pmatrix}=
 \begin{pmatrix}
 \Delta^*C\Gamma_{\rm pair}
 &-D_E^T(\mu)
 \\
 D_E(\mu)
 &\Delta\Gamma_{\rm pair}C
 \end{pmatrix}.
 \label{eq:appF-minimal-antisymmetric-matrix}
\end{align}
The antisymmetry condition in
Eq.~\eqref{eq:bcs-pair-matrix-definition}, together with the
conventions above, also gives
\((\Gamma_{\rm pair}C)^T=-\Gamma_{\rm pair}C\).  When the spacetime
arguments of Eq.~\eqref{eq:appF-minimal-antisymmetric-matrix} are
restored, the upper-right entry is
\(-D_E^T(y,x;\mu)\), the lower-left entry is
\(D_E(x,y;\mu)\), and each pairing entry contains
\(\delta_E(x-y)\).  The resulting kernel is antisymmetric under matrix
transposition together with \(x\leftrightarrow y\), and the exponent
in Eq.~\eqref{eq:appF-minimal-fixed-background-partition-function}
becomes
\begin{align}
 \frac12\int_0^\beta d\tau_x\int_{\cal V}d^3x
 \int_0^\beta d\tau_y\int_{\cal V}d^3y\,
 \begin{pmatrix}\psi^T(x)&\bar\psi(x)\end{pmatrix}
 \begin{pmatrix}
 \Delta^*C\Gamma_{\rm pair}\delta_E(x-y)
 &-D_E^T(y,x;\mu)
 \\
 D_E(x,y;\mu)
 &\Delta\Gamma_{\rm pair}C\delta_E(x-y)
 \end{pmatrix}
 \begin{pmatrix}\psi(y)\\\bar\psi^T(y)\end{pmatrix}.
 \label{eq:appF-minimal-independent-quadratic-form}
\end{align}
The mean-field partition function $Z_{\rm MF}(\Delta;\beta,\mu)$ is therefore written in terms of the Euclidean Nambu--Gorkov
 action of Sec.~\ref{subsec:finite-density-bcs-saddle}, as an integral over
independent Grassmann variables. Next, we proceed to evaluate it explicitly.

\paragraph{Dimensionally regulated determinant and Matsubara sum:} We consider the system in the finite volume \({\cal V}\) and
temporarily retain a finite set of modes that is invariant under
\((\omega_n,\boldsymbol k)\mapsto(-\omega_n,-\boldsymbol k)\).  This
step is used only to define the sign of the Grassmann Gaussian.  It is
removed before the continuum determinant is dimensionally continued
and does not impose a physical momentum boundary.  The functional
integral is then an ordinary finite-dimensional Berezin integral
\cite{Berezin1966}.  For independent Grassmann components \(\chi_i\)
and any antisymmetric matrix \(\mathsf A\),
\begin{align}
 \int d\chi_{2N}\cdots d\chi_1\,
 \exp\left(-\frac12\chi_i\mathsf A_{ij}\chi_j\right)\propto\operatorname{Pf}\mathsf A,
 \qquad
 (\operatorname{Pf}\mathsf A)^2=\det\mathsf A.
 \label{eq:appF-minimal-pfaffian-identity}
\end{align}
The proportionality sign allows for the orientation chosen for the
Grassmann measure. To compare every finite-dimensional Pfaffian with
the same regulated configuration \((\Delta,\mu)=(0,0)\), we consider the ratio $Z_{\rm MF}(\Delta;\beta,\mu)/Z_{\rm MF}(0;\beta,0)$. The
orientation sign is independent of \(\Delta\) and \(\mu\) and cancels
in this comparison.  Choosing the Pfaffian branch continuously from
the reference configuration gives
\begin{equation}
 \ln\operatorname{Pf}\mathsf A
 =\frac12\ln\det\mathsf A
 =\frac12\operatorname{Tr}\ln\mathsf A.
 \label{eq:appF-minimal-pfaffian-trace-log}
\end{equation}
This finite-dimensional identity is the origin of the factor
\(1/2\) in the fermionic trace logarithm.

The determinants of the two fixed matrices in
Eq.~\eqref{eq:appF-minimal-independent-variable-map} do not depend on
\(\Delta\) or \(\mu\).  They therefore cancel when the determinant is
divided by the same determinant at \((\Delta,\mu)=(0,0)\).  Equations
\eqref{eq:appF-minimal-antisymmetric-matrix} and
\eqref{eq:appF-minimal-pfaffian-trace-log} then give, up to a
\(\Delta\)- and \(\mu\)-independent constant,
\begin{align}
 \ln Z_{\rm MF}(\Delta;\beta,\mu)
 &=-\beta{\cal V}\frac{|\Delta|^2}{G_{C,0}}+\frac{g_{\rm pair}}{2}
 \operatorname{Tr}_{\rm AP}
 \ln\left[\beta{\cal O}_{\rm BCS}^{E}\right].
 \label{eq:appF-minimal-trace-log}
\end{align}
The trace is over the particle and antiparticle Nambu blocks of one
physical spin/internal copy, the antiperiodic Euclidean frequencies,
and the full momentum space.  The factor \(g_{\rm pair}\) comes from the
direct sum over identical physical copies, which is excluded from the trace.  The auxiliary term occurs once because \(\Delta\)
is the collective field of the projected channel. The factor \(1/2\) assumes that the trace covers the full symmetric momentum domain.

Resolving the pairing projector reduces the trace to the blocks
\(\mathbb H_{k,s}\) already obtained in
Eq.~\eqref{eq:bcs-two-by-two-block-main}.  Modes on which the projector
vanishes are unpaired and are not counted by \(g_{\rm pair}\).
Antiperiodicity fixes the fermionic frequencies to
\begin{equation}
 \omega_n=\frac{(2n+1)\pi}{\beta}.
 \label{eq:appF-minimal-matsubara-frequency}
\end{equation}
For either \(s=\pm\), the determinant of the corresponding block is
\begin{align}
 \det\left[i\omega_n{\bf1}_2-\mathbb H_{k,s}\right]
 &=\det
 \begin{pmatrix}
 i\omega_n-\xi_{k,s}&-\Delta\\
 -\Delta^*&i\omega_n+\xi_{k,s}
 \end{pmatrix}
 \nonumber\\
 &=-\left[
 \omega_n^2+\xi_{k,s}^2+|\Delta|^2
 \right]
 \nonumber\\
 &=-\left[
 \omega_n^2+{\cal E}_s^2(k;\Delta)
 \right].
 \label{eq:appF-minimal-mode-determinant}
\end{align}
At \((\Delta,\mu)=(0,0)\), both branches satisfy
\({\cal E}_s=E_k\), so the same determinant is
\(-[\omega_n^2+E_k^2]\).  The overall minus sign is independent of
\(\Delta\) and \(\mu\) and cancels in the ratio.

Keeping the volume, temporary
mode regulator, and Grassmann measure identical, Eqs.
\eqref{eq:appF-minimal-trace-log} and
\eqref{eq:appF-minimal-mode-determinant} give
\begin{align}
 -\frac{1}{\beta{\cal V}}
 \ln\frac{Z_{\rm MF}(\Delta;\beta,\mu)}
          {Z_{\rm MF}(0;\beta,0)}
 &=\frac{|\Delta|^2}{G_{C,0}}
 -\frac{g_{\rm pair}}{2\beta{\cal V}}
 \sum_n
 \sum_{\boldsymbol k}
 \sum_{s=\pm}
 \ln
 \frac{\omega_n^2+{\cal E}_s^2(k;\Delta)}
      {\omega_n^2+E_k^2}.
 \label{eq:appF-minimal-regulated-partition-ratio}
\end{align}
All field-independent measure factors have canceled in
Eq.~\eqref{eq:appF-minimal-regulated-partition-ratio}.  We now remove
the temporary mode truncation and continue the original Euclidean
determinant to spacetime dimension \(d=4-\epsilon\), consistently with
the convention used in Sec.~\ref{sec:free-fields}
\cite{tHooftVeltman1972}.  Performing the exact frequency sum first
leaves the following spatial representation of that dimensionally
continued determinant.  The spatial integral has
\(d-1=3-\epsilon\) dimensions, and the scale \(\mu_{\rm DR}\) keeps
the grand-potential density at its four-dimensional mass dimension:
\begin{equation}
 \frac{1}{\cal V}\sum_{\boldsymbol k}
 \longrightarrow
 \mu_{\rm DR}^{4-d}
 \int_{\mathbb R^{d-1}}
 \frac{d^{d-1}k}{(2\pi)^{d-1}}.
 \label{eq:appF-minimal-thermodynamic-limit}
\end{equation}
The same factor \(\mu_{\rm DR}^{4-d}\) multiplies the bare local term
\(|\Delta|^2/G_{C,0}\).  The scale \(\mu_{\rm DR}\) is a scalar
subtraction scale needed for dimensional regularization, and should not be confused with the chemical potential $\mu$. Although dimensional regularization preserves restricted covariance, the thermal state and
the chemical potential already select the rest frame in which the
Matsubara representation is written.  This sequential evaluation is
sufficient for the flat homogeneous potential considered here.  A
stress-tensor calculation or a derivative expansion on a general
background must instead be renormalized at the level of the covariant
effective action before metric variation.

The Matsubara logarithmic difference in the second term in Eq. \eqref{eq:appF-minimal-regulated-partition-ratio} is convergent at each fixed momentum.
Differentiating it with respect to \({\cal E}_s(k;\Delta)\) gives
\begin{align}
 &\frac{\partial}{\partial{\cal E}_s(k;\Delta)}
 \left[
 \frac{1}{\beta}\sum_n
 \ln\frac{\omega_n^2+{\cal E}_s^2(k;\Delta)}
          {\omega_n^2+E_k^2}
 \right]=
 \frac{2{\cal E}_s(k;\Delta)}{\beta}
 \sum_n\frac{1}{\omega_n^2+{\cal E}_s^2(k;\Delta)}
 =\tanh\left[
 \frac{\beta{\cal E}_s(k;\Delta)}{2}
 \right].
 \label{eq:appF-minimal-matsubara-derivative}
\end{align}
In the last step we used the fermionic contour sum
\begin{equation}
 \frac{1}{\beta}\sum_n
 \frac{1}{\omega_n^2+{\cal E}_s^2}
 =\frac{1}{2{\cal E}_s}
 \tanh\left(\frac{\beta{\cal E}_s}{2}\right).
 \label{eq:appF-minimal-elementary-matsubara-sum}
\end{equation}
This follows from the two simple poles at
\(\pm{\cal E}_s\); their thermal weights combine to
\((e^{\beta{\cal E}_s}-1)/(e^{\beta{\cal E}_s}+1)\). The logarithmic difference vanishes when
\({\cal E}_s(k;\Delta)=E_k\).  Integrating
Eq.~\eqref{eq:appF-minimal-matsubara-derivative} between these two
values therefore gives 
\begin{align}
 \frac{1}{\beta}\sum_n
 \ln\frac{\omega_n^2+{\cal E}_s^2(k;\Delta)}
          {\omega_n^2+E_k^2}
 =\int_{E_k}^{{\cal E}_s(k;\Delta)}d\varepsilon\,
 \tanh\left(\frac{\beta\varepsilon}{2}\right) ={\cal E}_s(k;\Delta)-E_k
 +\frac{2}{\beta}
 \ln\frac{1+e^{-\beta{\cal E}_s(k;\Delta)}}
          {1+e^{-\beta E_k}}.
 \label{eq:appF-minimal-matsubara-result}
\end{align}
Substituting this result into
Eq.~\eqref{eq:appF-minimal-regulated-partition-ratio}, taking the
continuum limit, and then applying the dimensional continuation
gives
\begin{align}
 -\frac{\mu_{\rm DR}^{4-d}}{\beta{\cal V}}
 \ln\frac{Z_{\rm MF}(\Delta;\beta,\mu)}
          {Z_{\rm MF}(0;\beta,0)}
 &=\mu_{\rm DR}^{4-d}\frac{|\Delta|^2}{G_{C,0}}
 -\frac{g_{\rm pair}}{2}
 \mu_{\rm DR}^{4-d}
 \int_{\mathbb R^{d-1}}\frac{d^{d-1}k}{(2\pi)^{d-1}}
 \sum_{s=\pm}
 \Bigg[
 {\cal E}_s(k;\Delta)-E_k +\frac{2}{\beta}
 \ln\frac{1+e^{-\beta{\cal E}_s(k;\Delta)}}
          {1+e^{-\beta E_k}}
 \Bigg].
 \label{eq:appF-minimal-finite-temperature-difference}
\end{align}

Equation~\eqref{eq:appF-minimal-finite-temperature-difference} fixes
the complete regulated dependence on \(\Delta\) and \(\mu\), up to an additive zero-density
normalization.  More importantly, the momentum integral still contains local
ultraviolet poles. In the following, we determine those poles before imposing the
vacuum matching condition.

\paragraph{Local poles and subtraction scheme:} Taking \(\beta\to\infty\) in
Eq.~\eqref{eq:appF-minimal-finite-temperature-difference} removes the
thermal logarithms.  The remaining determinant difference contains
the combination
\begin{equation}
 {\cal E}_-(k;\Delta)+{\cal E}_+(k;\Delta)-2E_k.
 \label{eq:appF-minimal-zero-temperature-integrand}
\end{equation}
At large \(E_k\), it has the expansion
\begin{align}
 {\cal E}_-+{\cal E}_+-2E_k
 &=\frac{|\Delta|^2}{E_k}
 +\frac{|\Delta|^2\mu^2}{E_k^3}
 -\frac{|\Delta|^4}{4E_k^3}
 +O(E_k^{-5})
 \nonumber\\
 &=\frac{|\Delta|^2}{k}
 +\frac{|\Delta|^2(\mu^2-m_{\rm eff}^2/2)}{k^3}
 -\frac{|\Delta|^4}{4k^3}
 +O(k^{-5}).
 \label{eq:appF-minimal-large-momentum-expansion}
\end{align}
We now comment on the subtraction scheme required by this ultraviolet divergence. Note that the theory we are considering is non-renormalizable from the start, so our focus is on identifying the origin of the running of $G_C$; we comment further on renormalizability at the end of this appendix.

To isolate the divergent structure in Eq. \eqref{eq:appF-minimal-large-momentum-expansion}, we add and subtract the first two terms in the expansion in inverse
powers of \(E_k\).  In the limit \(d\to4\), this gives
\begin{align}
 \mu_{\rm DR}^{4-d}
 \int_{\mathbb R^{d-1}}\frac{d^{d-1}k}{(2\pi)^{d-1}}
 \left[{\cal E}_-+{\cal E}_+-2E_k\right]&=
 \int_{\mathbb R^3}\frac{d^3k}{(2\pi)^3}
 \left[
 {\cal E}_-+{\cal E}_+-2E_k
 -\frac{|\Delta|^2}{E_k}
 -\frac{|\Delta|^2\mu^2-|\Delta|^4/4}{E_k^3}
 \right]
 \nonumber\\
 &\quad
 +|\Delta|^2\mu_{\rm DR}^{4-d}
 \int_{\mathbb R^{d-1}}\frac{d^{d-1}k}{(2\pi)^{d-1}}
 \frac{1}{E_k}
 +\left(|\Delta|^2\mu^2-\frac{|\Delta|^4}{4}\right)
 \mu_{\rm DR}^{4-d}
 \int_{\mathbb R^{d-1}}\frac{d^{d-1}k}{(2\pi)^{d-1}}
 \frac{1}{E_k^3}.
 \label{eq:appF-minimal-dimensional-split}
\end{align}
The first integral on the right-hand side now falls as \(k^{-5}\)
and can be evaluated directly in three dimensions.  The two remaining
integrals follow from the same dimensional master integral used in
Sec.~\ref{sec:free-fields}.  For \(d=4-\epsilon\), define
\begin{equation}
 L_\epsilon\equiv
 \frac{2}{\epsilon}-\gamma_E+\ln(4\pi).
 \label{eq:appF-minimal-pole-definition}
\end{equation}
Then
\begin{align}
 \mu_{\rm DR}^{4-d}
 \int\frac{d^{d-1}k}{(2\pi)^{d-1}}\frac{1}{E_k} &=
 \frac{m_{\rm eff}^2}{8\pi^2}
 \left[
 -L_\epsilon
 +\ln\left(\frac{m_{\rm eff}^2}{\mu_{\rm DR}^2}\right)-1
 \right]+O(\epsilon),
 \label{eq:appF-minimal-DR-integral-one}
 \\
 \mu_{\rm DR}^{4-d}
 \int\frac{d^{d-1}k}{(2\pi)^{d-1}}\frac{1}{E_k^3}
 &=
 \frac{1}{4\pi^2}
 \left[
 L_\epsilon
 +\ln\left(\frac{\mu_{\rm DR}^2}{m_{\rm eff}^2}\right)
 \right]+O(\epsilon).
 \label{eq:appF-minimal-DR-integral-three}
\end{align}
These equations show explicitly how the power and logarithmic
sensitivity found with a sharp spatial cutoff are represented by local poles
in dimensional regularization.

Multiplication by the fermionic factor \(-g_{\rm pair}/2\) shows that
the pole in the determinant contribution to the grand potential is
\begin{equation}
 \left.\Omega_{\rm det}\right|_{\rm pole}
 =\frac{g_{\rm pair}L_\epsilon}{16\pi^2}
 \left[
 (m_{\rm eff}^2-2\mu^2)|\Delta|^2
 +\frac{|\Delta|^4}{2}
 \right].
 \label{eq:appF-minimal-determinant-pole}
\end{equation}
Accordingly, the minimal subtraction (\(\overline{\rm MS}\)) counterterm appearing in the
homogeneous potential is
\begin{equation}
 \delta\Omega_{\rm ct}^{(0)}(\Delta;\mu)
 =-\frac{g_{\rm pair}L_\epsilon}{16\pi^2}
 \left[
 (m_{\rm eff}^2-2\mu^2)|\Delta|^2
 +\frac{|\Delta|^4}{2}
 \right].
 \label{eq:appF-minimal-explicit-counterterm}
\end{equation}
It vanishes at \(\Delta=0\), as required by the state-independent
vacuum normalization in the main text. These terms induce new local terms in the regularized action. Moreover, we know that the chemical potential can
enter the action only through a
background-covariant derivative and that the condensate field $\Delta$ carries twice the fermion number. Using these facts, we reconstruct the local
pair-background action before imposing homogeneity. In flat spacetime, and treating
\(m_{\rm eff}\) as a fixed renormalized parameter, the renormalized
local terms required through fourth order in \(\Delta\) and second
order in derivatives are
\begin{align}
 {\cal L}_{E,\rm loc}^{\rm pair,ren}
 &=\frac{|\Delta|^2}{G_C(\mu_{\rm DR})}
 +Z_\Delta(\mu_{\rm DR})
 (\partial_\alpha+2\mu\delta_{\alpha0})\Delta^*
 (\partial^\alpha-2\mu\delta^{\alpha}_0)\Delta
 +\lambda_\Delta(\mu_{\rm DR})|\Delta|^4.
 \label{eq:appF-renormalized-local-pair-lagrangian}
\end{align}
Here \(Z_\Delta\) is the coefficient of the derivative term, while
\(\lambda_\Delta\) is defined as the coefficient of \(|\Delta|^4\),
without an additional numerical factor. The normalization of
\(\Delta\) remains the one fixed by the Hubbard--Stratonovich identity,
so the fermion-pair vertex is unchanged. Thus
Eq.~\eqref{eq:appF-renormalized-local-pair-lagrangian} gives the local part of the effective action for the composite
background \(\Delta\), whose fluctuations are neglected in the present
mean-field calculation.

The Gaussian Hubbard--Stratonovich identity produces only the first
term in Eq.~\eqref{eq:appF-renormalized-local-pair-lagrangian} at the
classical level.  The determinant in Eq.~\eqref{eq:appF-minimal-determinant-pole}
shows, however, that the one-coupling local action does not absorb all
the ultraviolet poles of the composite background functional.  At the
order considered here, the bare local coefficients must therefore be
written as
\begin{align}
 \frac{1}{G_{C,0}}
 =\mu_{\rm DR}^{-\epsilon}
 \left[
 \frac{1}{G_C(\mu_{\rm DR})}
 +\delta\!\left(\frac{1}{G_C}\right)
 \right], \quad
 Z_{\Delta,0}
 =\mu_{\rm DR}^{-\epsilon}
 \left[Z_\Delta(\mu_{\rm DR})+\delta Z_\Delta\right],\quad
 \lambda_{\Delta,0}
 =\mu_{\rm DR}^{-\epsilon}
 \left[\lambda_\Delta(\mu_{\rm DR})
 +\delta\lambda_\Delta\right].
 \label{eq:appF-bare-renormalized-pair-coefficients}
\end{align}
The common factor \(\mu_{\rm DR}^{-\epsilon}\) converts these
coefficients to the four-dimensional normalization used in
Eq.~\eqref{eq:appF-minimal-finite-temperature-difference}. To find $\delta(G_{C})$, $\delta Z_{\Delta}$, and $\delta \lambda_{\Delta}$, we compare the pole structure of the counterterms with that in Eq. \eqref{eq:appF-minimal-explicit-counterterm}. Before imposing homogeneity, the pole is
the constant-background limit of
\begin{align}
 \delta{\cal L}_{E,\rm ct}^{\rm pair}
 =-\frac{g_{\rm pair}L_\epsilon}{32\pi^2}
 \Big[&
 (\partial_\alpha+2\mu\delta_{\alpha0})\Delta^*
 (\partial_\alpha-2\mu\delta_{\alpha0})\Delta+2m_{\rm eff}^2|\Delta|^2+|\Delta|^4
 \Big].
 \label{eq:appF-minimal-covariant-counterterm-origin}
\end{align}
Comparison with
Eq.~\eqref{eq:appF-bare-renormalized-pair-coefficients} gives
\begin{align}
 \delta\!\left(\frac{1}{G_C}\right)
 =-\frac{g_{\rm pair}m_{\rm eff}^2L_\epsilon}{16\pi^2},\quad 
 \delta Z_\Delta
 =\delta\lambda_\Delta
 =-\frac{g_{\rm pair}L_\epsilon}{32\pi^2}.
 \label{eq:appF-pair-counterterm-coefficients}
\end{align}
For a static homogeneous field, the derivative operator in
Eq.~\eqref{eq:appF-minimal-covariant-counterterm-origin} becomes
\(-4\mu^2|\Delta|^2\).  Equation
\eqref{eq:appF-pair-counterterm-coefficients} then reproduces
Eq.~\eqref{eq:appF-minimal-explicit-counterterm} and cancels the pole in
Eq.~\eqref{eq:appF-minimal-determinant-pole}.

The coefficient
\(G_C^{-1}\) may be fixed by the vacuum inverse pair susceptibility at
a specified reference momentum, \(Z_\Delta\) by its momentum derivative
there, and \(\lambda_\Delta\) by a specified one-particle-irreducible
four-point vertex of the pair field.  These are independent matching
data, even though their one-loop pole residues occur in the correlated
combination displayed above. At this order, the pole residues also determine the subtraction-scale
dependence.  Holding \(m_{\rm eff}\) and \(g_{\rm pair}\) fixed,
requiring
the bare coefficients in
Eq.~\eqref{eq:appF-bare-renormalized-pair-coefficients} to be independent
of \(\mu_{\rm DR}\), and then taking \(\epsilon\to0\), gives
\begin{align}
 \mu_{\rm DR}\frac{d}{d\mu_{\rm DR}}
 \left(\frac{1}{G_C}\right)
 &=-\frac{g_{\rm pair}m_{\rm eff}^2}{8\pi^2},
 \nonumber\\
 \mu_{\rm DR}\frac{dZ_\Delta}{d\mu_{\rm DR}}
 &=\mu_{\rm DR}\frac{d\lambda_\Delta}{d\mu_{\rm DR}}
 =-\frac{g_{\rm pair}}{16\pi^2},
 \nonumber\\
 \mu_{\rm DR}\frac{dG_C}{d\mu_{\rm DR}}
 &=\frac{g_{\rm pair}m_{\rm eff}^2}{8\pi^2}G_C^2.
 \label{eq:appF-pair-beta-functions}
\end{align}
If the finite coefficients are matched at a reference subtraction
scale \(\mu_0\), the solutions are
\begin{align}
 \frac{1}{G_C(\mu_{\rm DR})}
 &=\frac{1}{G_C(\mu_0)}
 -\frac{g_{\rm pair}m_{\rm eff}^2}{8\pi^2}
 \ln\!\left(\frac{\mu_{\rm DR}}{\mu_0}\right),
 \nonumber\\
 Z_\Delta(\mu_{\rm DR})
 &=Z_\Delta(\mu_0)
 -\frac{g_{\rm pair}}{16\pi^2}
 \ln\!\left(\frac{\mu_{\rm DR}}{\mu_0}\right),
 \nonumber\\
 \lambda_\Delta(\mu_{\rm DR})
 &=\lambda_\Delta(\mu_0)
 -\frac{g_{\rm pair}}{16\pi^2}
 \ln\!\left(\frac{\mu_{\rm DR}}{\mu_0}\right).
 \label{eq:appF-pair-running-solutions}
\end{align}
We use the matching convention
\begin{equation}
 Z_\Delta(\mu_0)=\lambda_\Delta(\mu_0)=0.
 \label{eq:appF-pair-matching-condition}
\end{equation}
This condition can be imposed at one scale, but the two coefficients
do not vanish at other scales.  For a static homogeneous background,
the local part of the potential at a general subtraction scale contains
\begin{equation}
 \frac{|\Delta|^2}{G_C(\mu_{\rm DR})}
 -4\mu^2Z_\Delta(\mu_{\rm DR})|\Delta|^2
 +\lambda_\Delta(\mu_{\rm DR})|\Delta|^4.
 \label{eq:appF-general-scale-homogeneous-local-potential}
\end{equation}
The running in Eq.~\eqref{eq:appF-pair-beta-functions} makes the sum of
these terms and the finite determinant independent of
\(\mu_{\rm DR}\) to the order retained.  In order to display the same
form of the potential used in the main text, all thermodynamic
expressions below are evaluated at \(\mu_{\rm DR}=\mu_0\), where
Eq.~\eqref{eq:appF-pair-matching-condition} applies.  At another scale,
the last two terms in
Eq.~\eqref{eq:appF-general-scale-homogeneous-local-potential} and their
derivatives must be retained in the potential, gap equation, and number
density.

\paragraph{Grand potential and gap equation:} We now fix the additive normalization by the same zero-density
condition used in the main text,
\begin{equation}
 -\lim_{\beta\to\infty}
 \frac{1}{\beta{\cal V}}
 \ln Z_{\rm MF}(0;\beta,0)
 \stackrel{\rm match}{=}
 \Omega_{\rm vac}(m_{\rm eff}).
 \label{eq:appF-minimal-vacuum-matching}
\end{equation}
Here \(\Omega_{\rm vac}\) is already renormalized according to the
vacuum prescription adopted in the main text; its dependence on the
vacuum renormalization scale and the corresponding renormalized vacuum
couplings is implicit.  The same paired-sector restriction applies to
\(\Omega_{\rm vac}\), while unpaired vacuum and medium terms are added
separately.  We also assume that the zero-density saddle of the
specified Cooper channel lies at \(\Delta=0\).  If that channel already
condenses at \(\mu=0\), the denominator of the partition-function
ratio and the matching condition must instead be evaluated at the
nonzero zero-density saddle.

Combining Eqs.~\eqref{eq:appF-minimal-dimensional-split}--
\eqref{eq:appF-pair-counterterm-coefficients}, imposing
Eq.~\eqref{eq:appF-pair-matching-condition}, and taking
\(\epsilon\to0\) yields
\begin{align}
 \overline\Omega_{\rm BCS}^{\rm MF}(\Delta;\mu)
 &=\Omega_{\rm vac}(m_{\rm eff})
 +\frac{|\Delta|^2}{G_C(\mu_0)}
 \nonumber\\
 &\quad-
 \frac{g_{\rm pair}}{2}
 \Bigg\{
 \int_{\mathbb R^3}\frac{d^3k}{(2\pi)^3}
 \left[
 {\cal E}_-(k;\Delta)+{\cal E}_+(k;\Delta)-2E_k
 -\frac{|\Delta|^2}{E_k}
 -\frac{|\Delta|^2\mu^2-|\Delta|^4/4}{E_k^3}
 \right]
 \nonumber\\
 &\hspace{24mm}
 +\frac{|\Delta|^2m_{\rm eff}^2}{8\pi^2}
 \left[
 \ln\left(\frac{m_{\rm eff}^2}{\mu_0^2}\right)-1
 \right]
 +\frac{|\Delta|^2\mu^2-|\Delta|^4/4}{4\pi^2}
 \ln\left(\frac{\mu_0^2}{m_{\rm eff}^2}\right)
 \Bigg\}.
 \label{eq:appF-minimal-zero-temperature-potential}
\end{align}
The coupling \(G_C(\mu_0)\) is the renormalized Cooper-channel
coupling fixed by the matching condition described above; its scale
argument is suppressed in the main text.  Equation
\eqref{eq:appF-minimal-zero-temperature-potential} is the explicitly
finite form of Eq.~\eqref{eq:bcs-absolute-potential-main} at the chosen
matching scale. The subtraction and logarithmic terms
are the finite remnants of the renormalized local pair-background
action in this matching convention, denoted $\delta \Omega_{\rm finite}$ in Sec. \ref{subsec:finite-density-bcs-saddle}. For \(m_{\rm eff}=0\), the separate subtraction
integrals in Eqs.~\eqref{eq:appF-minimal-DR-integral-one} and
\eqref{eq:appF-minimal-DR-integral-three} develop artificial infrared
singularities.  The massless limit must then be taken directly in the
dimensionally regulated expression before performing the split in
Eq.~\eqref{eq:appF-minimal-dimensional-split}.

The gap equation in Sec.~\ref{subsec:finite-density-bcs-saddle} is reproduced from the stationary point of the off-shell grand potential functional. Varying
Eq.~\eqref{eq:appF-minimal-zero-temperature-potential} with respect to
\(\Delta^*\) gives
\begin{align}
 0
 &=\frac{\Delta_\star}{G_C(\mu_0)}
 -\frac{g_{\rm pair}\Delta_\star}{2}
 \Bigg\{
 \int_{\mathbb R^3}\frac{d^3k}{(2\pi)^3}
 \left[
 \frac12\sum_{s=\pm}\frac{1}{{\cal E}_s(k;\Delta_\star)}
 -\frac{1}{E_k}
 -\frac{\mu^2-|\Delta_\star|^2/2}{E_k^3}
 \right]
 \nonumber\\
 &\hspace{25mm}
 +\frac{m_{\rm eff}^2}{8\pi^2}
 \left[
 \ln\left(\frac{m_{\rm eff}^2}{\mu_0^2}\right)-1
 \right]
 +\frac{\mu^2-|\Delta_\star|^2/2}{4\pi^2}
 \ln\left(\frac{\mu_0^2}{m_{\rm eff}^2}\right)
 \Bigg\}.
 \label{eq:appF-minimal-gap-equation}
\end{align}
Every integral in this equation is ultraviolet convergent.  It is the
explicitly subtracted form of Eq.~\eqref{eq:bcs-gap-equation-main} at
the matching scale \(\mu_0\). The equilibrium phase is selected
by the global minimum of the grand potential; a local minimum represents
at most a metastable branch. At finite volume, a broken-symmetry saddle
is selected by adding an infinitesimal homogeneous source for the pair
operator, taking the continuum limit, and then removing the source.

At the stationary point, it follows from the chain rule that the chemical-potential derivative is
simplified by the gap equation:
\begin{align}
 \frac{d}{d\mu}
 \overline\Omega_{\rm BCS}^{\rm MF}
 (\Delta_\star(\mu);\mu)
 &=\left.
 \frac{\partial\overline\Omega_{\rm BCS}^{\rm MF}}
 {\partial\mu}\right|_{\Delta_\star}
 +\left.
 \frac{\partial\overline\Omega_{\rm BCS}^{\rm MF}}
 {\partial\Delta}\right|_{\Delta_\star}
 \frac{d\Delta_\star}{d\mu}
 +\left.
 \frac{\partial\overline\Omega_{\rm BCS}^{\rm MF}}
 {\partial\Delta^*}\right|_{\Delta_\star}
 \frac{d\Delta_\star^*}{d\mu}
 \nonumber\\
 &=\left.
 \frac{\partial\overline\Omega_{\rm BCS}^{\rm MF}}
 {\partial\mu}\right|_{\Delta_\star}.
 \label{eq:appF-minimal-envelope-theorem}
\end{align}
At fixed \(m_{\rm eff}\), \(\mu_0\), and matching data,
Eqs.~\eqref{eq:grand-potential-ground-state} and
\eqref{eq:appF-minimal-zero-temperature-potential} then give
\begin{align}
 n_{\rm BCS}^{\rm paired}
 &=\frac{g_{\rm pair}}{2}
 \Bigg\{
 \int_{\mathbb R^3}\frac{d^3k}{(2\pi)^3}
 \left[
 -\frac{\xi_{k,-}}{{\cal E}_-(k;\Delta_\star)}
 +\frac{\xi_{k,+}}{{\cal E}_+(k;\Delta_\star)}
 -\frac{2\mu|\Delta_\star|^2}{E_k^3}
 \right]
 +\frac{\mu|\Delta_\star|^2}{2\pi^2}
 \ln\left(\frac{\mu_0^2}{m_{\rm eff}^2}\right)
 \Bigg\}.
 \label{eq:appF-minimal-number-density}
\end{align}
The subtraction and logarithmic terms are required for the density
obtained from the renormalized potential to be finite.  At a general
subtraction scale, differentiating the \(Z_\Delta\) term in
Eq.~\eqref{eq:appF-general-scale-homogeneous-local-potential} supplies
the additional contribution needed to keep the density independent of
that scale.

We finish this appendix with a discussion of why the finite system of running coefficients in
Eq.~\eqref{eq:appF-pair-beta-functions} does not make the underlying
four-dimensional contact theory renormalizable.  The coupling \(G_C\)
has mass dimension minus two.  For a connected graph constructed from
\(V\) four-fermion vertices, with \(I_F\) internal fermion lines,
\(L\) loops, and \(E_F\) external fermion legs, the superficial degree
of divergence satisfies
\begin{align}
 \omega=4L-I_F
 =4+2V-\frac{3}{2}E_F,\quad 4V=2I_F+E_F,
 \quad
 L=I_F-V+1.
 \label{eq:appF-four-fermion-power-counting}
\end{align}
This is an upper bound on the overall divergence, since symmetries and
Dirac algebra can reduce individual contributions.  Nevertheless, at
fixed \(E_F\), the bound increases by two whenever another
four-fermion vertex is added.  For \(V=2\) and \(E_F=4\), for example,
\(\omega=2\), so a generic one-loop amplitude can contain divergences
proportional to the external momenta squared.  Such terms require
derivative four-fermion counterterms, while higher orders allow
progressively more derivatives and higher-multiplicity fermion
operators.  No finite operator basis closes the perturbative EFT
expansion about the pointlike interaction to all orders
\cite{Buballa2005, Alford:2007xm}.

The determinant evaluated in this appendix is a restricted part of
that theory: it contains one fermion loop in the fixed background
\(\Delta\).  A loop with \(n\) external \(\Delta\) insertions has
superficial degree \(4-n\).  Fermion-number symmetry therefore leaves,
among the divergent \(\Delta\)-dependent flat-space terms at this
order, only the quadratic, two-derivative, and quartic local structures
collected in
Eq.~\eqref{eq:appF-renormalized-local-pair-lagrangian}.  This is why the
three running equations close within the present one-fermion-loop
background calculation.  In fermionic variables, the terms
proportional to \(Z_\Delta\) and \(\lambda_\Delta\) correspond
schematically to derivative four-fermion and eight-fermion
interactions.  Pair-field fluctuations and general multiloop graphs
require the usual tower of higher-dimensional operators.

Dimensional regularization does not provide any information about the coefficients
of the higher-dimensional operators nor an ultraviolet
completion of the contact interaction. The running in
Eq.~\eqref{eq:appF-pair-beta-functions} only relates values at
different subtraction scales within the EFT range and the mean-field approximation. Moreover, note that on a curved background,
covariance also permits a local term proportional to
\(R|\Delta|^2\), whose finite coefficient is not determined by the
flat homogeneous calculation.

\section{FLRW Geometry and Higher-Curvature Variations}
\label{app:flrw-geometry}

This appendix collects the geometric identities used in
Sec.~\ref{sec:flrw}. 

\paragraph{Connection and curvature tensor:} For the spatially flat metric
\begin{equation}
ds^2=-dt^2+a(t)^2\delta_{ij}dx^idx^j,
\qquad
H=\frac{\dot a}{a},
\end{equation}
we have
\begin{equation}
g_{00}=-1,
\quad
g_{ij}=a^2\delta_{ij},
\quad
g^{00}=-1,
\quad
g^{ij}=a^{-2}\delta^{ij},
\quad
\sqrt{-g}=a^3.
\end{equation}
The Christoffel symbols are
\begin{align}
\Gamma^0_{ij}
&=
\frac12\partial_0g_{ij}
=a\dot a\,\delta_{ij}=Hg_{ij},
\nonumber\\
\Gamma^i_{0j}=\Gamma^i_{j0}
&=
\frac12g^{ik}\partial_0g_{kj}
=H\delta^i{}_j,
\label{eq:appG-christoffels}
\end{align}
and all other independent components vanish.

Using
\begin{equation}
R^\rho{}_{\sigma\mu\nu}
=
\partial_\mu\Gamma^\rho_{\nu\sigma}
-\partial_\nu\Gamma^\rho_{\mu\sigma}
+\Gamma^\rho_{\mu\lambda}\Gamma^\lambda_{\nu\sigma}
-\Gamma^\rho_{\nu\lambda}\Gamma^\lambda_{\mu\sigma},
\end{equation}
the independent Riemann components are
\begin{align}
R^0{}_{i0j}
&=
\partial_0(Hg_{ij})-H^2g_{ij}
=(\dot H+H^2)g_{ij},
\nonumber\\
R_{0i0j}
&=-(\dot H+H^2)g_{ij},
\nonumber\\
R_{ijkl}
&=
H^2\left(g_{ik}g_{jl}-g_{il}g_{jk}\right).
\label{eq:appG-riemann-components}
\end{align}
Contracting the first and third indices gives
\begin{align}
R_{00}&=-3(\dot H+H^2),
\nonumber\\
R_{ij}&=(\dot H+3H^2)g_{ij},
\nonumber\\
R&=6(\dot H+2H^2),
\nonumber\\
G_{00}&=3H^2,
\qquad
G_{ij}=-(2\dot H+3H^2)g_{ij}.
\label{eq:appG-ricci-einstein}
\end{align}

\paragraph{Quadratic invariants and the traced spinor coefficient:}

The Ricci contraction follows directly from
Eq.~\eqref{eq:appG-ricci-einstein}:
\begin{align}
R_{\mu\nu}R^{\mu\nu}
&=
9(\dot H+H^2)^2
+3(\dot H+3H^2)^2=
12\left(\dot H^2+3H^2\dot H+3H^4\right).
\end{align}
Similarly, the temporal and spatial Riemann components give
\begin{align}
R_{\mu\nu\rho\sigma}R^{\mu\nu\rho\sigma}
&=
12(\dot H+H^2)^2+12H^4=
12\left(\dot H^2+2H^2\dot H+2H^4\right),
\nonumber\\
R^2
&=
36\left(\dot H^2+4H^2\dot H+4H^4\right).
\label{eq:appendix-FLRW-invariants}
\end{align}
These combinations imply
\begin{equation}
C_{\mu\nu\rho\sigma}C^{\mu\nu\rho\sigma}=0,
\qquad
E_4=24H^2(\dot H+H^2).
\label{eq:appG-Weyl-Euler}
\end{equation}

For a homogeneous scalar \(F(t)\),
\begin{align}
\Box F
&=
\frac1{a^3}\frac{d}{dt}
\left(a^3g^{00}\dot F\right)=-\ddot F-3H\dot F.
\label{eq:appG-boxF}
\end{align}
With \(R=6(\dot H+2H^2)\), this gives
\begin{align}
\Box R
&=
-6\dddot H-24\dot H^2-42H\ddot H-72H^2\dot H.
\label{eq:appG-boxR}
\end{align}

Substitution of Eq.~\eqref{eq:appendix-FLRW-invariants} into the traced
spinor coefficient in Eq.~\eqref{eq:appB-spinor-coefficients} gives
\begin{align}
f_2^\psi\big|_{\rm FLRW}
&=
\frac{11}{15}H^2(\dot H+H^2)-\frac1{30}\Box R
\nonumber\\
&=
\frac{11}{360}E_4-\frac1{30}\Box R.
\label{eq:appG-A2spin-FLRW}
\end{align}
This calculation makes the four-dimensional bulk statement in the main
text explicit: after subtraction, a constant-coefficient \(E_4\) term
is topological and a constant-coefficient \(\Box R\) term is a boundary
term under the stated boundary conditions.  Neither contributes to an
independent spatially flat FLRW bulk energy-momentum tensor.

\paragraph{Variation of an independent \texorpdfstring{\(R^2\)}{R-squared} term:}
For comparison, the metric variation of a separately renormalized
\(R^2\) coupling is
\begin{equation}
\mathcal H_{\mu\nu}^{(1)}
=
2RR_{\mu\nu}-\frac12g_{\mu\nu}R^2
+2\left(g_{\mu\nu}\Box-\nabla_\mu\nabla_\nu\right)R.
\label{eq:appG-H1-definition}
\end{equation}
With the manuscript convention
\(T_{\mu\nu}=-2(\sqrt{-g})^{-1}\delta W/\delta g^{\mu\nu}\), a term
\(+\beta_R\int\sqrt{-g}\,R^2\) therefore contributes
\(T_{\mu\nu}^{R^2}=-2\beta_R\mathcal H_{\mu\nu}^{(1)}\).
For a homogeneous scalar,
\begin{equation}
\nabla_0\nabla_0R=\ddot R,
\qquad
\nabla_i\nabla_jR=-Hg_{ij}\dot R.
\end{equation}
The \(00\) component is therefore
\begin{align}
\mathcal H_{00}^{(1)}
&=
-6R(\dot H+H^2)+\frac12R^2
+2(-\Box R-\ddot R)
\nonumber\\
&=
-6R(\dot H+H^2)+\frac12R^2+6H\dot R
\nonumber\\
&=
18\left(2H\ddot H+6H^2\dot H-\dot H^2\right).
\label{eq:appG-H100}
\end{align}
The spatial component is
\begin{align}
\mathcal H_{ij}^{(1)}
&=\left[
2R(\dot H+3H^2)-\frac12R^2
+2(\Box R+H\dot R)\right]g_{ij}
\nonumber\\
&=
-6\left(
2\dddot H+12H\ddot H+9\dot H^2+18H^2\dot H
\right)g_{ij}.
\label{eq:appG-H1ij}
\end{align}
Equations~\eqref{eq:appG-H100} and \eqref{eq:appG-H1ij} give rise to the energy density and pressure contributions of a finite \(R^2\) coupling, respectively.

\end{document}